\documentclass[nolineno]{aastex61}

\usepackage{graphicx}
\usepackage{verbatim,xspace,booktabs,xcolor}
\usepackage{natbib}
\usepackage{amsmath,amssymb,upgreek,mathtools}
\usepackage{wasysym}   
\let\splitbox\relax
\usepackage[Export,splitbox=false]{adjustbox}
\usepackage{tabularx}

\DeclareMathOperator\atan{atan}

\usepackage{tikz}
\usetikzlibrary{arrows.meta, positioning, shapes.geometric}
\usepackage{hyperref} 
\tikzset{
every node/.style={font=\small}, 
  block/.style={
    rectangle, draw, rounded corners, align=center,
    text width=3.8cm, 
    inner sep=2pt,    
    minimum height=7mm
  },
  decision/.style={
    diamond, draw, align=center, aspect=2.5,
    text width=4.0cm,
    inner sep=1.5pt
  },
  arrow/.style={-{Stealth[length=2mm,width=1.4mm]}, thick}
}

\definecolor{amethyst}{rgb}{0.6, 0.4, 0.8}
\definecolor{green}{rgb}{0.55, 0.71, 0.0}
\definecolor{apricot}{rgb}{0.98, 0.81, 0.69}
\definecolor{auburn}{rgb}{0.43, 0.21,0.1}
\definecolor{babyblueeyes}{rgb}{0.63, 0.79, 0.95}
\definecolor{bittersweet}{rgb}{1.0, 0.44, 0.37}

\newcommand{\ccol}[1]{\textcolor{black}{#1}}

\newcommand\ddfrac[2]{\ensuremath{\frac{\displaystyle #1}{\displaystyle #2}}}  

\submitjournal{ApJS}
\shorttitle{Analytical Solution for Muon Calibration with Dual-Mirror Telescopes}
\shortauthors{M. Gaug, V. Gir\'aldez-Segal\`as, F. Redmen}

\begin{document}

\title{Using Muon Rings for the Calibration of the Cherenkov Telescope Array: An Analytical Solution for the Dual-Mirror Telescope Using Vector Geometry }

\correspondingauthor{Markus Gaug}
\email{markus.gaug@uab.cat}

\author{Markus~Gaug}
\affiliation{Unitat de F\'isica de les Radiacions, Departament de F\'isica, and CERES-IEEC, Universitat Aut\`onoma de Barcelona, E-08193 Bellaterra, Spain}
\author{Víctor~Gir\'aldez-Segal\`as}
\affiliation{Unitat de F\'isica de les Radiacions, Departament de F\'isica, Universitat Aut\`onoma de Barcelona, E-08193 Bellaterra, Spain}
\author{Fiona Redmen}
\affiliation{Unitat de F\'isica de les Radiacions, Departament de F\'isica, and CERES-IEEC, Universitat Aut\`onoma de Barcelona, E-08193 Bellaterra, Spain}





\begin{abstract}
The analysis of ring images produced by muons in Imaging Atmospheric Cherenkov Telescopes (IACTs) provides a powerful and precise method for calibrating the optical throughput of the instrument and monitoring its optical point-spread function. To date, analytical solutions have been derived for single–mirror telescopes with reflectors assumed flat.
However, a complete analytical description of the Cherenkov light produced by muons and detected by a dual-mirror telescope — accounting for both the secondary mirror and the camera — has remained elusive, owing to the complexity of the problem.


In this work, we derive such a solution using a vector-geometry formalism supported by symbolic manipulation and Taylor expansions performed with the computer algebra system \textit{SageMath}. We validate the formalism against known analytical solutions in simpler configurations and, for more complex terms, against limiting cases exhibiting the expected physical behavior. The behavior of the full solution is illustrated visually by varying the relevant parameters. 

The largest effects were found in the shadowing of Cherenkov light produced by inclined muons in dual-mirror telescopes,  particularly for the Schwarzschild-Couder Telescope (SCT) design with baffles surrounding the secondary mirror. Deviations of up to 40\% are observed relative to  previously employed methods. 


As a by-product, we derive the first-order correction to the maximum emission height of Cherenkov photons emitted by a muon, arising from the curvature of the primary mirror — an effect neglected in previous studies - as well as the impact of coma aberration on the muon rings in single-mirror parabolic telescopes. Our results are directly applicable to muon-based calibration of the Cherenkov Telescope Array Observatory (CTAO).

\end{abstract}
%
%


\section{Introduction}

Muon ring calibration for Imaging Atmospheric Cherenkov Telescopes (IACTs) was first proposed by \citet{HILLAS:JPGPP1990a} and \citet{rowell}, and later developed in detail by \citet{vacanti}.
A comprehensive review of the method, including its accuracy and limitations in the context of the Cherenkov Telescope Array Observatory (CTAO), was presented in \citet{GaugMuons:2019}.

The Cherenkov Telescope Array Observatory (CTAO)\footnote{see also \url{https://www.ctao.org}}~\citep{cta,ctaconcept} is the next-generation ground-based gamma ray observatory, comprising more than seventy IACTs distributed over two sites: CTAO-South at Atacama desert near Paranal, Chile, and CTAO-North at the Observatorio del Roque de los Muchachos (ORM) on the Canary Island of La Palma, Spain. Both sites are located at altitudes of approximately 2200~m above sea level. CTAO will deliver a substantial performance improvement over current facilities~\citep{bernloehr2013,Hassan:2017,Maier:2019}. 

More than half of the CTAO telescopes will employ dual-mirror optical designs inspired by the Schwarzschild-Couder aplanatic configuration~\citep{Schwarzschild:1905,Couder:1926}, which has been adapted for IACT applications~\citep{Vassiliev2007,canestrari2013,Rulten:2016,Sironi:2017kjv,White:2021}. In addition, upgrades of some single-mirror telescopes to dual-mirror configurations are under consideration~\citep{vasiliev2013,DiVenere:2023}. 
\ccol{The performance of dual-mirror telescopes has been recently reported for the ASTRI-1
telescope within the ASTRI Mini-Array~\citep{Crestan:2025VJ}.}

Dual-mirror IACTs offer wide fields-of-view approaching 10$^\circ$, reduced plate scale, and improved optical performance, enabling the efficient use of cameras equipped with Geiger-mode avalanche photodiodes~\citep[G-APDs, or SiPMs;][]{biland2014,White:2017ji,Lombardi:2020}.  

To meet the  CTAO requirement of a 10\% accuracy on the global energy scale, the total transmission of Cherenkov photons through the optical system and their conversion to photoelectrons in the camera must be known to better than 5\%, given that systematic uncertainties related to atmospheric transmission are unlikely to be reduced below approximately 8\% within reasonable efforts. Cherenkov light emitted by local muons provides a continuously available and well-understood calibration source, with an intrinsic intensity known to the precision with which the Cherenkov angle can be determined~\citep{Navas:2024}.
Since muon calibration can be performed concurrently with regular science observations, it has been selected as the primary method for monitoring the optical bandwidth of all CTAO telescopes~\citep{gaugSPIE2014}.

Classical muon calibration of IACTs has so far been applied to single-mirror telescopes~\citep{HILLAS:JPGPP1990a,rowell,vacanti,rose,jiang:1993,shayduk,meyermuons,goebel,Tyler:2003,bolzphd,chalmecalvet2014,Noethe:2016}. 
\ccol{Preliminary 
results on a
muon-based calibration for the ASTRI-1 dual-mirror telescope have been presented recently in ~\citet{Mineo:2025}}. Its use for dual-mirror telescopes has been, however, hampered by the absence of a precise analytical prediction for the amount of light received along the ring recorded by the camera, in particular for inclined muons and for muons with non-zero impact parameters relative to the center of the primary mirror. In the single-mirror case, this prediction is provided by the simple Eqs.~5 and~6 of \citet{vacanti}; \citep[see also Eqs.~7 and~8 in][]{GaugMuons:2019}, for which no direct analogue has previously existed for dual-mirror optical systems. 
An initial attempt in that direction was presented by~\citet{mitchellphd}, who introduced an average shadowing factor for all muons, independent of their inclination or impact distance.  
Notably, for the  \ccol{first}  operational dual-mirror \ccol{IACT}, the ASTRI-Horn telescope~\citep{Lombardi:2020}, no explicit correction for the shadowing introduced by the secondary mirror has been applied to muon Cherenkov light~\citep{Strazzeri:2013,mineo2016,Mineo:2019}. Instead, the formalism of \citet{vacanti} was adopted without modification as if no secondary mirror was present, due to the difficulty of deriving a suitable analytical description.

In this work, we derive such an analytical solution, based on vector geometry and the extensive use of the computer algebra system \textit{SageMath}~\citep{sagemath} to simplify and Taylor-expand the derived mathematical
expressions. 
\ccol{When necessary, the expression were expanded with respect to the small quantities: muon incidence and Cherenkov angles in radians,  and the mirror curvature. Expansions are kept up to first or second order in these parameters, ensuring that the expanded expression deviates less than 1\% from the full solution.}
The formalism is validated against known analytical solutions in limiting single-mirror cases, and for more complex solutions, 
through consistency checks in parameter limits where the expected physical behavior is recovered.

\clearpage
\section{Definition of the parameters and the problem}
\label{sec:definition}

\begin{figure}
    \centering
    \includegraphics[width=0.4\linewidth,trim={0cm 0cm 0cm 0cm},clip]{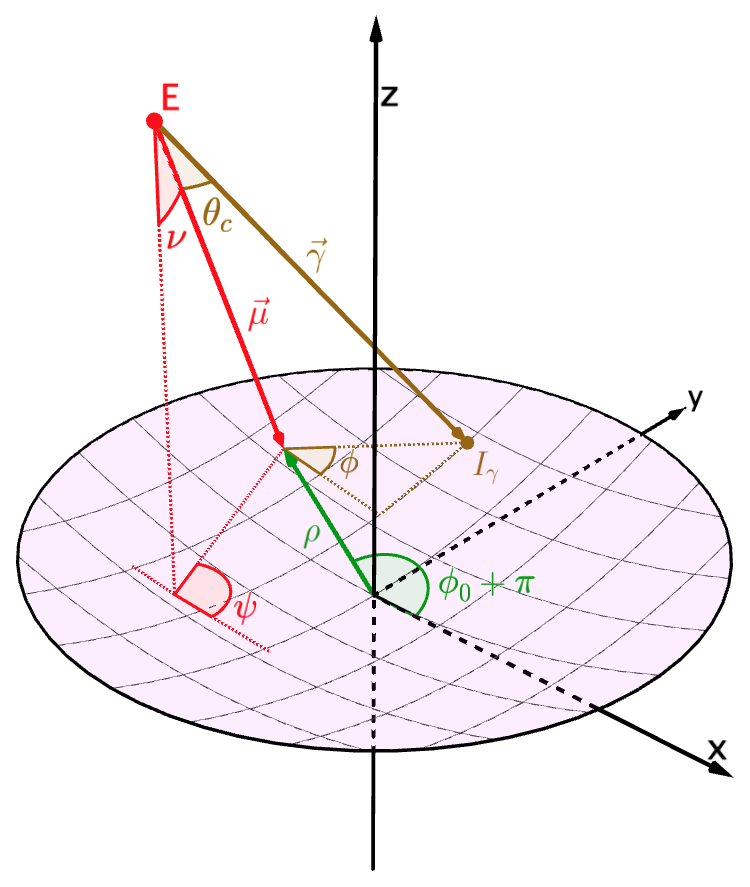}
    \caption{Sketch of the parameters used.}
    \label{fig:descriptionmethod}
\end{figure}

We consider an inclined muon characterized by an inclination angle $\nu$ and an azimuthal angle $\psi$, defined as the projection of the muon directional vector onto the ground plane.  The normalized velocity vector of the muon is then given by:
\begin{equation}
  \boldsymbol{\vec{\mu}} = \left(
  \begin{array}{c}
    \sin(\nu)\cdot\cos(\psi) \\
    \sin(\nu)\cdot\sin(\psi)\\
    -\cos(\nu) \\
    \end{array}
  \right)  \quad,  \label{eq:M}
  \end{equation}
where a coordinate system originating at the pole of the primary mirror is used. See also Fig.~\ref{fig:descriptionmethod} for a sketch of the vectors, angles and distances employed in the following.\\

The velocity vector $\boldsymbol{\vec{\gamma}}$ of the Cherenkov photon is constructed in three steps. 
First, a light ray is generated by a vertically incident muon travelling along the $x$-axis emitting radiation with the Cherenkov angle $\theta_c$: 
\begin{equation}
\boldsymbol{\vec{\gamma}_x}  = 
  \begin{pmatrix}
    {}~\sin(\theta_c) \\
    0 \\
    -\cos(\theta_c)
  \end{pmatrix} \quad.
\end{equation}
This light ray is then rotated around the $y$-axis to match the muon incidence angle $\nu$, and subsequently rotated around the $z$-axis by the azimuthal component $\psi$ of the muon incidence angle:

\begin{equation}
\boldsymbol{\vec{\gamma}_0}  = 
\boldsymbol{\vec{\gamma}_x }
\times
  \begin{pmatrix}
    \cos(\nu) & 0 & -\sin(\nu) \\
    0 & 1 & 0 \\
    \sin(\nu) & 0 & {}~\cos(\nu) 
  \end{pmatrix}
   \times
  \begin{pmatrix}
    \cos(\psi) & -\sin(\psi) &  0 \\
    \sin(\psi) & {}~\cos(\psi) & 0\\
    0 & 0 & 1 
  \end{pmatrix}
   \quad.
  \end{equation}
Finally, we apply Rodrigues' rotation formula to rotate $\boldsymbol{\vec{\gamma}_0}$ around $\boldsymbol{\vec{\mu}}$ by an angle ($\phi-\psi$)  in clockwise direction, as seen from the telescope: 

\begin{equation}
\boldsymbol{\vec{\gamma}} = 
\boldsymbol{\vec{\gamma}_0} \cdot \cos(\phi-\psi) 
- \boldsymbol{\vec{\mu}} \times   
  \boldsymbol{\vec{\gamma}_0} \sin(\phi-\psi) 
  + \boldsymbol{\vec{\mu}} \cdot \left( \boldsymbol{\vec{\mu}} \cdot \boldsymbol{\vec{\gamma}_0} \right) 
  \cdot \big( 1 - \cos\left(\phi-\psi\right)\big) \quad.
\end{equation}
Note that $\boldsymbol{\vec{\mu}}$ was defined as a unit vector in Eq.~\ref{eq:M}.
The resulting vector $\boldsymbol{\vec{\gamma}}$ is then given by: 
\begin{subequations}
\begin{align}
\boldsymbol{\vec{\gamma}}  &= 
\begin{pmatrix}
\cos\left(\psi\right) \cos\left(\theta_{c}\right) \sin\left(\nu\right) + \frac{1}{2} \, {\left({\left(\cos\left(\nu\right) - 1\right)} \cos\left(\phi\right) \cos\left(2 \, \psi\right) + {\left(\cos\left(\nu\right) - 1\right)} \sin\left(\phi\right) \sin\left(2 \, \psi\right) + {\left(\cos\left(\nu\right) + 1\right)} \cos\left(\phi\right)\right)} \sin\left(\theta_{c}\right) \\
\cos\left(\theta_{c}\right) \sin\left(\nu\right) \sin\left(\psi\right) - \frac{1}{2} \, {\left({\left(\cos\left(\nu\right) - 1\right)} \cos\left(2 \, \psi\right) \sin\left(\phi\right) - {\left(\cos\left(\nu\right) - 1\right)} \cos\left(\phi\right) \sin\left(2 \, \psi\right) - {\left(\cos\left(\nu\right) + 1\right)} \sin\left(\phi\right)\right)} \sin\left(\theta_{c}\right)\\
-\cos\left(\nu\right) \cos\left(\theta_{c}\right) + {\left(\cos\left(\phi\right) \cos\left(\psi\right) \sin\left(\nu\right) + \sin\left(\nu\right) \sin\left(\phi\right) \sin\left(\psi\right)\right)} \sin\left(\theta_{c}\right)
\end{pmatrix}   \\
&= \quad \begin{pmatrix}
\theta_c \cdot \cos(\phi) + \nu \cdot \cos(\psi)  \\
\theta_c \cdot \sin(\phi) + \nu \cdot \sin(\psi) \\
 - 1
 \end{pmatrix}   + \dfrac{1}{2}
 \begin{pmatrix}
 0 \\
 0 \\ 
\nu^2 + \theta_c^2  + 2\,\nu \cdot \theta_c \cdot \cos(\phi - \psi)
 \end{pmatrix}  + 
 \mathrm{h.c.} ~\left( O(\nu^3,\theta_c^3) \right)  \label{eq:Gamma}  \\ 
&:= \quad 
 \begin{pmatrix}
 \quad  \gamma_x \quad\quad  \\
 \quad  \gamma_y \quad\quad  \\ 
 \quad  \gamma_z \quad \quad
 \end{pmatrix}
 \quad,  \label{eq:Gamma_abbr}
\end{align}
\end{subequations}
\noindent
where, in the second line, a Taylor expansion up to the second order around $\theta_c=0$ and $\nu=0$ has been performed, and, in the third line, the three components of the Taylor-expanded vector are defined. \\

In a next step, we introduce the primary mirror of the telescope. The muon impact parameter on the primary mirror is defined in polar coordinates by $\rho$ and $\pi+\phi_0$\,\footnote{Recall the convention that the impact angle $\phi_0$ is defined as the angle corresponding to the longest chord on the telescope mirror~\citep[see Fig.~1 of][]{GaugMuons:2019} and the definition originally proposed in \citet{vacanti}.}. The adopted coordinate conventions are illustrated in Fig.~\ref{fig:descriptionmethod}. The origin of the coordinate system is located at the center of the primary mirror. 
Using a function $z(\rho)$ to describe the curvature of the primary mirror, which to first order can be approximated to $z(\rho)\approx c\cdot \rho^2$, the muon impact point on the mirror, $\boldsymbol{I}_\mu$, has the coordinates:
\begin{equation}
  \boldsymbol{I}_\mu = 
  \begin{pmatrix}
    \rho \cdot \cos(\phi_0+\pi)\\
    \rho \cdot \sin(\phi_0+\pi)\\
    c\cdot \rho^2
  \end{pmatrix}
   \quad. \label{eq:Imu}
  \end{equation}
 We then denote by $l$ the distance between the emission point of the Cherenkov photon and the muon impact point on the mirror, and assign the following coordinates to the emission point $\boldsymbol{E}$ (recalling that $\boldsymbol{\vec{\mu}}$ is a unit vector):
 \begin{subequations}
    \begin{align}
\boldsymbol{E} &= \boldsymbol{I}_\mu - l \cdot \boldsymbol{\vec{\mu}}    \\
        &= 
  \begin{pmatrix}
    \rho \cdot \cos(\phi_0+\pi) - l\cdot     \sin(\nu)\cdot\cos(\psi) \\
    \rho \cdot \sin(\phi_0+\pi) - l\cdot     \sin(\nu)\cdot\sin(\psi)\\
    c\cdot \rho^2 + l\cdot  \cos(\nu) \\
  \end{pmatrix}
   \quad. \label{eq:O}
  \end{align}
  \end{subequations}
  The Cherenkov photon is then emitted at point $\boldsymbol{E}$, propagates a given distance $m$ along the unit vector $\boldsymbol{\vec{\gamma}}$, and finally intersects the primary mirror at the impact point $\boldsymbol{I}_\gamma = (x,y,c\cdot(x^2+y^2))$.

\clearpage
\section{Solution for the single-mirror telescope}

\noindent
 For a single-mirror telescope, we must solve the following vector equation: 
 \begin{equation}
\boldsymbol{E} + m \cdot \boldsymbol{\vec{\gamma}} = \boldsymbol{I}_\gamma 
  \end{equation}
for the free parameters: photon propagation length $m$ and photon impact point coordinates $\boldsymbol{I}_\gamma = \left(x,y,c\cdot x^2+ c\cdot y^2\right)$. 
The solution for the two parameters of interest,  $x$ and $y$, is then: 
\begin{subequations}
\label{eq:x}
 \begin{align}
x &= \Big( l \cdot\theta_{c} \cdot\cos\left(\phi\right) - \rho \cdot\cos\left(\phi_{0}\right) \Big) 
+ \frac{l \cdot \theta_c}{2} \cdot \Big( 2\,\nu \cdot \theta_c \cdot \cos(\psi) + \theta_c^2\cdot\cos(\phi) 
+ \nu \cdot \theta_c \cdot \cos(2\,\phi - \psi) 
\Big) 
+ {} \label{eq:x_zeroth}\\ 
 & \qquad {} + c \cdot l \cdot\theta_c\cdot 
   \Big( 
2\,\rho \cdot\cos\left(\phi - \phi_{0}\right)  -l \cdot \theta_{c}  
 \Big) \cdot \Big( \nu  \cdot\cos\left(\psi\right)  + \theta_c \cdot \cos\left(\phi\right) \Big) +  {} \label{eq:x_first}\\ 
& \qquad {}+  \mathrm{h.c.} ~\left( O(c^2,\nu^3,\theta_c^3) \right) \quad,
\end{align}
\end{subequations}
\vspace{-7mm}
\begin{subequations}
\label{eq:y}
\begin{align}
y &= \Big( l \cdot\theta_{c} \cdot\sin\left(\phi\right) - \rho \cdot\sin\left(\phi_{0}\right) \Big) 
+ \frac{l \cdot \theta_c}{2} \cdot \Big(2\, \nu \cdot \theta_c \cdot \sin(\psi) + \theta_c^2\cdot\sin(\phi) 
+ \nu \cdot \theta_c \cdot \sin(2\,\phi - \psi)
\Big) 
+ {} \label{eq:y_zeroth}\\ 
& \qquad {} + c \cdot l \cdot \theta_c\cdot   
\Big(
2\, \rho \cdot \cos\left(\phi - \phi_{0}\right)  -l \cdot \theta_{c}  
 \Big) \cdot \Big( \nu  \cdot\sin\left(\psi\right) +   
 \theta_c \cdot \sin\left(\phi\right) \Big ) + {} \label{eq:y_first}\\
 & \qquad {} +  \mathrm{h.c.} ~\left( O(c^2,\nu^3,\theta_c^3) \right) \quad. 
\end{align}
\end{subequations}
Here, the zeroth-order solutions are given in the first lines, \ccol{Eqs~\ref{eq:x_zeroth} and~\ref{eq:y_zeroth}}, respectively. No first-order corrections arise; the next-to-leading-order corrections, \ccol{Eqs~\ref{eq:x_first} and~\ref{eq:y_first},} appear in the second lines.  \\

The photon with direction $\boldsymbol{\vec{\gamma}}$ is then reflected at the mirror at its impact point using the surface normal vector $\boldsymbol{\vec{n}}$ at that point, $\boldsymbol{\vec{n}} = \left(-2\,c \cdot x,-2\,c \cdot y,1\right)$. After expanding to first order in $c,\nu$, and $\theta$, the direction vector of the reflected photon is given by: 
\begin{subequations}
    \begin{align}
\boldsymbol{\vec{\gamma}}_\mathrm{refl} &= 
\boldsymbol{\vec{\gamma}} - 2\,\frac{(\boldsymbol{\vec{\gamma}}\cdot\boldsymbol{\vec{n}})}{|\boldsymbol{n}|^2}\cdot\boldsymbol{\vec{n}} \\
&=
  \begin{pmatrix}
   \gamma_x\\
   \gamma_y\\
   1
  \end{pmatrix} 
  + 4\,c \cdot 
   \begin{pmatrix}
   -x   \\
   -y   \\
  x \cdot \gamma_x
  + 
  y \cdot \gamma_y
  \end{pmatrix}  
  - 8\,c^2 \cdot
   \begin{pmatrix}
   x^2\cdot\gamma_x + x\,y\,\gamma_y   \\
    y^2\cdot\gamma_y + x\,y\,\gamma_x  \\
  x^2+y^2
  \end{pmatrix}  
  + \mathrm{h.c.} ~\left( O(c^3,\nu^3,\theta_c^3,c^2\theta_c,c^2\nu) \right)
   \quad.
  \end{align}
  \end{subequations}
Finally, the reflected photon is propagated to intersect a focal plane, for example the camera plane of a Large-Sized-Telescope (LST). To this end, we solve the following vector equation: 
 \begin{equation}
\boldsymbol{I}_\gamma + p \cdot \boldsymbol{\vec{\gamma}_\mathrm{refl}} = \boldsymbol{C}_\gamma 
  \end{equation}
\noindent 
for the free parameters: photon propagation length $p$, and the photon impact point coordinates on the camera, $\boldsymbol{C}_\gamma = \left(x_\mathrm{cam},y_\mathrm{cam},1/(4\,c)\right)$, \ccol{coinciding, in this study, with the primary mirror focal plane,} with the result: 
\begin{subequations}
\begin{align}
 \begin{pmatrix}
x_\mathrm{cam} \\
y_\mathrm{cam} 
\end{pmatrix} &= \frac{1}{4\,c} \cdot 
\begin{pmatrix}
\gamma_x \\ 
\gamma_y
\end{pmatrix} + \label{eq:muonring} \\ 
{} & {} + c \cdot 
\begin{pmatrix}
 2\,x^2\cdot\gamma_x+2\,x\,y\cdot\gamma_y + r^2\cdot\gamma_x\\ 
 2\,y^2\cdot\gamma_y+2\,x\,y\cdot\gamma_x + r^2\cdot\gamma_y
\end{pmatrix} 
 + \label{eq:muonring:coma} \\[1mm] 
{} & {} \qquad + \mathrm{h.c.} ~\left( O(c^2,\nu^3,\theta_c^3) 
\right) \qquad.
\end{align}
\end{subequations}
\noindent
Here, $x_\mathrm{cam}$ and $y_\mathrm{cam}$ are calculated in the telescope coordinate system, with the first line of Eq.~\ref{eq:muonring} corresponding to the expected leading-order Gaussian optics approximation. Note that the first-order corrections of $O(\nu^2/c,\theta^2/c,(\nu\cdot\theta_c)/c)$  cancel identically, as expected for a parabolic mirror. The third-order correction in Eq.~\ref{eq:muonring:coma} represents  the contribution from coma aberration. To verify whether this correction exhibits the expected behaviour for coma in a parabolic mirror, we assume, without loss of generality, a muon incident along the azimuthal reference direction, i.e. $\phi=0$ and $\psi=0$. In this case, $\gamma_y=0$ and $\gamma_x=\theta_c+\nu$. Defining $\alpha=\atan(y/x)$, we obtain for the displacement $(\Delta  x_\mathrm{cam},\Delta y_\mathrm{cam})$ of the photon in the camera with respect to its Gaussian optics expectation $\left((\theta_c+\nu)/(4c),0\right)$: 
\begin{subequations}
\begin{align}
 \begin{pmatrix}
\Delta  x_\mathrm{cam}    \\
\Delta y_\mathrm{cam} 
\end{pmatrix} & \overset{(\phi=0, \psi=0)}{=}  \quad c\cdot 
\begin{pmatrix}
 2\,x^2+ r^2 \\ 
 2\,x\,y
\end{pmatrix} \cdot (\theta_c+\nu)  \\ 
{} &\quad~  = \quad \quad ~{} 
c \cdot r^2 \cdot 
\begin{pmatrix}
 \cos(2\alpha)+2 \\ 
 \sin(2\alpha)
\end{pmatrix}  \cdot (\theta_c+\nu) \label{eq:coma_traditional} \\ 
{} &\quad~  = \quad \quad ~{} 
c \cdot 
\begin{pmatrix}
\rho^2\,\sin^2(\phi_0) +  3\,\left(l^2\,\theta_c^2+\rho^2-2\,l\,\theta_c\,\rho\cdot\cos(\phi_0)\right)  \\ 
 \rho^2\,\sin(2\phi_0) - 2\,l\,\theta_c\,\rho\cdot\sin(\phi_0)
\end{pmatrix}  \cdot (\theta_c+\nu) \quad, 
\label{eq:coma_inserted}
\end{align} 
\end{subequations}
and, correspondingly, for $\phi=\pi$ and $\psi=0$ with respect to its Gaussian optics expectation $\left((\theta_c-\nu)/(4c),0\right)$:
\begin{subequations}
\begin{align}
 \begin{pmatrix}
\Delta  x_\mathrm{cam}    \\
\Delta y_\mathrm{cam} 
\end{pmatrix} & \overset{(\phi=\pi, \psi=0)}{=} 
- c \cdot r^2 \cdot 
\begin{pmatrix}
 \cos(2\alpha)+2 \\ 
 \sin(2\alpha)
\end{pmatrix}  \cdot (\theta_c-\nu) \label{eq:coma_traditional_pi} \\ 
{} &\quad~  = \quad \quad ~{} 
- c \cdot 
\begin{pmatrix}
\rho^2\,\sin^2(\phi_0) +  3\,\left(l^2\,\theta_c^2+\rho^2+2\,l\,\theta_c\,\rho\cdot\cos(\phi_0)\right)  \\ 
 \rho^2\,\sin(2\phi_0) + 2\,l\,\theta_c\,\rho\cdot\sin(\phi_0)
\end{pmatrix}  \cdot (\theta_c-\nu) \quad, 
\label{eq:coma_inserted_pi}
\end{align} 
\end{subequations}
where Eqs.~\ref{eq:coma_traditional} and~\ref{eq:coma_traditional_pi} reproduce the tangential and sagittal image-plane manifestations of third-order (Seidel) coma aberrations for a parabolic mirror. Eqs.~\ref{eq:coma_inserted} and~\ref{eq:coma_inserted_pi} are obtained by substituting the corresponding solutions for $x$ and $y$ from Eqs.~\ref{eq:x} and~\ref{eq:y}. 

The coma aberration for the general case of an arbitrarily inclined muon and photon, expressed in relative camera coordinates, is then given by: 
\begin{align}
 \begin{pmatrix}
\Delta  x_\mathrm{cam} / x_\mathrm{cam} \\
\Delta y_\mathrm{cam} / y_\mathrm{cam} 
\end{pmatrix} & = 4\, c^2 \cdot 
\begin{pmatrix}
\rho^{2} \cdot {\left(2 + \cos\left(2 \, \phi_{0}\right) + \gamma_y/\gamma_x \sin\left(2 \, \phi_{0}\right) \right)}  
 \\
\rho^{2}\cdot {\left(2 -\cos\left(2 \, \phi_{0}\right) + \gamma_x/\gamma_y \sin\left(2 \, \phi_{0}\right)  \, \right)}  
\end{pmatrix} +  \nonumber\\
& + 
\begin{pmatrix}
- 2 \, l \cdot \rho \cdot \theta_{c} \cdot {\left(2\,\cos\left(\phi - \phi_{0}\right) + \cos\left(\phi + \phi_{0}\right) + \gamma_y/\gamma_x \sin\left(\phi + \phi_{0}\right)\right)}  \\
- 2 \,l\,\cdot\rho \cdot\theta_{c}\cdot {\left(2 \, \cos\left(\phi - \phi_{0}\right) -\cos\left(\phi + \phi_{0}\right) + \gamma_x/\gamma_y \sin\left(\phi + \phi_{0}\right)\right)} 
\end{pmatrix} +  \nonumber\\
& + 
\begin{pmatrix}
 l^2\cdot\theta_{c}^{2}\cdot {\left(2 + \cos\left(2 \, \phi\right) +  \gamma_y/\gamma_x \sin\left(2 \, \phi\right)    \right)} \\
 l^2\cdot \theta_{c}^{2}\cdot{\left(2 -\cos\left(2 \, \phi\right) + \gamma_x/\gamma_y \sin\left(2 \, \phi\right)  \, \right)} 
\end{pmatrix} 
\quad, \label{eq:coma_full}
\end{align}

We now compute the squared distance of the incident Cherenkov photon from the origin in order to compare it, in a subsequent step, with the mirror boundaries. 
\begin{subequations}
\begin{align}
r^2 &= x^2+y^2 \\[0.3cm]
{} &= \rho^{2} + l^{2}\cdot  \theta_{c}^{2} - 2 \, l \cdot \theta_{c} \cdot \rho\cdot \cos\left(\phi-\phi_{0}  \right) +  \label{eq:r2_leading_order}\\[0.3cm]
& {}\quad + l\cdot \theta_c^3 \cdot \Big( l\cdot \theta_c
   - \rho\cdot \cos(\phi - \phi_0) \Big) 
+ {} \label{eq:r2_first_order_flat1}\\
& {}\quad + \nu \cdot l \cdot \theta_c^2 \cdot  \Bigg(3\,l \cdot \theta_c \cdot \cos\left(\phi - \psi\right) 
- \rho \cdot \Big( \cos(2\phi - \phi_0 - \psi) + 2\, \cos(\psi -\phi_0)\Big)\Bigg) -{} 
   \label{eq:r2_first_order_flat2}\\
 & {} \quad - 2\,c\cdot l \cdot \theta_c \cdot \Bigg(  
 \theta_c \cdot \Big( 
l^2\cdot \theta_c^2 + 2\, \rho^{2}\cdot   
 \cos\left(\phi- \phi_{0} \right)^2
-3\,l\cdot \theta_c  \cdot \rho \cdot \cos(\phi -\phi_0)  \Big) + {} 
 \nonumber\\
& {}\qquad\qquad\qquad\quad + \nu \cdot \Big( l^2\cdot \theta_c^2  \cdot \cos(\phi - \psi)  + 2\,\rho^{2}\cdot
  \cos\left(\phi - \phi_{0}\right) \cdot 
  \cos\left(\phi_0 - \psi\right) - {} \nonumber\\
 & {}\qquad\qquad\qquad\quad\qquad 
  - l \cdot \theta_c\cdot\rho\cdot\Big(
  \cos\left(2\,\phi - \phi_0 -  \psi\right) 
  + 2\,\cos\left(\psi-\phi_{0} \right) \Big) \Big)  
 \Bigg)    + {} \label{eq:r2_first_order_curved}\\ 
 & \quad {} +  \mathrm{h.c.} ~\left( O(c^2,\nu^3,\theta_c^3) \right) \quad. \nonumber
\end{align}
\end{subequations}
As before, the first line of Eq.~\ref{eq:r2_leading_order} represents the leading-order solution. Since $l$ may be large and $\theta_c$ is always small, all three leading terms  are of order $\lesssim R_1^2$, where $R_1$ denotes the radius of the primary mirror. The subsequent \ccol{two lines}, \ccol{Eq.~\ref{eq:r2_first_order_flat1} and~\ref{eq:r2_first_order_flat2}}, give the leading-order corrections for a flat mirror. These terms are suppressed by factors of order $\nu\cdot\theta_c$; consequently, all linear corrections in $\theta_c$ and $\nu$ alone cancel identically, in agreement with previous results~\citep{vacanti,GaugMuons:2019}. Recall that $\theta_c < 0.03$~\ccol{rad} under all atmospheric conditions and telescope altitudes, and that $\nu < 0.05$~\ccol{rad} even for the widest-angle IACTs proposed for the CTAO, for fully contained muon ring images. 
Similarly, the corrections for the parabolic mirror (the following three lines,   Eq.~\ref{eq:r2_first_order_curved}) are of order $c\cdot\theta\cdot\rho$ or $c\cdot\nu\cdot\rho$. These terms are therefore only linearly suppressed provided that $c \cdot \rho  \lesssim 1$, which is the case for Schwarzschild-Couder telescopes with small magnification. This constitutes a first key result of our study: previous analyses neglected the curvature of the mirror and consequently missed the first-order correction to the maximum emission height of the Cherenkov photons produced by a muon. 

Finally, by setting both $c$ and $\nu$ to zero and substituting $l \cdot \theta_c$ by the chord $D$ -- defined as the distance between the muon impact point and the boundary of a flat mirror --, setting $r^2 = R_1^2$ and solving for $D$, we recover the well-known solution for a flat mirror and non-inclined muon~\citep{vacanti}: 
\begin{equation}
D = \rho\cdot  \cos\left(\phi - \phi_{0}\right) \pm \sqrt{R_{1}^{2} - \rho^{2} \cdot\sin\left(\phi - \phi_{0}\right)^{2} }  \quad. \label{eq:vacanti}
\end{equation}
Both solutions are real only if the expression under the square root is positive, leading to the requirement
\begin{equation}
\sin\left(\phi - \phi_{0}\right)^{2} \overset{!}{<} 1 - \frac{\rho^2}{R_1^2} \quad,
\label{eq:vacanticondition}
\end{equation}
which is always satisfied for a muon impacting the mirror ($\rho<R_1$), but constitutes a non-trivial condition otherwise~\citep{vacanti}. We will use this condition later in the treatment of light subtraction due to the central hole in the mirror. 
One can also observe that the positive sign in Eq.~\ref{eq:vacanti} always yields a positive chord, whereas the negative sign results in an unphysical negative chord for a muon impacting the mirror $(\rho<R_1)$. In the opposite case $(\rho>R_1)$, two real and physically meaningful solutions are obtained whenever Eq.~\ref{eq:vacanticondition} is satisfied.

\begin{figure}
\centering
\includegraphics[width=0.43\textwidth]{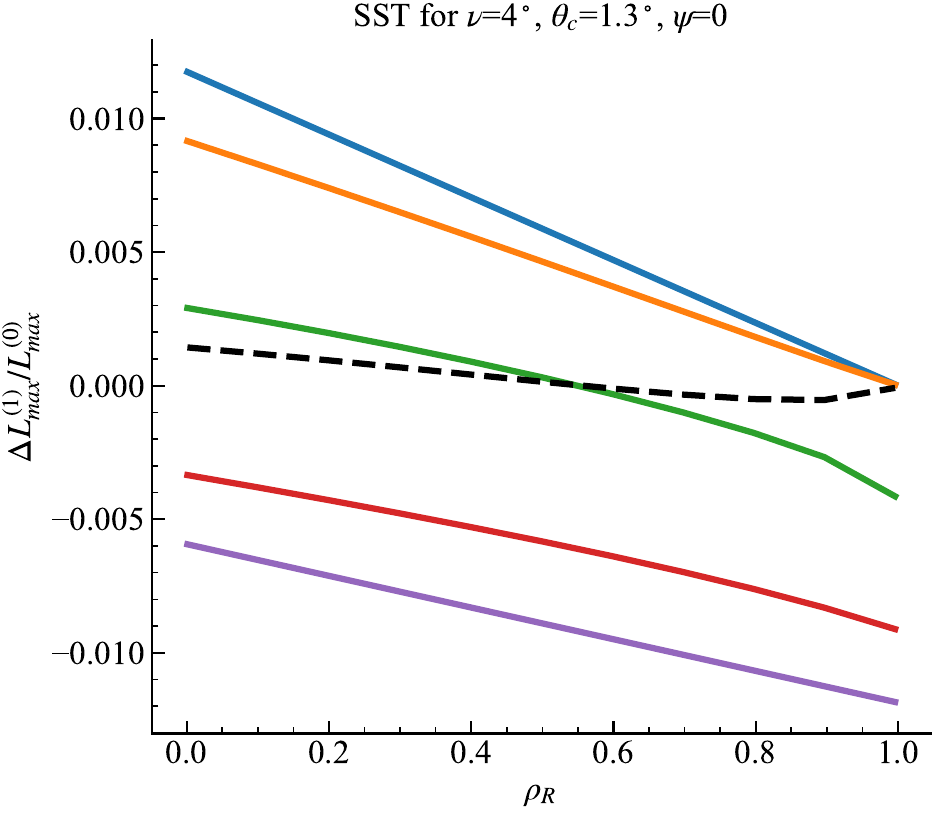}
\includegraphics[width=0.53\textwidth]{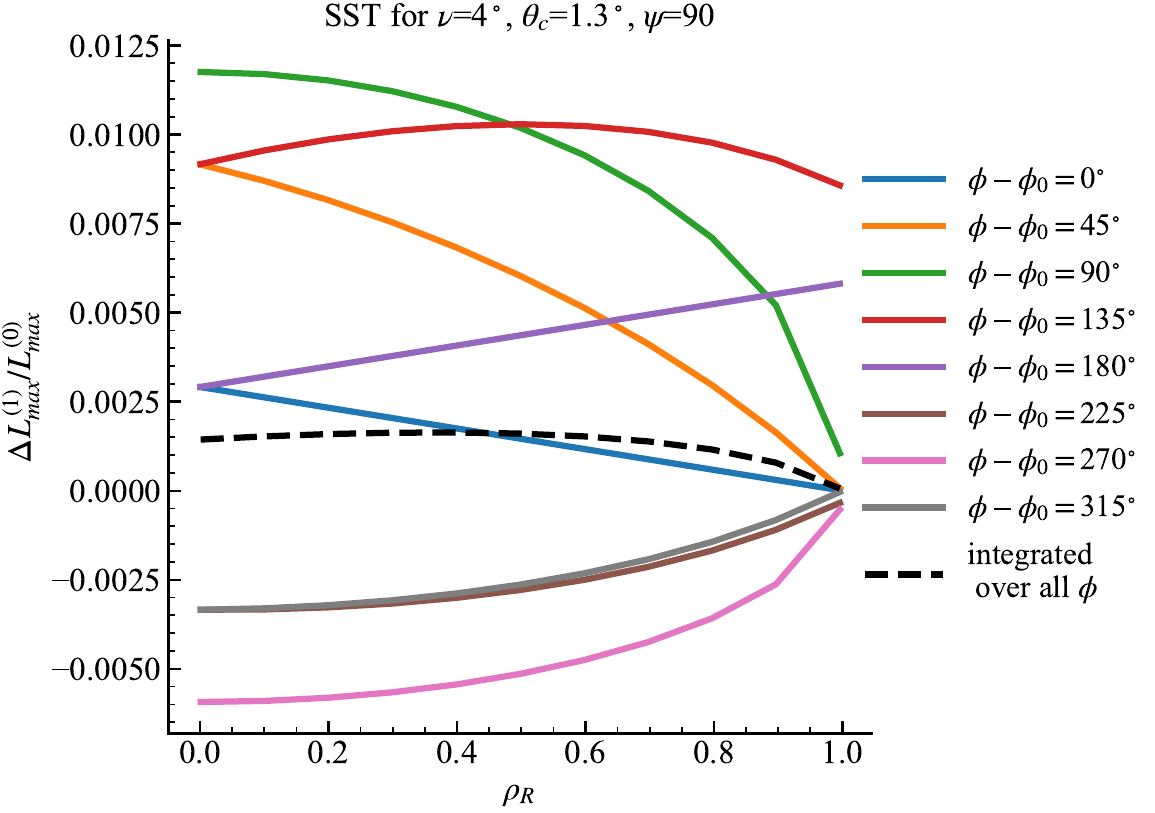}
\caption{Relative magnitudes of the first-order corrections $\Delta L_\mathrm{max}^{(1)}$ on the canonical version of $L_\mathrm{max}^{(0)}$~\citep{vacanti} for a maximally inclined muon ($\nu=4^\circ$) imaged by a Small-Sized-Telescope (SST) camera, for two different azimuthal projections $\psi$ of the muon incidence angle on ground and different photon emission angles with respect to the muon impact point on the mirror, $\phi-\phi_0$.  The primary mirror of the SST has been approximated to a parabolic shape. The color-coded legend on the right side applies to both figures, however, some angles show a degeneracy and are therefore not visible on the left-hand side. \label{fig:Lmaxfirstorder}}
\end{figure}

The successful recovery of Eq.~\ref{eq:vacanti} provides additional confidence in the calculations performed thus far (all of which were supported with the help of the computer algebra system \textit{SageMath}). We now proceed a step further and solve $r^2 = R_1^2$ for the general case, expressing it as a function of the (maximum) emission height $l$. In the following, we neglect the second-order corrections suppressed by $\nu\cdot\theta$, and solve $r^2 = R_1^2$ for $l$ to obtain the maximum emission height $l = L_\mathrm{max}(\rho,(\phi-\phi_0),\nu,\psi;c,\theta_c)$ of the Cherenkov photon: 
\begin{subequations}
\label{eq:Lmax}
\begin{align}
L_\mathrm{max} &= 
\frac{\rho \cdot  \cos\left(\phi - \phi_{0}\right) + \sqrt{R_{1}^{2} -\rho^{2} \cdot\sin\left(\phi - \phi_{0}\right)^{2}}}{\theta_{c}} + {} \label{eq:LmaxVacanti}\\
& {} + c \cdot (R_1^2 - \rho^2) \cdot \Bigg(1 + \frac{\nu}{\theta_c} \cdot \Big(\cos(\phi-\psi)-\frac{\rho\cdot \sin(\phi-\phi_0)}{2\,\sqrt{R_1^2-\rho^2\cdot \sin(\phi-\phi_0)^2}}\cdot\sin(\phi-\psi)\Big)\Bigg) + {} \label{eq:Lmax2ndOrder}\\
& {} + O(c^2,\nu^2,\nu\cdot\theta_c) \quad.
  \end{align}
  \end{subequations}
The first-order correction to $L_\mathrm{max}$, Eq.~\ref{eq:Lmax2ndOrder}, is of order $(\theta_c+\nu)/(8f_\#) \lesssim \textit{FOV}/(16f_\#) \approx 1\%$, where $f_\# = F/D = 1/(8\, c \cdot R_1)$ for a parabolic primary mirror and \textit{FOV} is the telescope's field-of-view in radians. Note that the function $(1-\rho_R^2)\cdot\rho_R/2\cdot\sin(\phi-\phi_0)/\sqrt{1-\rho_R\cdot\sin(\phi-\phi_0)^2}$ varies only between approximately -0.2 and 0.2.
Although the largest uncertainty in $L_\mathrm{max}$ arises from the deviations of a (possibly tessellated) primary mirror from a spherical shape~\citep{mitchell2015m}, 
it is nevertheless noteworthy that the relatively straightforward first-order correction term, $c\cdot(R_1^2-\rho^2)$, has been entirely neglected in the past. \ccol{Figure~\ref{fig:Lmaxfirstorder} shows the relative magnitude of the correction for a Small-Sized-Telescope (SST)}. 

The solution to $r^2=R_1^2$ yields a second physically meaningful case for a muon passing outside the primary mirror, corresponding to the minimum photon emission height 
\begin{equation}
L_\mathrm{min} = \left\{ 
\begin{array}{ll}
 \ddfrac{\rho \cdot  \cos\left(\phi - \phi_{0}\right) - \sqrt{R_{1}^{2} -\rho^{2} \cdot\sin\left(\phi - \phi_{0}\right)^{2}}}{\theta_{c}} + c \cdot (R_1^2 - \rho^2) + O(c^2,\nu^2,\nu\cdot\theta_c)     &  \mathrm{for}~\rho>R_1\\
  0   & \mathrm{else}
\end{array}
\right. \quad.\label{eq:Lmin}
\end{equation}
Note that $(L_\mathrm{max}-L_\mathrm{min})$ retrieves Vacanti's formula~\citep[Eq.~6 of~][]{vacanti}, except for the higher-order correction. 

\begin{figure}
\centering
\includegraphics[width=0.485\textwidth]{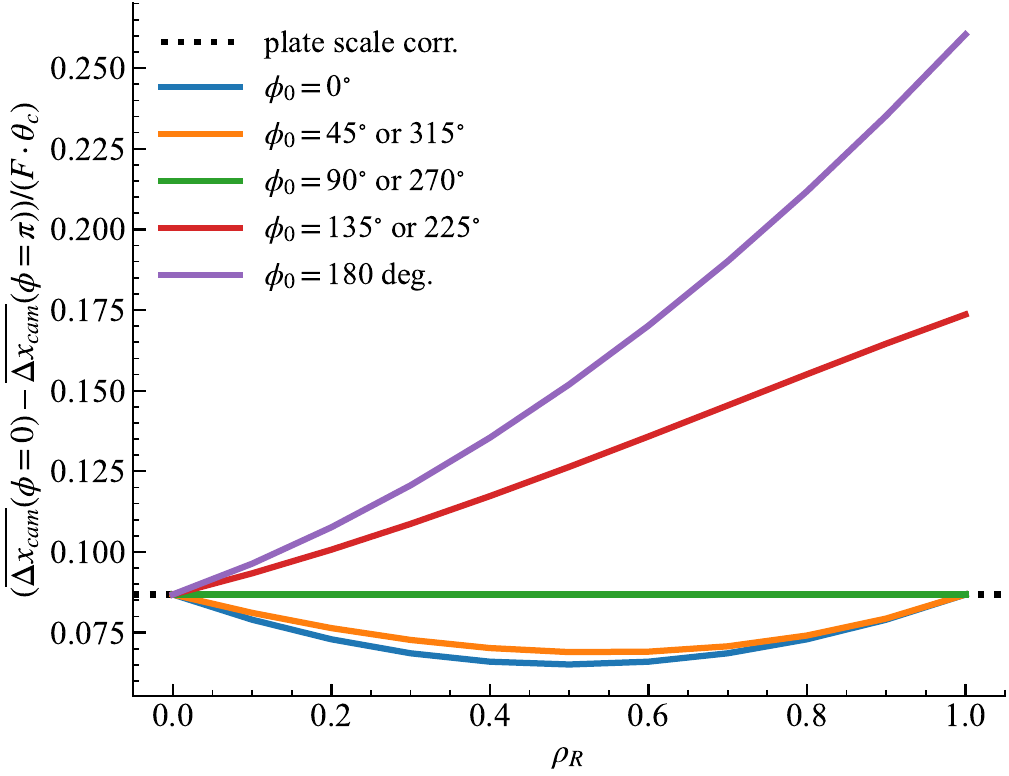}
\includegraphics[width=0.485\textwidth]{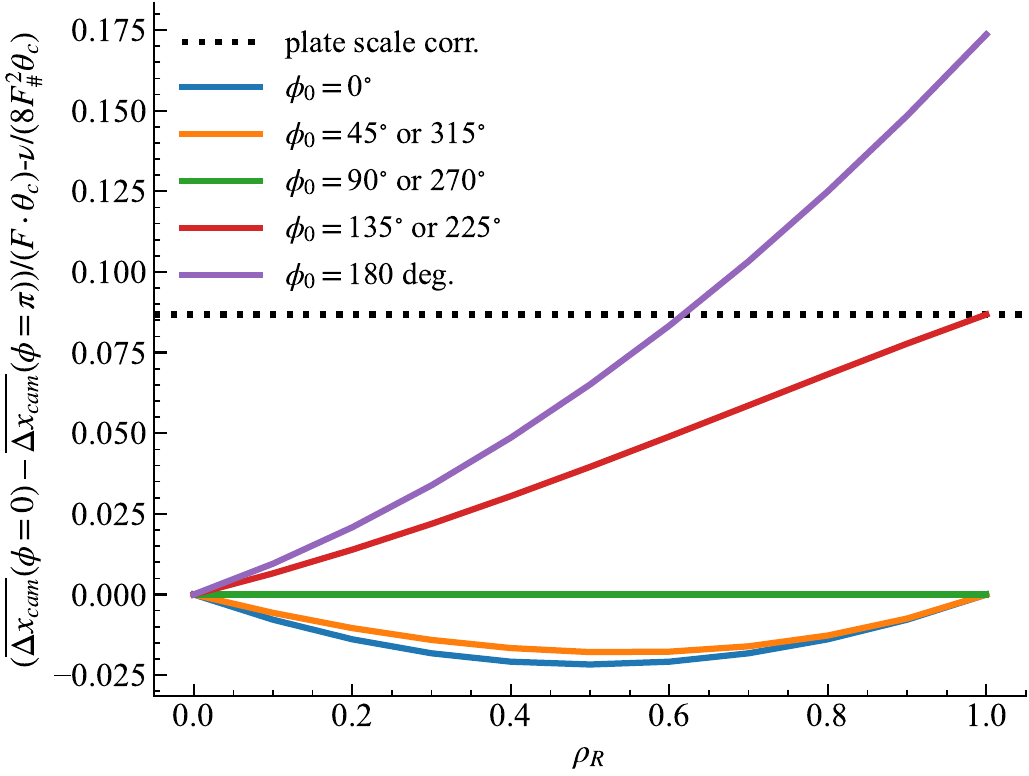}
\caption{Tangential-coma-induced bias between the outer and inner edges of the muon ring for a parabolic telescope with $f_\#=1.2$, muon Cherenkov angle $\theta_c=1^\circ$, and inclination angle $\nu=1^\circ$, shown without plate-scale correction (left) and with plate-scale correction (right). The expected bias for parallel light, typically used for a plate-scale correction, is indicated by the black dotted line. Note the different vertical-axis scales in the two panels. \label{fig:coma}}
\end{figure}

The difference in tangential coma aberration between the outer and inner parts of the muon ring can now be averaged over all $l$ from zero to $L_\mathrm{max}$ for rings that do not enclose the camera center ($\nu>\theta_c$) using Eqs.~\ref{eq:coma_inserted} and~\ref{eq:coma_inserted_pi}: 
\begin{subequations}
\begin{align}
\overline{\Delta x_\mathrm{cam}}(\phi=0)-\overline{\Delta x_\mathrm{cam}}(\phi=\pi) &\approx c \cdot \Bigg( 
2\,\theta_c \cdot \left(\rho^2 + 2\,\rho^2\cdot \cos^2(\phi_0) + L_\mathrm{max}^2\cdot\theta_c^2   \right)
-6 \, \nu  \cdot L_\mathrm{max}\cdot\theta_c \cdot \rho\cdot \cos(\phi_0) 
\Bigg)  \\
&\approx c\cdot \Bigg( 
2\,\theta_c \cdot \left(4\,\rho^2\cdot \cos^2(\phi_0) + R_1^2 + 2\,\rho \cdot \cos(\phi_0) \cdot \sqrt{R_{1}^{2} -\rho^{2} \cdot\sin^2\left( \phi_{0}\right)}\right) - \nonumber\\
& {} \qquad\quad - 6 \, \nu  \cdot \left(
\rho^2\cdot \cos^2(\phi_0) + \rho \cdot \cos(\phi_0) \cdot \sqrt{R_{1}^{2} -\rho^{2} \cdot\sin^2\left( \phi_{0}\right)}
\right) 
\Bigg) \quad,
\end{align}
\end{subequations}
or, expressed in relative parameters $\rho_R = \rho/R_1$, an upper limit on the coma-induced relative reconstruction bias of the muon ring radius is given by:
\begin{align}
\frac{\overline{\Delta x_\mathrm{cam}}(\phi=0)-\overline{\Delta x_\mathrm{cam}}(\phi=\pi)}{F\cdot\theta_c} &=  \frac{1}{8  f_\#^2 } \cdot \Bigg( 
 \left(4\,\rho_R^2\cdot \cos^2(\phi_0) + 1 
 + 2\,\rho_R \cdot \cos(\phi_0) \cdot \sqrt{1 -\rho_R^{2} \cdot\sin^2\left( \phi_{0}\right)}\right) - \nonumber\\
& {} \qquad\qquad\quad - \frac{3\nu}{\theta_c}  \cdot \left(
\rho_R^2\cdot \cos^2(\phi_0) + \rho_R \cdot \cos(\phi_0) \cdot \sqrt{1 -\rho_R^{2} \cdot\sin^2\left( \phi_{0}\right)}
\right) 
\Bigg) \quad. \label{eq:deltaxcam_relative}
\end{align}
We observe that for $\phi_0=90^\circ$ or $\phi_0=270^\circ$, the difference is exactly equal to the bias expected for parallel light incident a angles $\theta_c+\nu$ and $\theta_c-\nu$, respectively, namely $\nu/(8f_\#^2 \theta_c)$ for a parabolic mirror \citep[see, e.g., Eq.~4~of][]{Fegan:2024}. The behaviour of Eq.~\ref{eq:deltaxcam_relative} is illustrated in Fig.~\ref{fig:coma} for typical LST telescope parameters, shown without plate scale correction (left) and with plate scale correction (right). 
The situation is not as dire as might be inferred from Fig.~\ref{fig:coma}, because the ring-reconstruction algorithm will identify a shifted ring center and reconstruct a ring radius of approximately half the deviations shown in Fig.~\ref{fig:coma}. This  leads to a maximum error in the ring radius of 9\% for a muon impacting at the edge of the primary mirror and inclined towards the opposite opposite to the mirror. For random distributions of impact points on the mirror, the plate-scale corrected average relative error on reconstructed ring radius can be calculated as: 
\begin{align}
\frac{1}{2} \cdot \Big\langle \frac{\overline{\Delta x_\mathrm{cam}}(\phi=0)-\overline{\Delta x_\mathrm{cam}}(\phi=\pi)}{F\cdot\theta_c} - \frac{\nu}{8f_\#^2\theta_c} \Big\rangle &= \frac{1}{16 f_\#^2\pi}\int_0^{2\pi} \int_{\rho_{R,\mathrm{min}}}^1   
 4\,\rho_R^3\cdot \cos^2(\phi_0)  + \rho_R 
 + 2\,\rho_R^2 \cdot \cos(\phi_0) \cdot \sqrt{1 -\rho_R^{2} \cdot\sin^2\left( \phi_{0}\right)} - \nonumber\\
& {} \qquad\qquad\qquad - \frac{3\nu}{\theta_c}  \cdot \left(
\rho_R^3\cdot \cos^2(\phi_0) + \rho_R^2 \cdot \cos(\phi_0) \cdot \sqrt{1 -\rho_R^{2} \cdot\sin^2\left( \phi_{0}\right)}
\right) - \nonumber\\
& {} \qquad\qquad\qquad -\frac{\nu\cdot\rho_R}{\theta_c} \quad
\mathrm{d}\rho_R \mathrm{d}\phi_0 \nonumber\\
&= \frac{1}{8f_\#^2} \cdot \left(1-\rho_{R,\mathrm{min}}^2 + \frac{\nu}{\theta_c}\cdot \left( \rho_{R,\mathrm{min}}^2 - \frac{7}{8} \right)\right) \quad,
\end{align}
\noindent
which yields an $O(1\%)$ positive error contribution to the telescope's throughput calibration for $\nu \approx \theta_c$ and $f_\#=1.2$ for the LST, assuming a typical lower impact-distance analysis cut of $\rho_{R,\mathrm{min}}=0.2$.  For rings observed closer to the camera center ($\nu \lesssim \theta_c$), the effect is even smaller. These numbers apply, however, only to the case of a plate-scale correction properly carried out. 
 
\clearpage
\section{Solution for the dual-mirror telescope}

The secondary mirror may shadow  Cherenkov photons emitted above it, up to a maximum emission height $L_\mathrm{2,max}$ from the muon. This height may be greater or smaller than $L_\mathrm{max}$ and depends on the impact parameters, inclination angles, and the telescope geometry.


\begin{figure}
    \centering
    \includegraphics[width=0.5\linewidth]{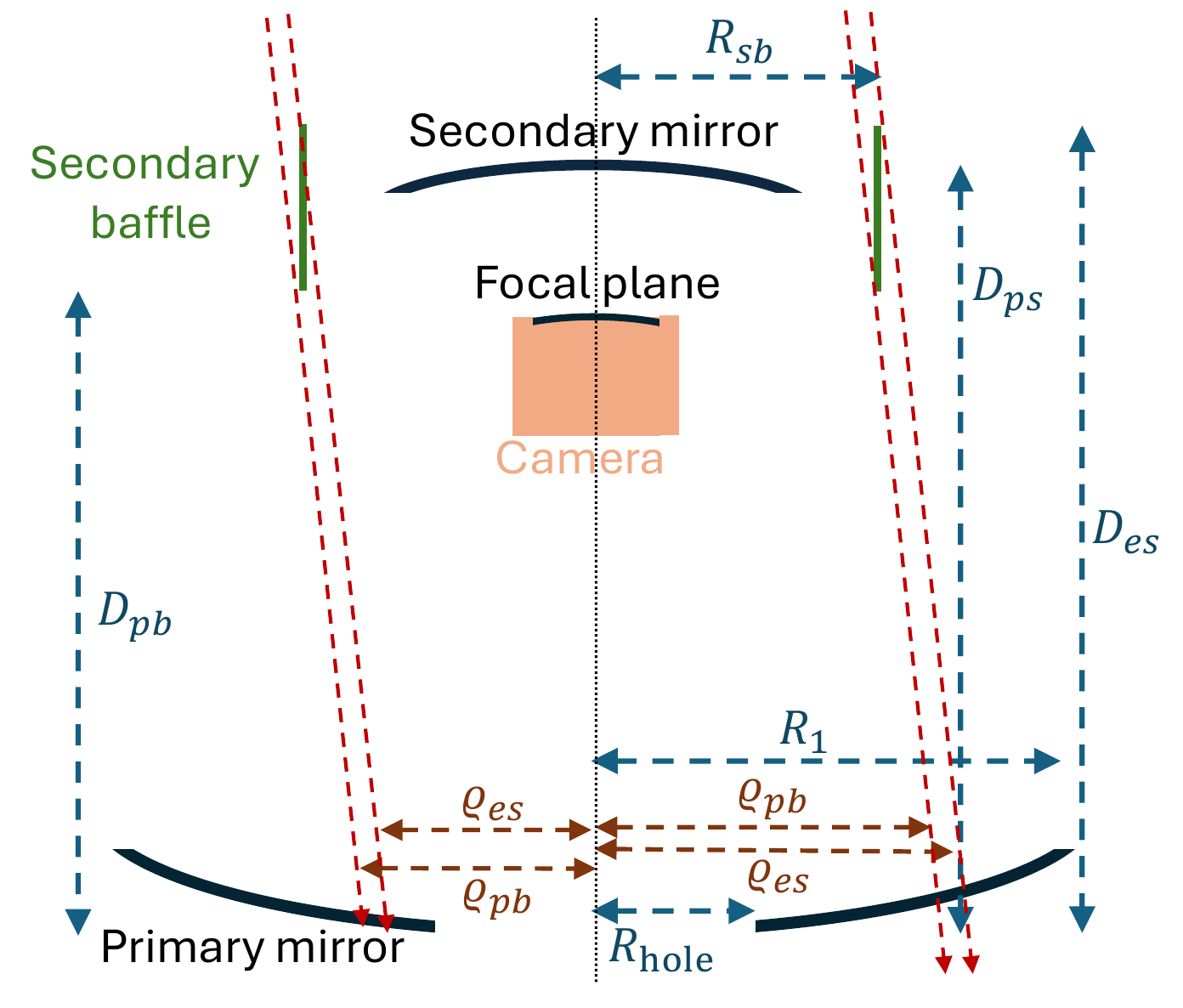}
    \caption{Schematic illustration of the parameters  used in this work for the dual-mirror telescope. The red arrows denote limiting cases of muons traversing the protective baffles of the secondary mirror.  \label{fig:dualmirror}}
\end{figure}

\begin{table}[h!]
\centering
\begin{tabular}{lccr@{.}lr@{.}l} 
\toprule
Parameter & Symbol & Unit & \multicolumn{2}{c}{SCT} 
& \multicolumn{2}{c}{SST} \\
\midrule \addlinespace[0.15cm]
Outer radius primary mirror & $R_1$ & m & 4&83  & 2&03\rlap{$^*$} 
\\
Inner radius primary mirror & $R_h$& m &  2&19  &  0&48\rlap{$^*$}  \\
Focal length primary mirror & $F_p$ & m & 12&57 & 3&96 \\
Radius secondary mirror   & $R_2$ & m & 2&71 & 0&90 \\
Radius secondary mirror incl. baffles or support  & $R_\textit{sb}$ & m & 2&80\rlap{$^\dagger$} & 1&07 \\
Focal length secondary mirror & $F_s$ & m & 3&35 & 1&14 \\
Effective focal length of telescope &  $F$  & m & 5&59  & 2&15 \\
Separation primary-secondary mirror & $D_\textit{ps}$ & m & 8&39 &  3&11  \\
Separation primary - end of support structure & $D_\textit{es}$ & m & \multicolumn{2}{c}{10\rlap{$^\ddagger$}}  &   3&00 \\
Separation primary - baffles/support structure & $D_\textit{pb}$ & m & 6&80 &  2&82 \\
Separation secondary-focal plane  & $D_\textit{sf}$ & m & 1&86  & 0&52 \\
Camera body radius & $R_\mathrm{cam}$ & m & 0&77 &  0&28 \\
Camera body depth & $D_\mathrm{cam}$ & m & 0&80\rlap{$^\dagger$} &  0&43 \\
Camera field-of-view diameter & FOV & deg & 8&2 & 10&5  \\
\addlinespace[0.1cm] 
\bottomrule 
\end{tabular}
\caption{\label{tab:tellist}Current CTAO dual-mirror telescope design parameters \citep{catalano2014,Byrum:2015,Rulten:2016,Adams:2020,White:2021} relevant for this work.  
\ccol{Note that the telescopes of the ASTRI mini-array~\citep{Leto:20239f,Crestan:2025VJ} have almost identical design parameters as the CTAO SSTs and are therefore not mentioned in a separate column.}
$^*$ Averaged over all azimuth angles, individual values vary around this number because of the hexagonal geometry of the tessellated mirror facets. $^\dagger$ best guess.  $^\ddagger$ approximate average of a structured surface. 
}
\end{table}

In the following, we denote by $D_\textit{es}$ the distance between the pole of the primary mirror and the upper part of the obstructing cylinder that houses the secondary mirror (see Fig.~\ref{fig:dualmirror} for a schematic illustration of the simplified geometry and Table~\ref{tab:tellist} for typical values of the introduced parameters). 
For simplicity, we approximate the shadow of the secondary mirror housing as two flat circular disks  of radius $R_2$, located at $z=D_\textit{es}$ and $z=D_\textit{pb}$, respectively. Here, $D_\textit{pb}$ denotes the distance between the pole of the primary mirror and the lower part of the protective baffles or support ring surrounding the secondary mirror (see~\citet{Byrum:2015,White:2021,Adams:2022} for a further illustration of the secondary baffles).  The current design of the Small-Sized Telescope for the CTAO~\citep{Trois:2024} \ccol{and the current ASTRI mini-array telescopes~\citep{Pareschi:2023ar}} do not include baffles and supports the secondary mirror via a ring structure only.  \ccol{Nevertheless, the ASTRI mini-array prototype telescope ASTRI-Horn does include secondary baffles as shown in~\citet{Leto:20239f}.} 

We first solve the vector equation for the impact point of the Cherenkov photon on the plane $z=D_\textit{es}$: 
 \begin{equation}
\boldsymbol{E} + m_2 \cdot \boldsymbol{\vec{\gamma}} = \boldsymbol{I}_{2} \label{eq:solver2}
  \end{equation}
for a photon travel length $m_2$ and photon impact point coordinates $\boldsymbol{I}_{2} = \left(x_2,y_2,D_\textit{es}\right)$.  
The solutions for $x_2$ and $y_2$ are given by: 
\begin{subequations}
 \begin{align}
x_2 &=  \left(l-D_\textit{es}\right) \cdot\theta_{c} \cdot\cos\left(\phi\right) - \rho \cdot\cos\left(\phi_{0}\right) - D_\textit{es}\cdot\nu\cdot \cos(\psi) 
+  \mathrm{h.c.}\Big( O(c\cdot\nu,\nu^2,c\cdot\theta_c) \Big) \quad, \label{eq:x2}\\
y_2 &= \left(l-D_\textit{es}\right) \cdot\theta_{c} \cdot\sin\left(\phi\right) - \rho \cdot\sin\left(\phi_{0}\right) - D_\textit{es}\cdot\nu\cdot \sin(\psi) 
+  \mathrm{h.c.}\Big( O(c\cdot\nu,\nu^2,c\cdot\theta_c) \Big) \quad, \label{eq:y2}  \\
r_2^2 = x_2^2+y_2^2 
&= \rho^2 + \left(l-D_\textit{es}\right)^2 \cdot \theta_c^2 
- 2 \, \left(l-D_\textit{es}\right) \cdot \theta_{c} \cdot \rho\cdot \cos\left(\phi-\phi_{0}  \right)  + \label{eq:r22_leadingorder} \\
&  {}  
+ 2 \cdot D_\textit{es} \cdot \nu \cdot \Big(  
\rho \cdot \cos\left(\psi-\phi_0\right) - l\cdot\theta_c \cdot \cos\left(\phi-\psi\right) 
\Big) + D_\textit{es}^2\cdot \nu^2 {} \label{eq:r22_secondorder} \\
 & \qquad {} +  \mathrm{h.c.} ~\left( O(c^2,\nu^3,\theta_c^3) \right) \quad. \nonumber
\end{align}
\end{subequations}
Note the similarity between these solutions and those for the impact point on the primary mirror (Eqs.~\ref{eq:x},~\ref{eq:y} and \ref{eq:r2_first_order_curved}), with only two additional terms proportional to $D_\textit{es}$. A residual dependence on $c$ arises through the definitions of the muon impact point on the primary mirror and the photon emission point (Eq.~\ref{eq:O}).

\begin{figure}
\centering
\includegraphics[width=0.485\textwidth]{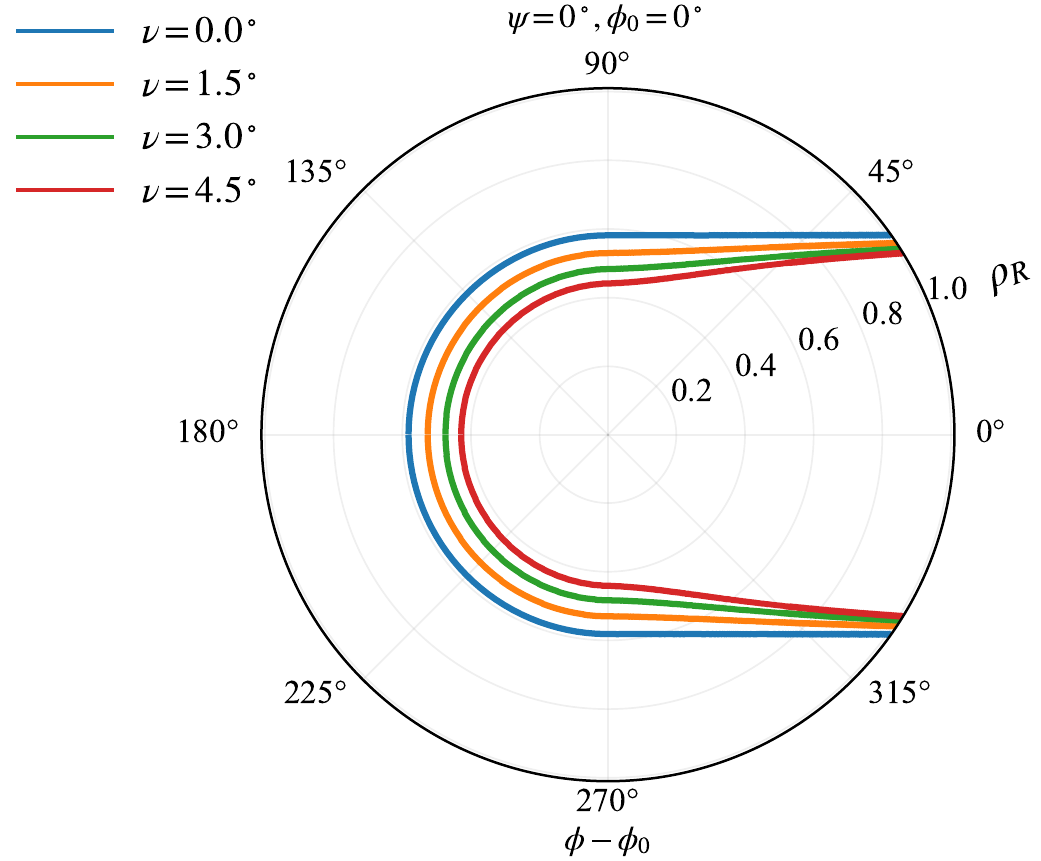}
\includegraphics[width=0.485\textwidth]{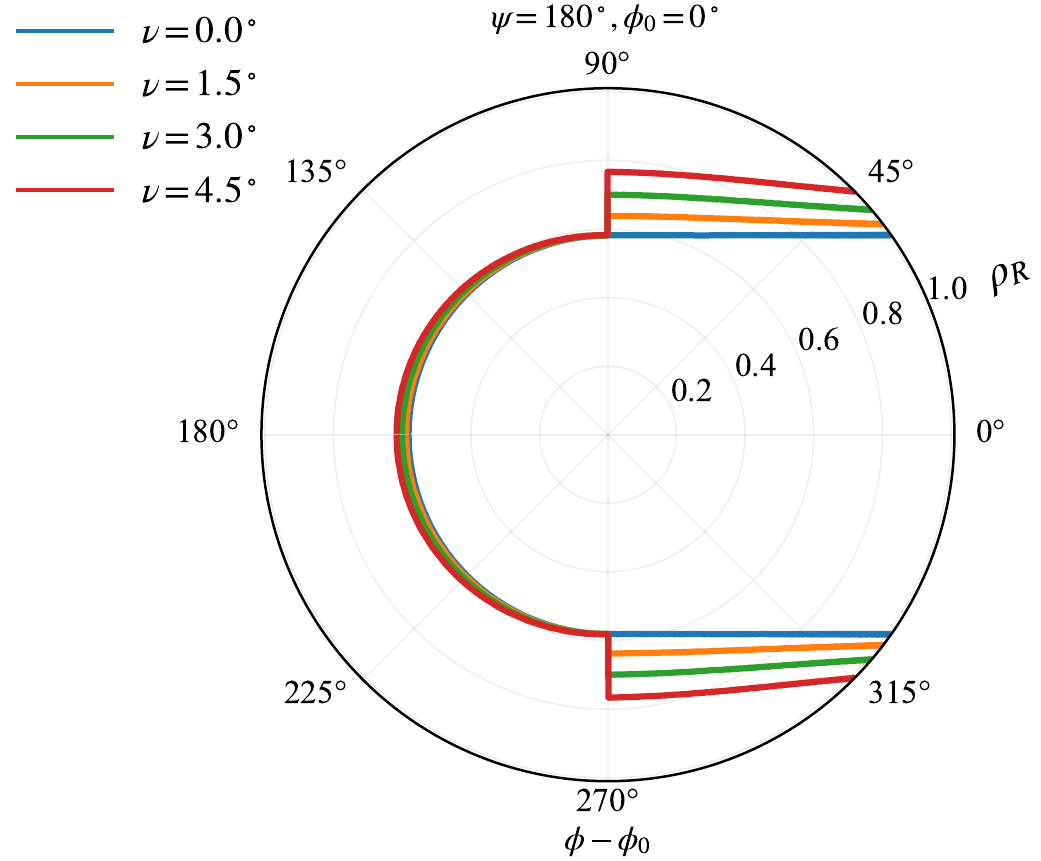}
\includegraphics[width=0.485\textwidth]{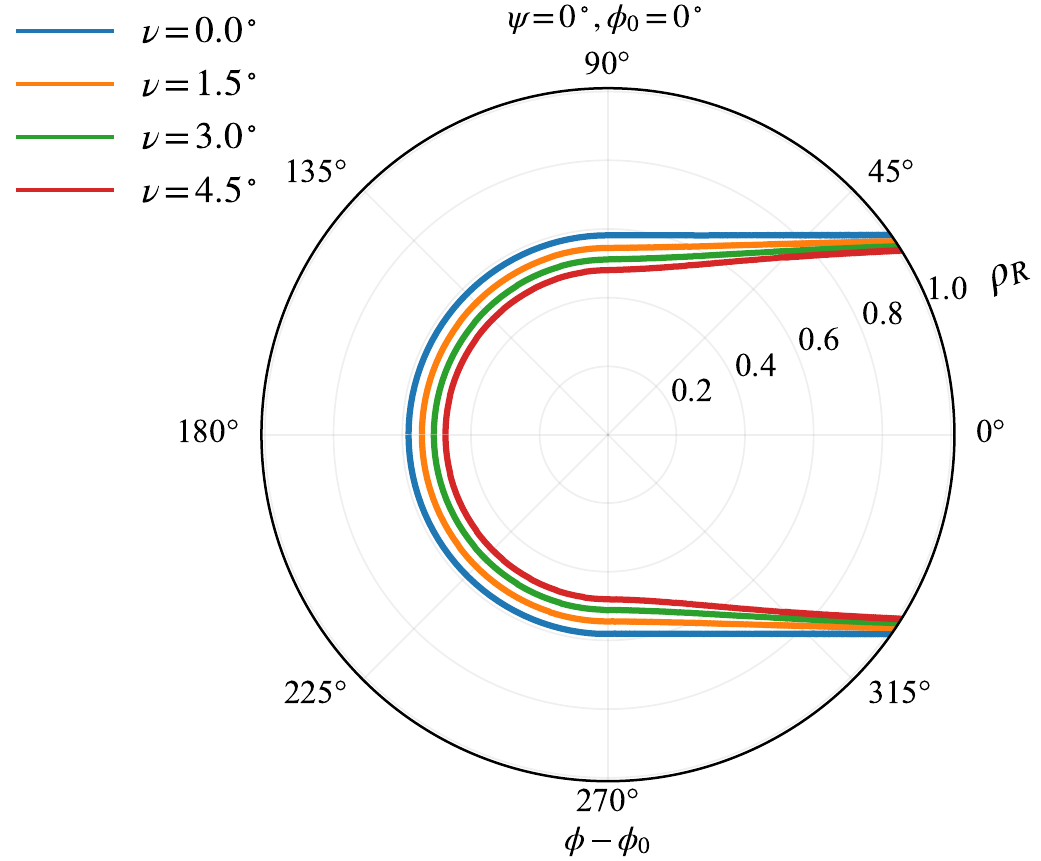}
\includegraphics[width=0.485\textwidth]{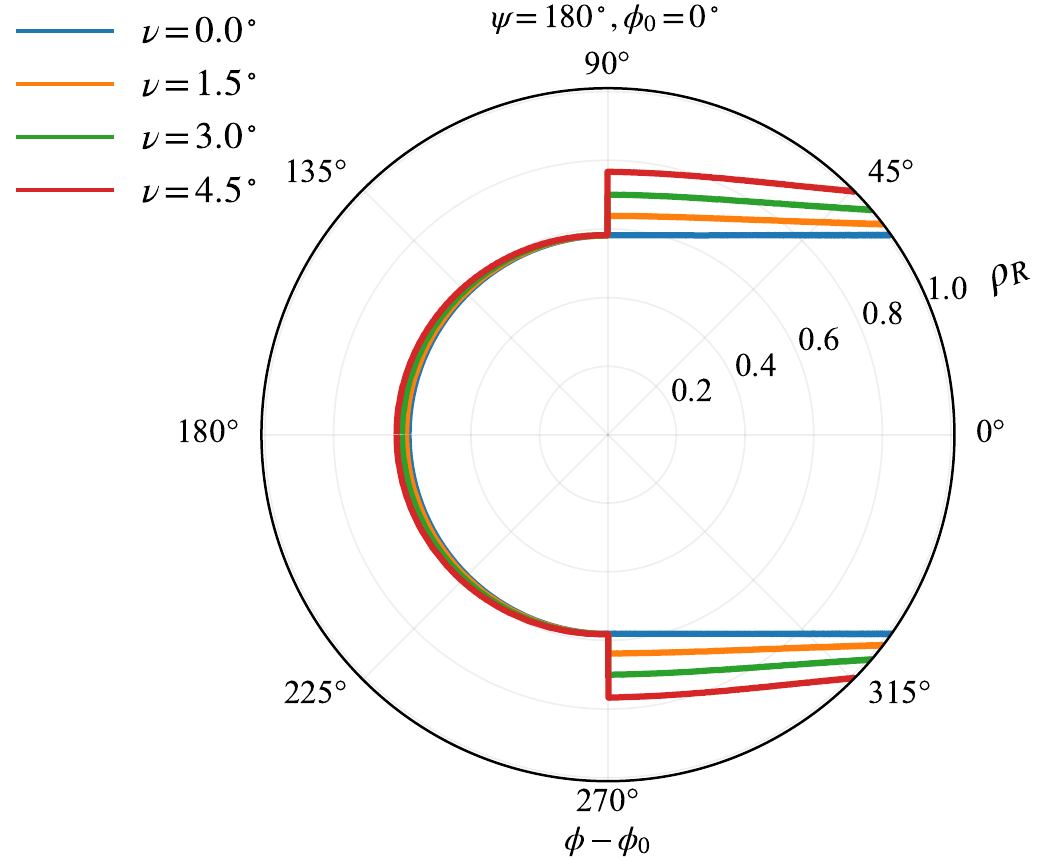}
\caption{Shadow condition (Eq.~\ref{eq:shadowcondition}) shown for muons with different normalized impact distances $\rho_R$, at fixed impact angle $\phi_0$, for the parameters of an SCT. Top figures neglect the effects of the M2 baffles, while the bottom figures include their contribution using Eq.~\ref{eq:shadowcondition2}. 
A Cherenkov angle of 1.3$^\circ$ is assumed.  The region enclosed by the colored curves corresponds to trajectories experiencing shadowing along the muon path.
\ccol{Different muon impact points have been simulated located on a line connecting the center with left-most part of the outer ring, corresponding to $\rho_R=1$. The radial axis $\rho_R$ denotes the impact distance on that line}. The azimuthal axis represents the photon emission angle ($\phi-\phi_0$). By convention, $(\phi-\phi_0=0^\circ)$ corresponds to the longest chord on the mirror; photons emitted in that direction are always shadowed, as they propagate toward the mirror center and thus toward the secondary mirror. Photons emitted in the opposite direction are shadowed only until the muon reaches $R_\mathrm{shadow}$. In the left-hand \ccol{panel}, the muon is inclined  toward the left with varying inclination angles $\nu$; in the right-hand panel, it is inclined toward the right. Sharp transitions are visible at $(\phi-\phi_0=90^\circ)$ and $(\phi-\phi_0=270^\circ)$ when the muon is inclined toward the secondary mirror, along with a slight increase of $R_\mathrm{shadow}$ for all directions of  $(\phi-\phi_0)$. \label{fig:Shadowcondition}}
\end{figure}

\begin{figure}
\centering
\includegraphics[width=0.485\textwidth]{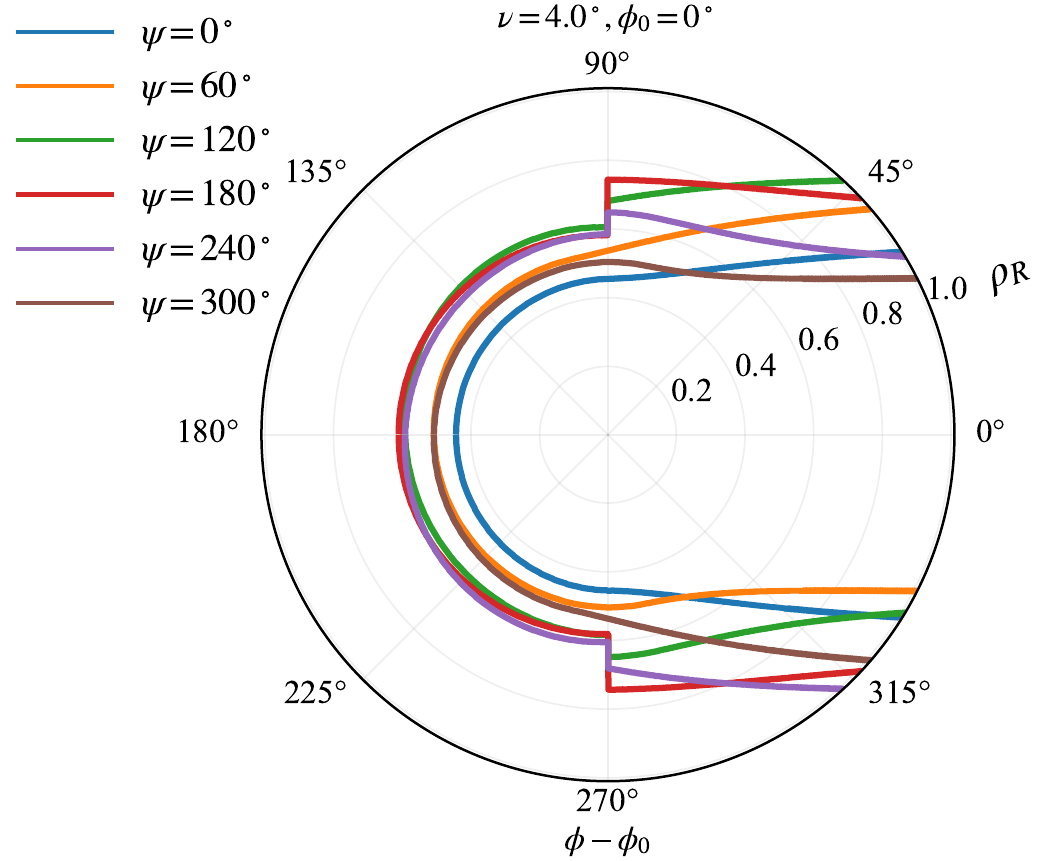}
\includegraphics[width=0.485\textwidth]{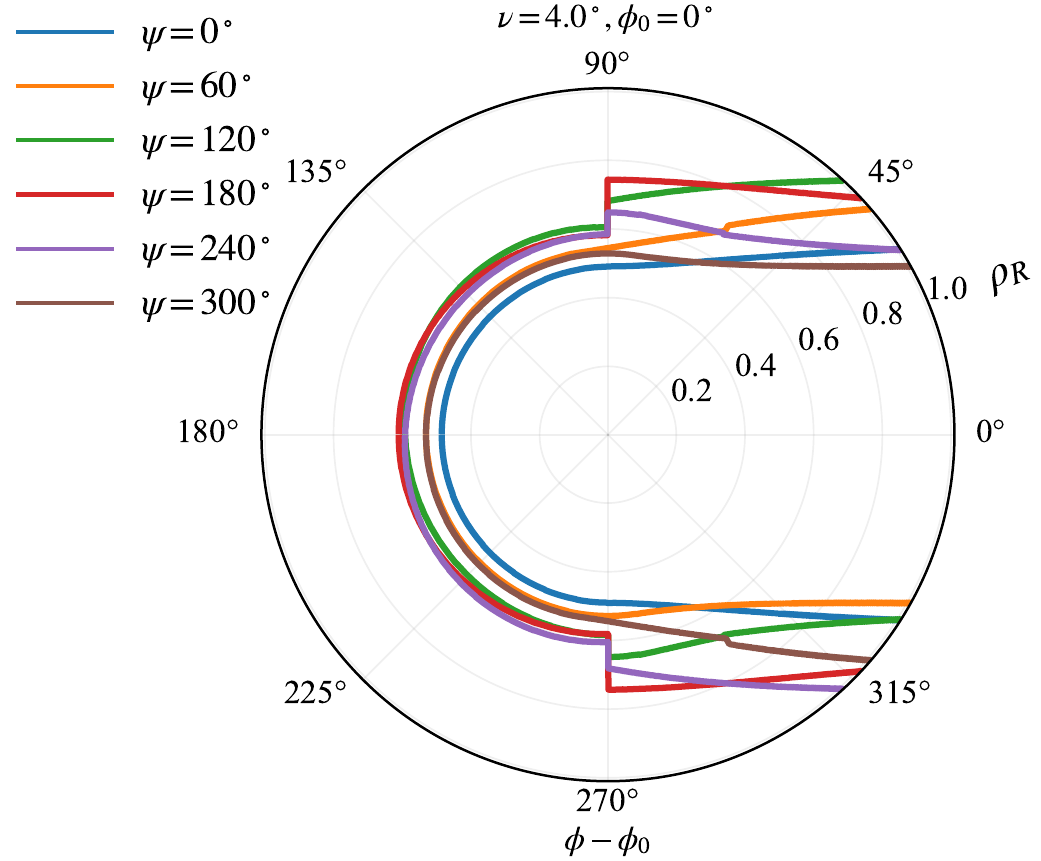}
\caption{Shadow condition (Eq.~\ref{eq:shadowcondition}) shown for muons with different normalized impact distances $\rho_R$, at fixed impact angle $\phi_0$, for the parameters of an SCT. The left panel neglects the effect of  the M2 baffles, while the right panel includes their contribution using Eq.~\ref{eq:shadowcondition2}.  Here, the muon has a fixed inclination angle of $\nu=4^\circ$, with varying azimuthal projections $\psi$. See also Figure~\ref{fig:Shadowcondition}.\label{fig:Shadowconditionpsi}}
\end{figure}

As in the single mirror case, we solve for the highest point along the muon track, $L_{2,\mathrm{max}}$, that  creates a shadow -- i.e.,  where $r_2^2 = R_\textit{sb}^2$ -- and obtain: 
\begin{subequations}
\label{eq:L2max}
\begin{align}
\mathrm{with} & \quad R_{\textit{es},\mathrm{proj}} := 
D_\textit{es} \cdot \left(\nu\cdot \cos\left(\phi - \psi\right) + \theta_c\right) \label{eq:Desproj} \\
L_{2,\mathrm{max}} &= \frac{1}{\theta_c} \cdot \Bigg(
R_{\textit{es},\mathrm{proj}} 
+ \rho \cdot \cos\left(\phi - \phi_{0}\right) 
+ {} \nonumber\\
& {} 
+ \sqrt{R_\textit{sb}^{2}
- \rho^{2} \sin^2\left(\phi - \phi_{0}\right)  
-2 D_\textit{es} \cdot\nu \cdot\rho\cdot \sin(\phi-\psi)\sin(\phi-\phi_0)
+R_{\textit{es},\mathrm{proj}}^2
} ~\Bigg)
\label{eq:L2maxradicand}\\
 & \qquad {} +  \mathrm{h.c.} ~\left( O(c,\nu^2,\theta_c) \right) \quad,\nonumber
\end{align}
\end{subequations}
where $R_{\textit{es},\mathrm{proj}}$ is the ground-projected distance of a photon emitted from height $D_{\textit{es}}$.
Note that in the limit $D_\textit{es} \rightarrow 0$, $L_{2,\mathrm{max}}$  correctly converges  to the solution of~\citet{vacanti}. 
A shadow exists if the radicand of Eq.~\ref{eq:L2maxradicand} is positive leading to the following shadow condition:
\begin{subequations}
\label{eq:shadowcondition}
\begin{align}
\mathrm{with} & \quad R_{\mathrm{shadow}} := 
\sqrt{R_\textit{sb}^2+R_{\textit{es},\mathrm{proj}}^2} \label{eq:Rshadow} \\[2mm]
 &  
 \mathrm{\quad for\quad} \rho < R_{\mathrm{shadow}} \mathrm{\quad or\quad} |\phi-\phi_0| < \pi/2 ~\mathrm{:} \nonumber\\[2mm]
-R_{\mathrm{shadow}} &
-D_\textit{es}\cdot\nu\cdot\sin\left(\phi-\psi\right) < \rho\cdot\sin\left(\phi-\phi_0\right) 
< 
R_{\mathrm{shadow}} -D_\textit{es}\cdot\nu\cdot\sin\left(\phi-\psi\right)
\quad. 
\end{align}
\end{subequations}
Eq.~\ref{eq:shadowcondition} is illustrated in Fig.~\ref{fig:Shadowcondition} for different muon inclination and light emission angles, \ccol{and in Fig.~\ref{fig:Shadowconditionpsi} for different azimuthal projections $\psi$ of the inclination angle}. \\

In the presence of a shadow, $L_{2,\mathrm{max}}$ is always \textit{smaller} than $L_{\mathrm{max}}$ of the unshadowed light (Eq.~\ref{eq:Lmax}), even in the limiting case of a highly inclined muon ($\nu_\mathrm{max} \approx \textit{FOV}/2 - \theta_c$), where \textit{FOV} is the telescope's field-of-view) impacting the telescope at the edge of the primary mirror ($\rho \rightarrow R_1$). This can be demonstrated for representative telescope parameters under worst-case conditions ($\phi-\phi_0=0$; $\cos(\phi-\psi)=1$):
\begin{subequations}
\begin{align}
R_1^2-R_\textit{sb}^2 & >  
2\, D_\textit{es} \cdot R_1 \cdot \left( \nu_\mathrm{max} + \theta_c \right) \approx  D_\textit{es}\cdot R_1 \cdot \textit{FOV} \\ 
\mathrm{or:~} R_\textit{sb} &< R_1 \cdot \sqrt{1-2\alpha\cdot f_\#\cdot \textit{FOV}}
\label{eq:Lmax2condition} \quad.
\end{align}
\end{subequations}
 With $\theta_c<0.03$ \ccol{rad} in air and the planned CTAO dual-mirror telescopes' fields-of-view (\textit{FOV}) of $\lesssim$11$^\circ \approx 0.2$~rad~\citep{White:2021,DiVenere:2023}, as well as $f$-ratios of the primary in the range of $1.4 \lesssim f_\# \lesssim 2.1$ and primary-secondary separations $0.5 \lesssim \alpha \lesssim 0.7$ in units of $f_\#$~\citep{Vassiliev2007}, this condition is always fulfilled by telescope design.

$L_{2,\mathrm{max}}$ is always \textit{larger} than the muon's maximum emission height $L_\mathrm{hole,max}$ at which the photon intersects the central mirror hole and is therefore  not reflected onto the camera. For the worst-case configuration ($\rho = R_1, \phi-\phi_0=0$; $\cos(\phi-\psi)=-1$, the corresponding emission height is $L_\mathrm{hole,max}=(R_1+R_h)/\theta_c$):
\begin{subequations}
\begin{align}
%
R_\textit{sb}^2-R_h^2 &>
D_\textit{es} \cdot R_h \cdot \left( \nu_\mathrm{max} - \theta_c \right) \approx  D_\textit{es}\cdot R_h \cdot \left(\textit{FOV}/2 - 2\,\theta_c\right) \\ 
\mathrm{or:~} R_\textit{sb} &> R_h \cdot \sqrt{1-\alpha\cdot f_\#/\eta\cdot \left(\textit{FOV}-4\,\theta_c\right)}
\label{eq:Lmax2holecondition} \quad,
\end{align}
\end{subequations}
where the central mirror hole fraction $\eta=R_h/R_1>0.5$ has been assumed.  This inequality is always satisfied for typical telescope designs.

In particular, secondary mirror baffle design of~\citet{Byrum:2015} modifies the shadow condition Eq.~\ref{eq:shadowcondition} such that $D_\textit{es}$ is replaced by the distance between the pole of the primary mirror and height of the lower edge of the protective baffles, $D_\textit{pb}$, thereby defining:
\begin{subequations}
\begin{align}
D_\textit{l} & := \left\{
\begin{array}{ll}
D_\textit{pb} & \mathrm{if~}\sin(\phi-\psi) \leq 0 \\
D_\textit{es} & \mathrm{\quad\quad else} \quad,\\
\end{array} \right.\\
D_\textit{r} & := \left\{
\begin{array}{ll}
D_\textit{es} & \mathrm{if~}\sin(\phi-\psi) \leq 0 \\
D_\textit{pb} & \mathrm{\quad\quad else}\quad, \\
\end{array} \right. \\
R_{\textit{l},\mathrm{proj}} & := \left\{
\begin{array}{ll}
R_{\textit{pb},\mathrm{proj}} & \mathrm{if~}\sin(\phi-\psi) \leq 0 \quad,\\
R_{\textit{es},\mathrm{proj}} & \mathrm{\quad\quad else} \\
\end{array} \right.\\
R_{\textit{r},\mathrm{proj}} & := \left\{
\begin{array}{ll}
R_{\textit{es},\mathrm{proj}} & \mathrm{if~}\sin(\phi-\psi) \leq 0 \\
R_{\textit{pb},\mathrm{proj}} & \mathrm{\quad\quad else} \\
\end{array} \right.
 \quad, \label{eq:definitionDlDr}
\end{align}
\end{subequations}
\noindent
where $R_{\textit{pb},\mathrm{proj}} := 
D_\textit{pb} \cdot \big(\nu\cdot \cos\left(\phi - \psi\right) + \theta_c\big)$ 
has been used.\\

\noindent
We arrive at a global shadow condition: 
\begin{align}
&
\mathrm{\quad for\quad} \rho < R_{\mathrm{shadow}} \mathrm{\quad or\quad} |\phi-\phi_0| < \pi/2 ~\mathrm{:} \nonumber\\[2mm]
-\sqrt{R_\textit{sb}^2+R_{\textit{l},\mathrm{proj}}^2}&-D_\textit{l}\cdot\nu\cdot\sin\left(\phi-\psi\right) < \rho\cdot\sin\left(\phi-\phi_0\right) 
< \sqrt{R_\textit{sb}^2+R_{\textit{r},\mathrm{proj}}^2}-D_\textit{r}\cdot\nu\cdot\sin\left(\phi-\psi\right) \quad.
\label{eq:shadowcondition2}
\end{align}

The effect of the baffles on the shadow condition is illustrated in Fig.~\ref{fig:Shadowcondition} by comparing the shadowed regions computed with and without the SCT baffles. Their impact is most pronounced when the muon is inclined away from the secondary mirror. 

If condition Eq.~\ref{eq:shadowcondition} is satisfied,  $L_\mathrm{2,max}$ is given by Eq.~\ref{eq:L2max}.  Otherwise, if 
\begin{align}
 -\sqrt{R_\textit{sb}^2+R_{\textit{r},\mathrm{proj}}^2}-D_\textit{r}\cdot\nu\cdot\sin\left(\phi-\psi\right) &< \rho\cdot\sin\left(\phi-\phi_0\right) 
< \quad\llap{--}\sqrt{R_\textit{sb}^2+R_{\textit{l},\mathrm{proj}}^2}-D_\textit{l}\cdot\nu\cdot\sin\left(\phi-\psi\right) \quad  \mathrm{or} \nonumber\\
\sqrt{R_\textit{sb}^2+R_{\textit{r},\mathrm{proj}}^2}-D_\textit{r}\cdot\nu\cdot\sin\left(\phi-\psi\right) &< \rho\cdot\sin\left(\phi-\phi_0\right) 
< \quad \sqrt{R_\textit{sb}^2+R_{\textit{l},\mathrm{proj}}^2}-D_\textit{l}\cdot\nu\cdot\sin\left(\phi-\psi\right) &  \quad,  \label{eq:condbaffle}
\end{align}
the Cherenkov light may get shadowed only by the protecting baffles and 
\begin{align}
L_\mathrm{2,max} &= \frac{1}{\theta_c} \cdot \Big(
R_{\textit{pb},\mathrm{proj}}  
+ \rho \cdot \cos\left(\phi - \phi_{0}\right) 
+ {} \nonumber\\
& {} \qquad
+ \sqrt{R_\textit{sb}^{2}
- \rho^{2} \sin^2\left(\phi - \phi_{0}\right)  
-2 D_\textit{pb} \cdot\nu \cdot\rho\cdot \sin(\phi-\psi)\sin(\phi-\phi_0) + R_{\textit{pb},\mathrm{proj}}^2
} ~\Big) \quad.
\label{eq:L2maxpb}
\end{align}
The behaviour of $L_{2,\mathrm{max}}$ is shown in the top rows of Figs.~\ref{fig:L2SCT} and~\ref{fig:L2SST} for an SCT and an SST, respectively.  \\

The lowest point of the muon track, $L_{2,\mathrm{min}}$, that still produces a shadow depends on whether the muon intersects the secondary mirror, the baffles, or neither. 
If the muon does \textit{not} intersect either the baffles or the secondary mirror, but the shadow condition is satisfied, then
$L_{2,\mathrm{min}}$ corresponds to the second solution of $r_2^2=R_\textit{sb}^2$; 
\begin{align}
L_{2,\mathrm{min}} &= \frac{1}{\theta_c} \cdot \Big(
R_{\textit{pb},\mathrm{proj}} + 
\rho \cdot \cos\left(\phi - \phi_{0}\right) 
- {} \nonumber\\
& {} 
- \sqrt{R_\textit{sb}^{2}
- \rho^{2} \sin^2\left(\phi - \phi_{0}\right)  
-2 D_\textit{es} \cdot\nu \cdot\rho\cdot \sin(\phi-\psi)\sin(\phi-\phi_0)
+R_{\textit{pb},\mathrm{proj}}^2
} ~\Big) \quad.
\label{eq:L2min}
\end{align}
The condition for the muon to intersect the secondary mirror can be retrieved as:
\begin{equation}
\rho < \rho_\textit{es}\quad,\quad\mathrm{with:\quad}
\rho_\textit{es} = 
R_\textit{sb} - D_\textit{es}\cdot \nu \cdot \cos(\phi_0-\psi) 
+  \mathrm{h.c.} ~\left( O(c,\nu^2,\theta_c^2,\nu\cdot\theta_c) \right)\quad, 
\label{eq:murhoes}
\end{equation}
\noindent
and to enter the baffle-surrounded region at its point of closest approach to the primary mirror: 
\begin{equation}
\rho < \rho_\textit{pb} \quad,\quad\mathrm{with:\quad}
\rho_\textit{pb} =  
R_\textit{sb} - D_\textit{pb}\cdot \nu \cdot \cos(\phi_0-\psi)  
+  \mathrm{h.c.} ~\left( O(c,\nu^2,\theta_c^2,\nu\cdot\theta_c) \right)\quad.
\label{eq:murhopb}
\end{equation}

\begin{figure}
\centering
\includegraphics[width=0.325\textwidth]{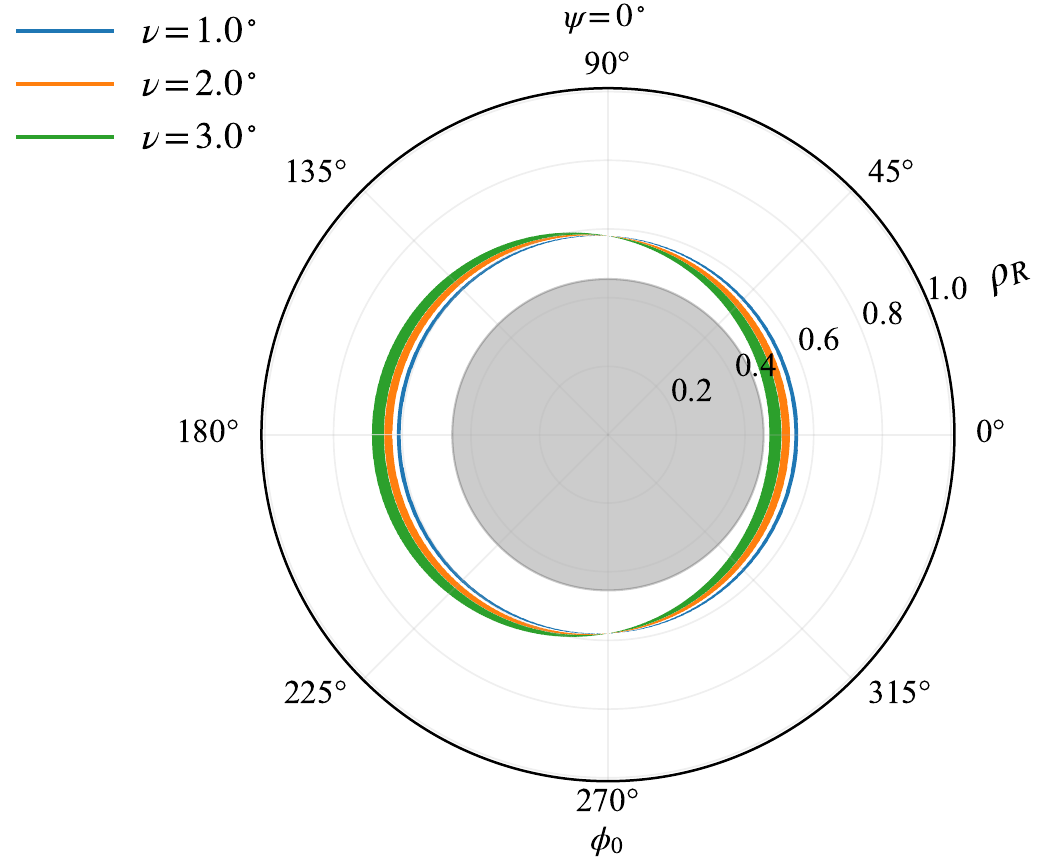}
\includegraphics[width=0.325\textwidth]{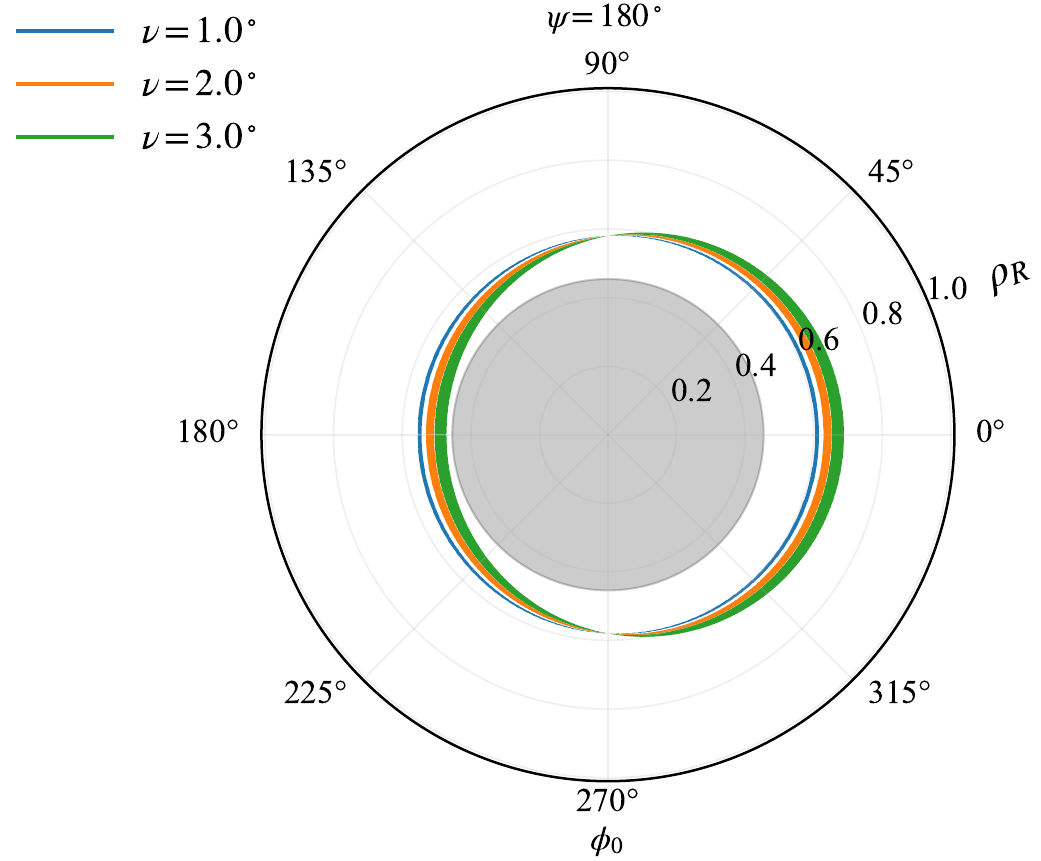}
\includegraphics[width=0.325\textwidth]{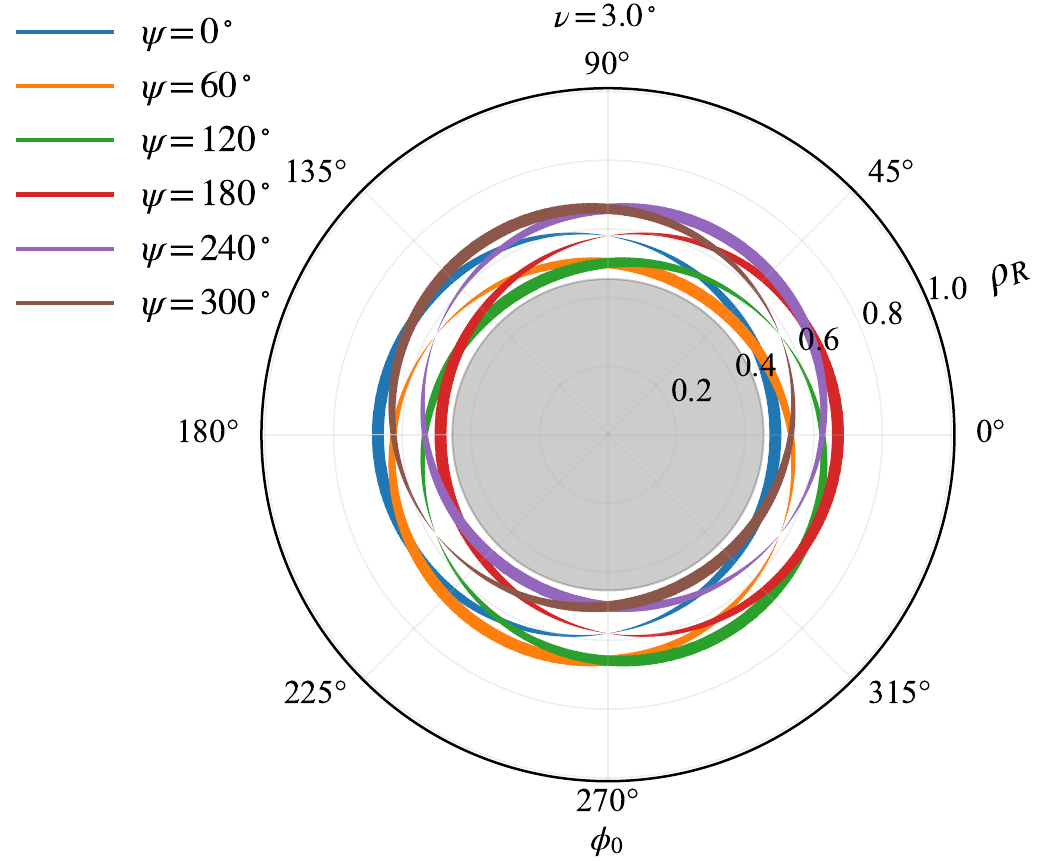}
\caption{Muon baffle-crossing conditions (Eq.~\ref{eq:condpassbafflefromoutside} and~\ref{eq:condpassbafflefrominside}) shown for muons with different normalized impact distances $\rho_R$ and impact angles $\phi_0$, for the parameters of an SCT. In the left panel, the muon is inclined toward the left ($\psi =0^\circ$) with varying inclination angles $\nu$, in the central panel, it is inclined toward the right ($\psi=180^\circ$), and in the right panel the inclination angle is fixed at $\nu=4^\circ$, while  the azimuthal projection $\psi$ varies. Note that the azimuthal axis represents here the muon angle $\phi_0$, whereas in the previous figures the photon emission angle $\phi-\phi_0$ was used. The grey shaded region in the center indicates the central hole in the primary mirror. \label{fig:bafflecondition}}
\end{figure}

There is an intermediate range in which the muon may traverse the protecting baffles, at distances from the primary mirror between 
$D_\textit{pb}$ and $D_\textit{es}$. This can occur in two ways: the muon may enter the baffle-covered region from outside, when:
\begin{align}
\rho_\textit{es} < \rho < \rho_\textit{pb}
& \quad\mathrm{and}\quad \cos(\phi_0-\psi) > 0 \quad, \label{eq:condpassbafflefromoutside}
\end{align}
\noindent
or it may cross the baffles from inside, when
\begin{align}
\rho_\textit{pb} < \rho < \rho_\textit{es}  & \quad\mathrm{and}\quad \cos(\phi_0-\psi) < 0 \quad. \label{eq:condpassbafflefrominside}
\end{align}
\noindent
See also Fig.~\ref{fig:dualmirror} for a visualization of the geometry. Figure~\ref{fig:bafflecondition} shows the baffle-crossing condition for different muon inclination angles $\nu$ and their azimuthal projections on ground $\psi$.  

In this case, the baffle-crossing distance, $D_c$, is given by: 
\begin{equation}
D_c = \left\{ 
\begin{array}{cc}
\frac{-\rho\cdot \cos(\phi_0-\psi)+\sqrt{R_\textit{sb}^2-\rho^2\cdot\sin^2(\phi_0-\psi)}}{\nu} \quad & \mathrm{for}\quad \rho <= R_\textit{sb} \\
\frac{-\rho\cdot \cos(\phi_0-\psi)-\sqrt{R_\textit{sb}^2-\rho^2\cdot\sin^2(\phi_0-\psi)}}{\nu} \quad & \mathrm{else}
 \end{array}
\right. 
\label{eq:Dc} \quad.
\end{equation}

We briefly demonstrate that, for realistic telescope designs, muons traversing the baffles cannot produce Cherenkov light emitted below the height of the secondary mirror that simultaneously intersects the central hole of the primary mirror. We consider a maximally inclined muon $\nu_\mathrm{max} \approx \textit{FOV}/2 - \theta_c$ traversing the upper edge of the baffles and directed toward the mirror center, hence $\cos(\phi_0-\psi) = 1$. In this case, the condition in Eq.~\ref{eq:condpassbafflefromoutside} becomes $\rho_\mathrm{min,baffle~crossing} > R_\textit{sb} - \nu_\mathrm{max}\cdot D_\textit{es} \approx R_\textit{sb} -  D_\textit{es}\cdot \left(\textit{FOV}/2 - \theta_c\right)$. Considering a Cherenkov photon emitted exactly toward the mirror center, we further require  $R_\mathrm{hole} \lesssim \rho_\mathrm{min,baffle~crossing} - \theta_c\cdot D_\mathrm{es} \approx R_\textit{sb} - D_\textit{es}\cdot \left(\textit{FOV}/2 - 2\,\theta_c\right)$, which is comfortably satisfied. 

By contrast, a muon traversing the secondary mirror can readily produce Cherenkov light that intersects the central hole and is therefore not imaged onto the camera, even if the muon itself intersects the the primary mirror. 

If the muon passes through the hole, $\rho < R_\mathrm{hole}$, Cherenkov light is imaged only if: 
\begin{align}
L_\mathrm{max,hole} &= 
\frac{\rho \cdot  \cos\left(\phi - \phi_{0}\right) + \sqrt{R_\mathrm{hole}^{2} -\rho^{2} \cdot\sin\left(\phi - \phi_{0}\right)^{2}}}{\theta_{c}} 
+ c \cdot (R_\mathrm{hole}^2 - \rho^2) \cdot \Big(1 + \frac{\nu}{\theta_c} \cdot \cos(\phi-\psi)
\Big) 
< D_\textit{es} \quad. \label{eq:conditionhole}
\end{align}
In that case, 
\begin{align}
L_{2,\mathrm{min}} &= D_\textit{es} - L_\mathrm{max,hole}\quad.
\label{eq:Lminhole}
\end{align}
If the muon does not traverse the central mirror hole, we test for central hole losses as:
\begin{align}
R_\mathrm{hole}^2 &> \rho^{2} \cdot\sin\left(\phi - \phi_{0}\right)^{2} \quad \mathrm{and}\quad
|\phi-\phi_0| < \pi/2 \quad \mathrm{and}\quad
\nonumber\\
L_\mathrm{min,hole} &= 
\frac{\rho \cdot  \cos\left(\phi - \phi_{0}\right) - \sqrt{R_\mathrm{hole}^{2} -\rho^{2} \cdot\sin\left(\phi - \phi_{0}\right)^{2}}}{\theta_{c}} < D_\textit{es}
\label{eq:conditionhole2}
\end{align}
In that case, 
\begin{align}
L_{2,\mathrm{min}} &= L_\mathrm{min,hole}
\label{eq:Lminhole2}
\end{align}

\begin{figure}[h]
\centering
\resizebox{0.97\textwidth}{!}{%
\begin{tikzpicture}[
  node distance=7mm and 7mm
]
\node (sph) [decision, align=center] {
M1 has spherical shape\\ if seen from above?};
\node (sphyes) [block, below left=of sph, xshift=-1.2cm] {$L_\mathrm{max} =$~Eq.~\ref{eq:Lmax}};
\node (sphno) [block, below right=of sph, xshift=1.2cm, align=center] {$L_\mathrm{max}= D(\phi-\phi_0)/\theta_c + $\\ 2$^\mathrm{nd}$ order corrections Eq.~\ref{eq:Lmax2ndOrder}};
\node (shadowsec) [decision, below=of sph, align=center] {Shadow from M2 or baffles?\\ Condition~\ref{eq:shadowcondition2} fulfilled ?};
\node (shadowsecyes) [decision, below left=of shadowsec, xshift=-1.2cm] {Shadow only from baffles?\\ Condition~\ref{eq:condbaffle} fulfilled ?};
\node (shadowsecno) [block, below=of sphno, xshift=0cm, yshift=-2.5cm, align=center] {$L_{2,\mathrm{max}} = 0$\\
$L_{2,\mathrm{min}}=0$};
\node (shadowbaffleno) [block, below right=of shadowsecyes, xshift=1.2cm] { $L_{2,\mathrm{max}} =$~Eq.~\ref{eq:L2max}};
\node (shadowbaffleyes) [block, below left=of shadowsecyes, xshift=-1.2cm] { $L_{2,\mathrm{max}} =$~Eq.~\ref{eq:L2maxpb}};
\node (muonbaffle) [decision, below=of shadowsecyes, yshift=0.2cm, align=center] {Muon traverses baffle?\\Conditions ~\ref{eq:condpassbafflefromoutside} or~\ref{eq:condpassbafflefrominside} fulfilled?};
\node (muonbaffleyes) [block, below left=of muonbaffle, xshift=-1.2cm] { $L_{2,\mathrm{min}} = D_c$, Eq.~\ref{eq:Dc}};
\node (muonmirror) [decision, below right=of muonbaffle, xshift=-0.2cm, align=center] {Muon traverses M2, but no baffle?\\Condition~\ref{eq:murhoes} fulfilled?};
\node (muonmirrorno) [block, below right=of muonmirror, xshift=-0.2cm] { $L_{2,\mathrm{min}} = $ Eq.~\ref{eq:L2min}};
\node (muonmirroryes) [decision, below left=of muonmirror, xshift=-0.8cm] {Muon traverses central hole?\\
Condition~\ref{eq:conditionhole} fulfilled?};
\node (muonholeyes) [block, below left=of muonmirroryes, xshift=2.2cm,yshift=-0.35cm] { $L_{2,\mathrm{min}} = $ Eq.~\ref{eq:Lminhole}};
\node (muonholeno) [decision, below right=of muonmirroryes, xshift=-0.2cm] { Other light losses from central hole? 
Condition~\ref{eq:conditionhole2} fulfilled?};
\node (lightlossyes) [block, below left=of muonholeno, xshift=1.2cm] { $L_{2,\mathrm{min}} = $ Eq.~\ref{eq:Lminhole2}};
\node (lightlossno) [block, below right=of muonholeno, xshift=-1.2cm] { $L_{2,\mathrm{min}} = D_\textit{ps}$ };
\node (end) [block, below=of lightlossyes,xshift=4cm] {$L=L_\mathrm{max}-L_{2,\mathrm{max}}+L_{2,\mathrm{min}}$};
\draw [arrow] (sph) -- node[left, sloped, above] {Yes} (sphyes);
\draw [arrow] (sph) -- node[right,sloped, above] {No} (sphno);
\draw [arrow] (sphyes) --  (shadowsec);
\draw [arrow] (sphno) --  (shadowsec);
\draw [arrow] (shadowsec) -- node[left,sloped, above] {Yes} (shadowsecyes);
\draw [arrow] (shadowsec) -- node[right,sloped, above] {No} (shadowsecno);
\draw [arrow] (shadowsecyes) -- node[left,sloped, above] {Yes} (shadowbaffleyes);
\draw [arrow] (shadowsecyes) -- node[right,sloped, above] {No} (shadowbaffleno);
\draw [arrow] (shadowbaffleyes) -- (muonbaffle);
\draw [arrow] (shadowbaffleno) -- (muonbaffle);
\draw [arrow] (muonbaffle) -- node[left,sloped, above] {Yes} (muonbaffleyes);
\draw [arrow] (muonbaffle) -- node[right,sloped, above] {No} (muonmirror);
\draw [arrow] (muonmirror) -- node[right,sloped, above] {No} (muonmirrorno);
\draw [arrow] (muonmirror) -- node[left,sloped, above] {Yes} (muonmirroryes);
\draw [arrow] (muonmirroryes) -- node[left,sloped, above] {Yes} (muonholeyes);
\draw [arrow] (muonmirroryes) -- node[right,sloped, above] {No} (muonholeno);
\draw [arrow] (muonholeno) -- node[right,sloped, above] {No} (lightlossno);
\draw [arrow] (muonholeno) -- node[left,sloped, above] {Yes} (lightlossyes);
\draw [arrow] (shadowsecno) |- (end);
\draw [arrow] (muonbaffleyes) |- (end);
\draw [arrow] (muonholeyes) |- (end);
\draw [arrow] (muonmirrorno) |- (end);
\draw [arrow] (lightlossno) |- (end);
\draw [arrow] (lightlossyes) |- (end);
\end{tikzpicture}
}
\caption{Flowchart of the computation of the Cherenkov light received in the presence of shadowing losses in a dual-mirror telescope.}
\label{fig:flow}
\end{figure}
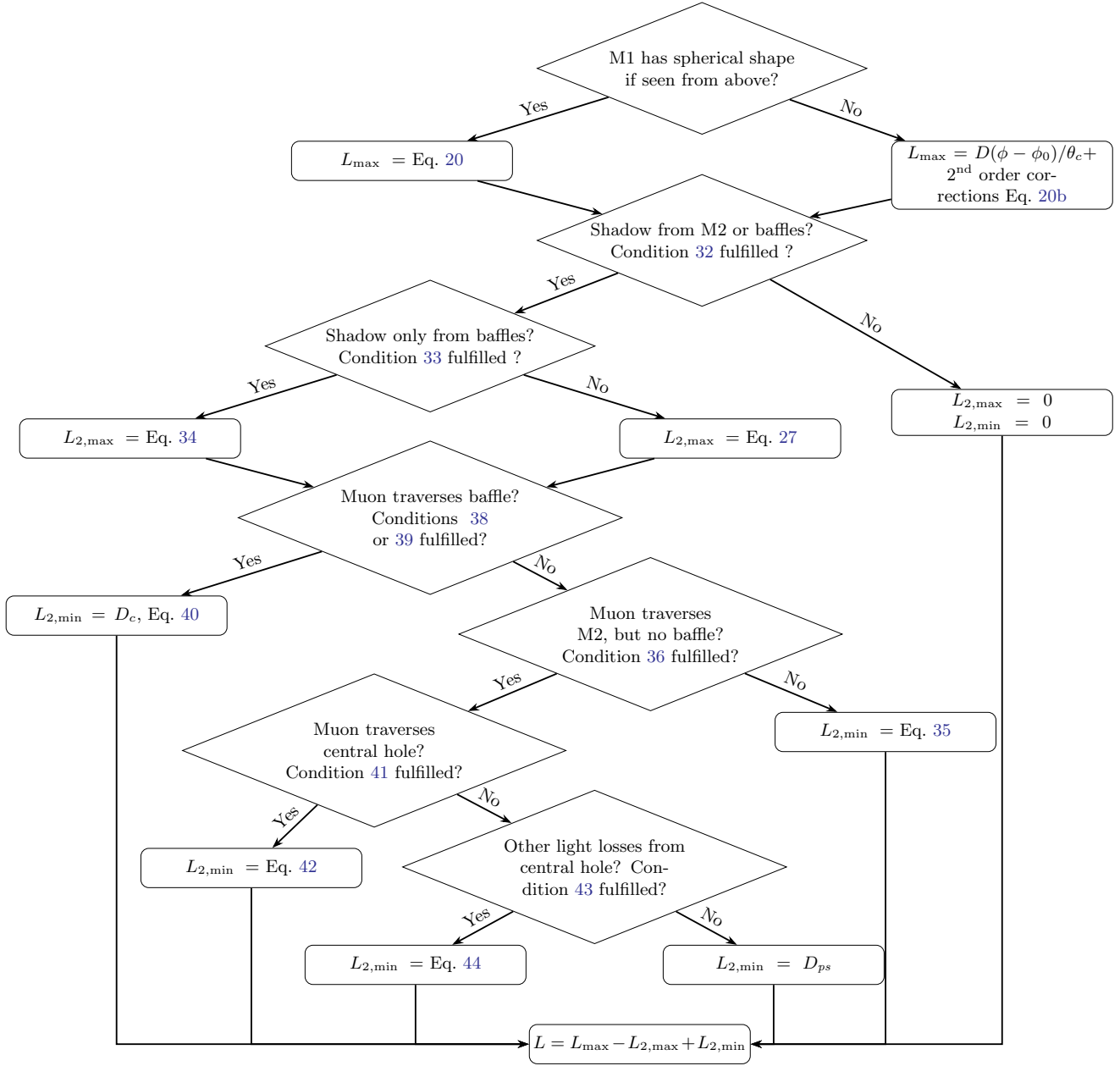

\begin{figure}[tbh]
\setlength\tabcolsep{6pt}
  \adjustboxset{width=0.95\linewidth,valign=c}
\centering
\begin{tabularx}{1.0\linewidth}
{@{}c@{\hspace{6pt}}>{\centering\arraybackslash}X
      @{\hspace{6pt}}>{\centering\arraybackslash}X
      @{\hspace{6pt}}>{\centering\arraybackslash}X@{}}
\rotatebox[origin=c]{90}{\small Track length shadow start}
& \includegraphics[width=0.28\textwidth]{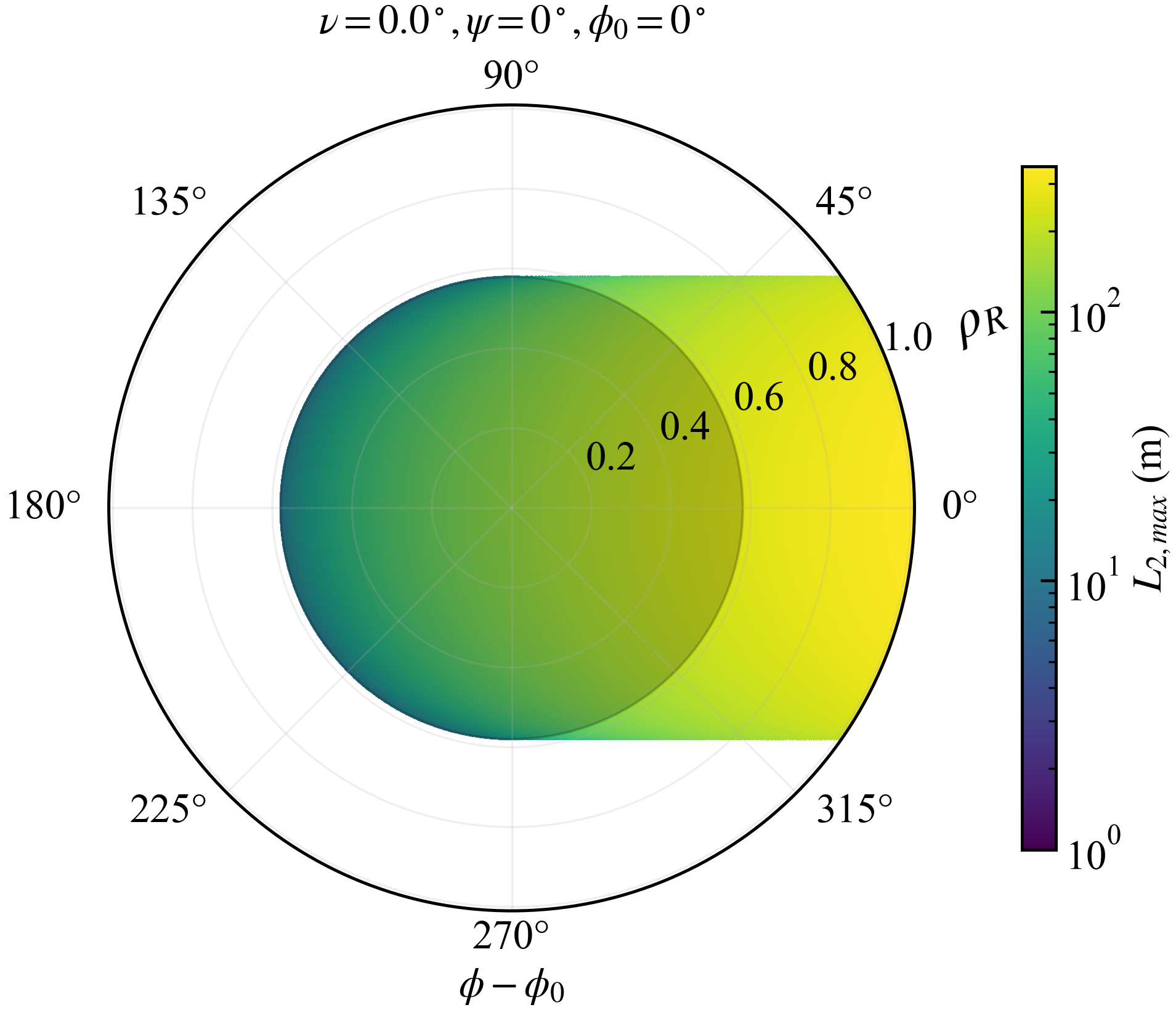}
& \includegraphics[width=0.28\textwidth]{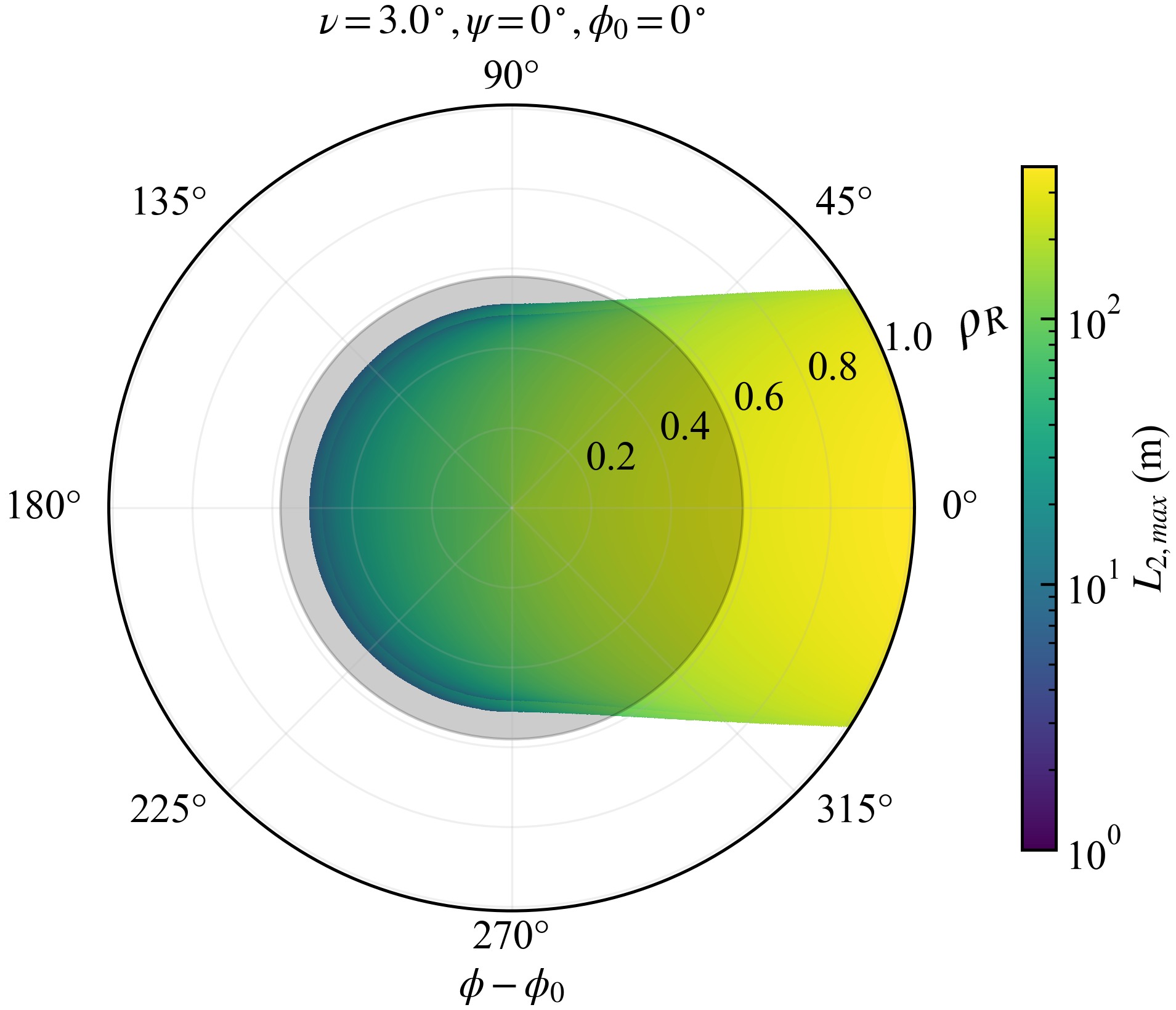}
& \includegraphics[width=0.28\textwidth]{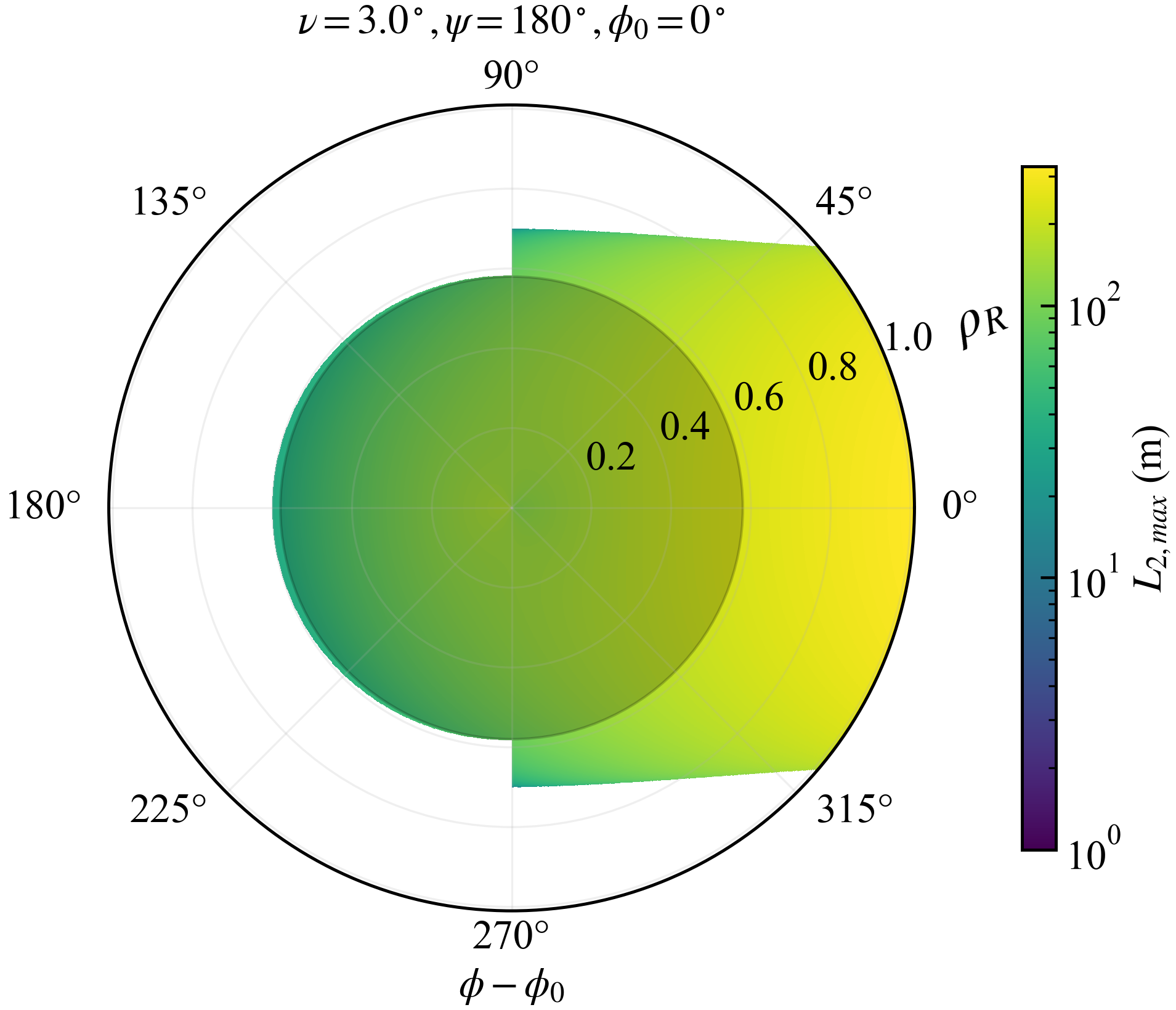} \\
\rotatebox[origin=c]{90}{\small Track length below shadow}
& \includegraphics[width=0.28\textwidth]{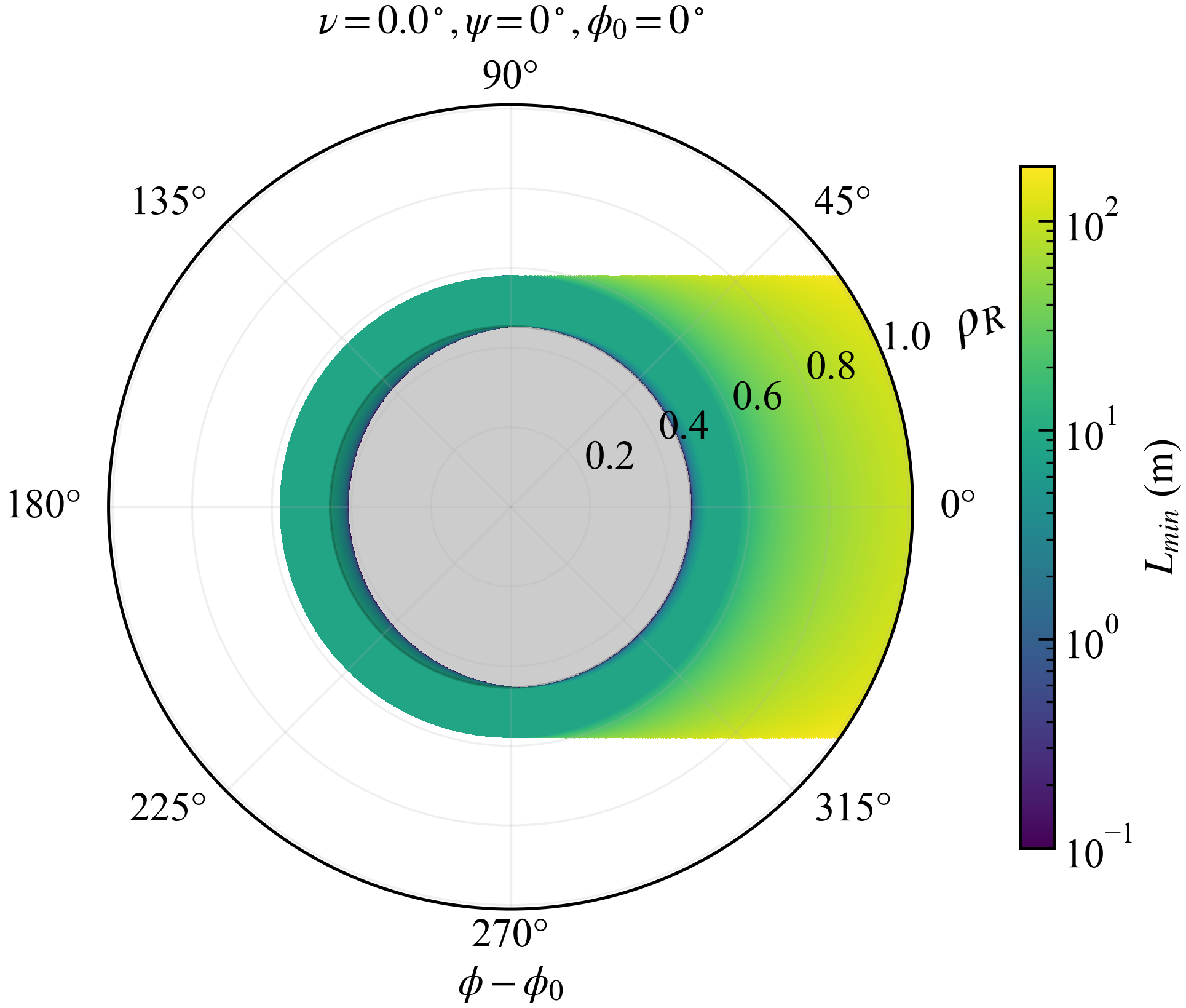}
& \includegraphics[width=0.28\textwidth]{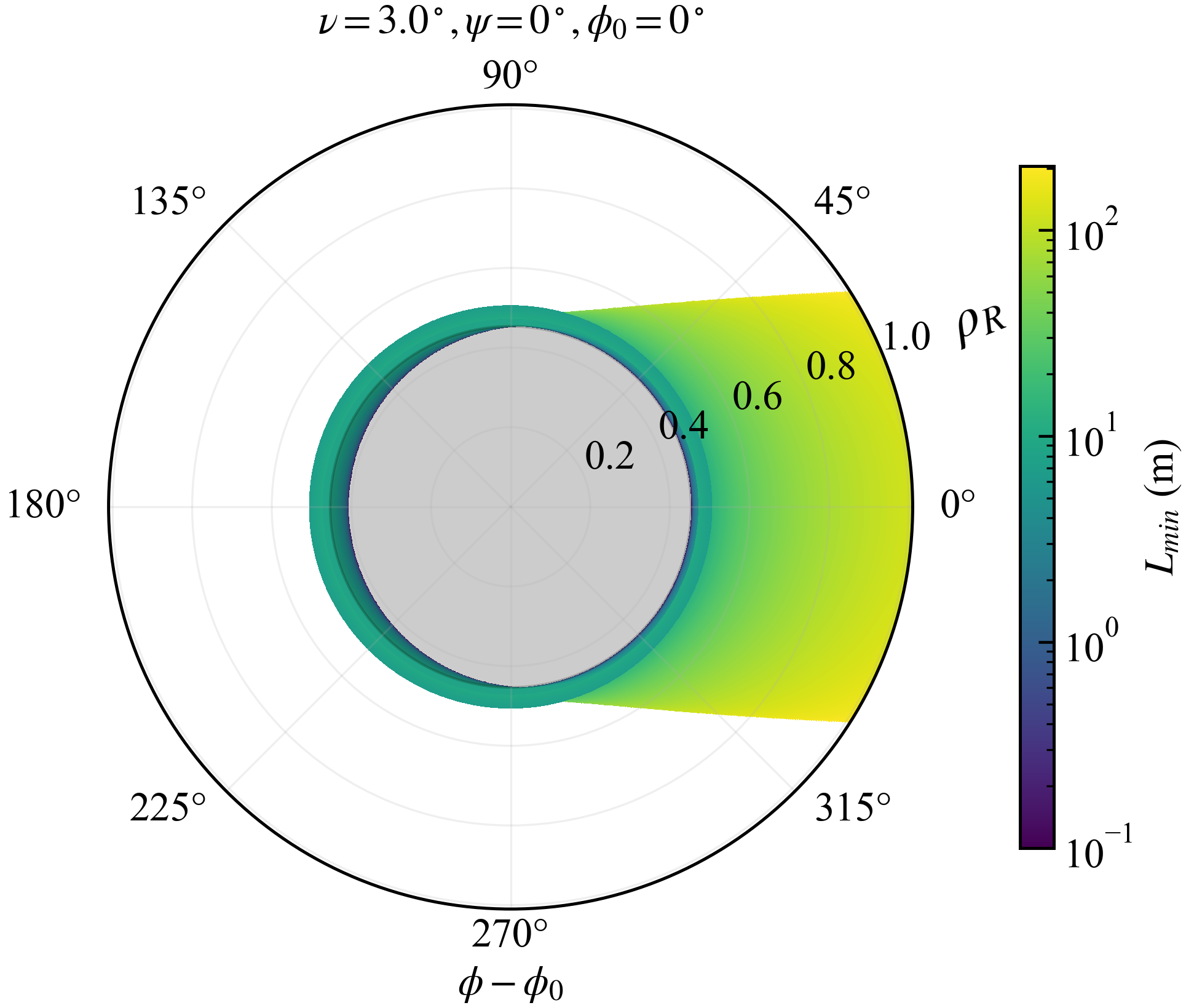}
& \includegraphics[width=0.28\textwidth]{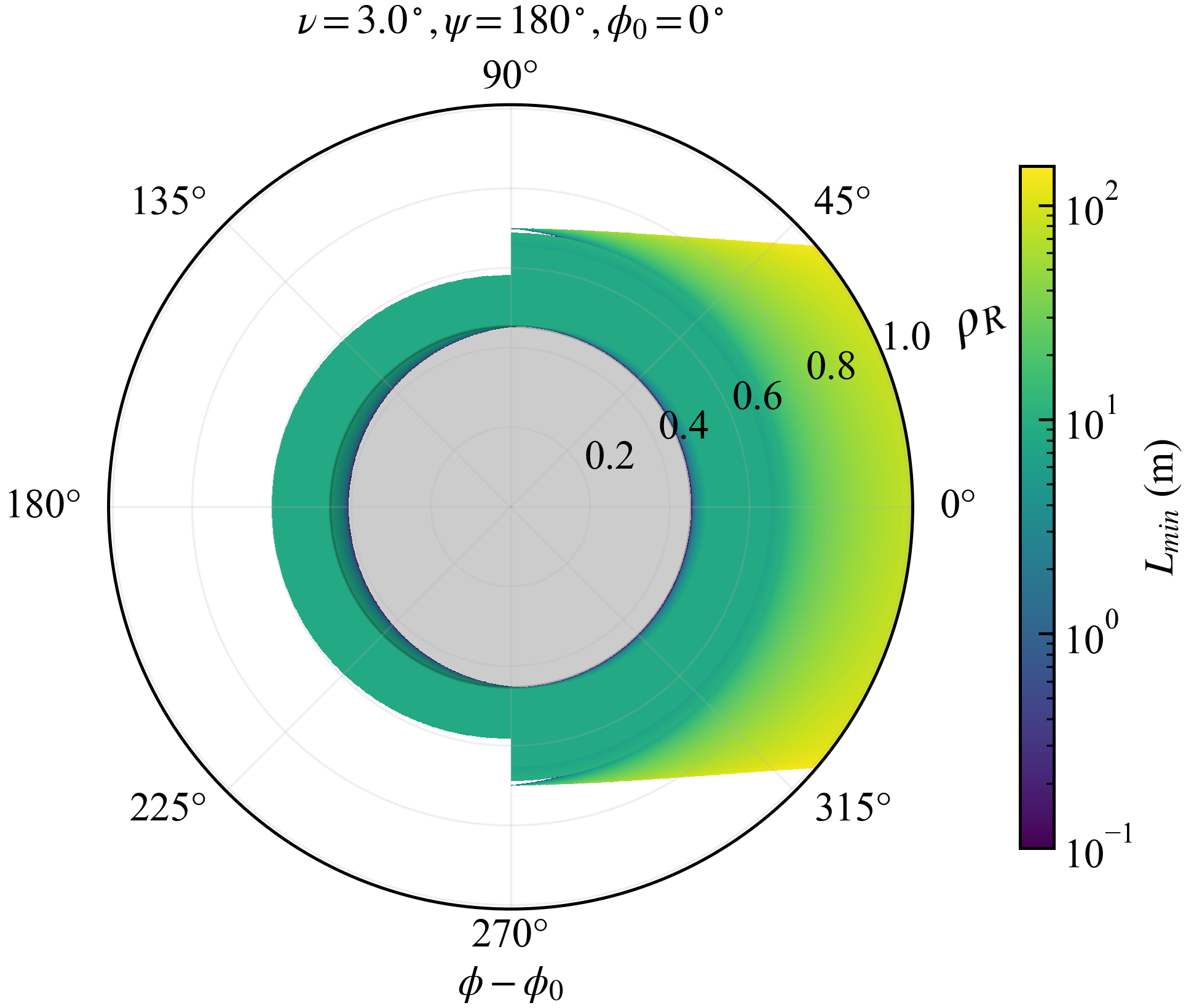} \\
\rotatebox[origin=c]{90}{{\small Imaged track length}}
& \includegraphics[width=0.28\textwidth]{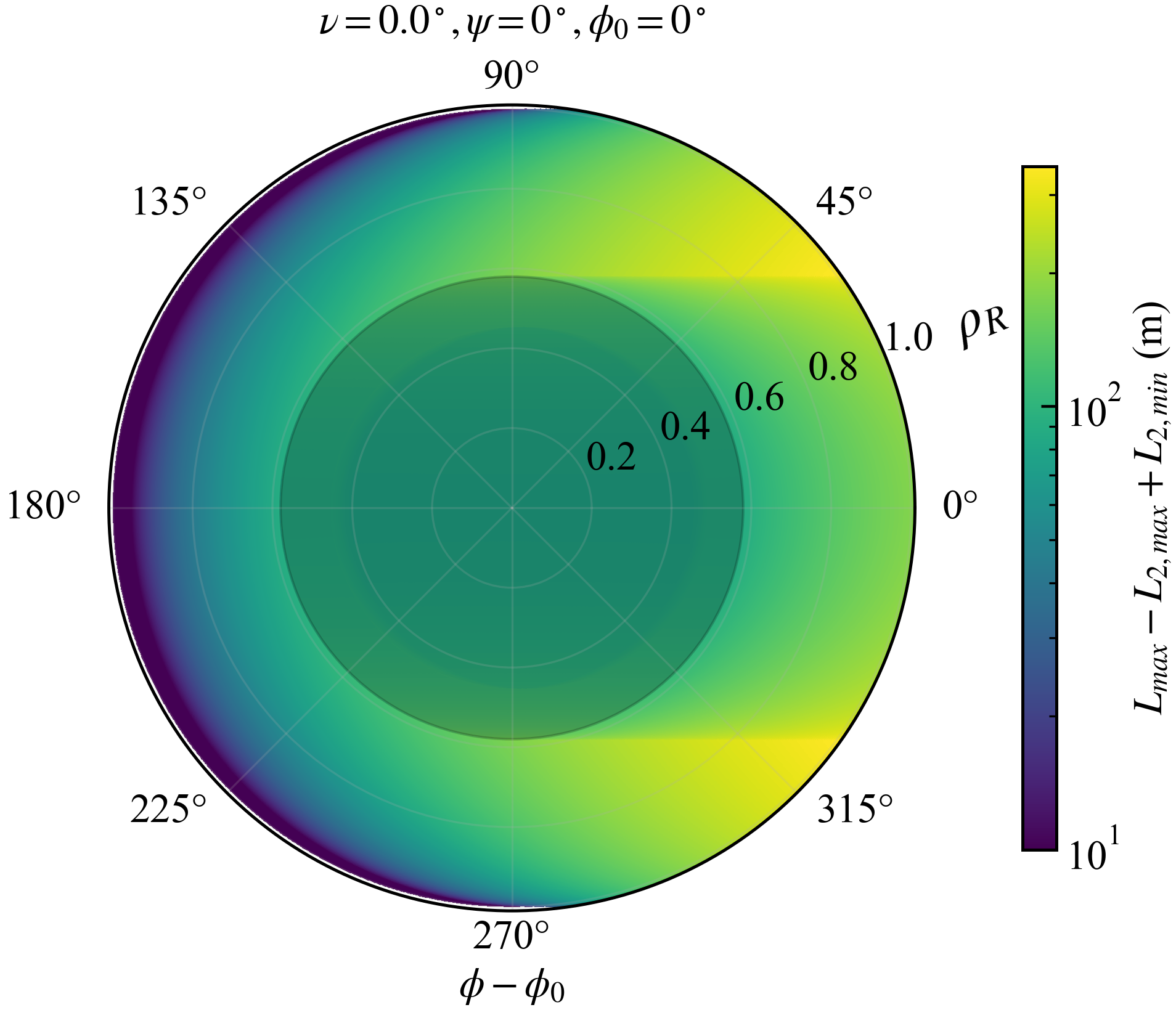} 
& \includegraphics[width=0.28\textwidth]{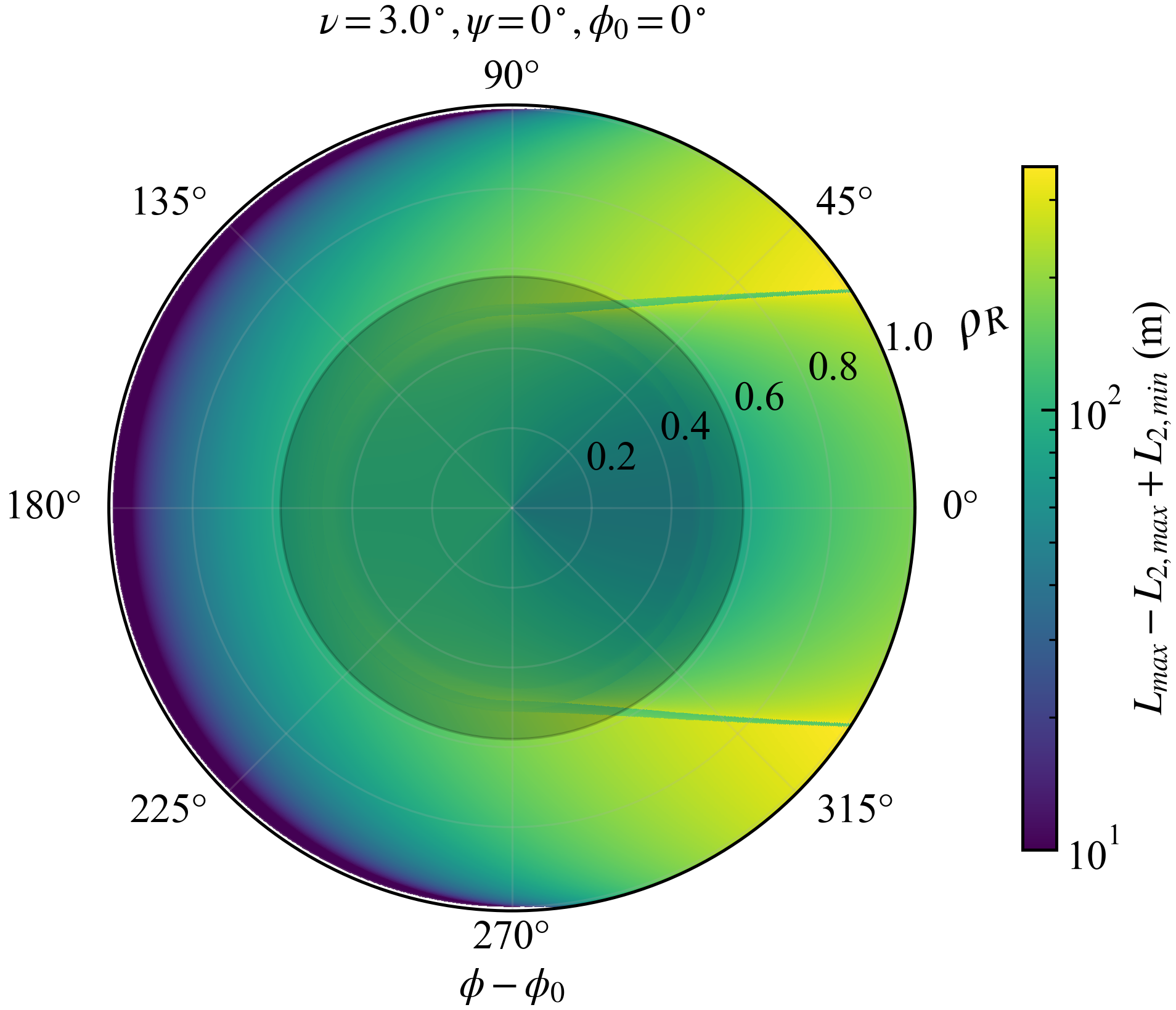}
& \includegraphics[width=0.28\textwidth]{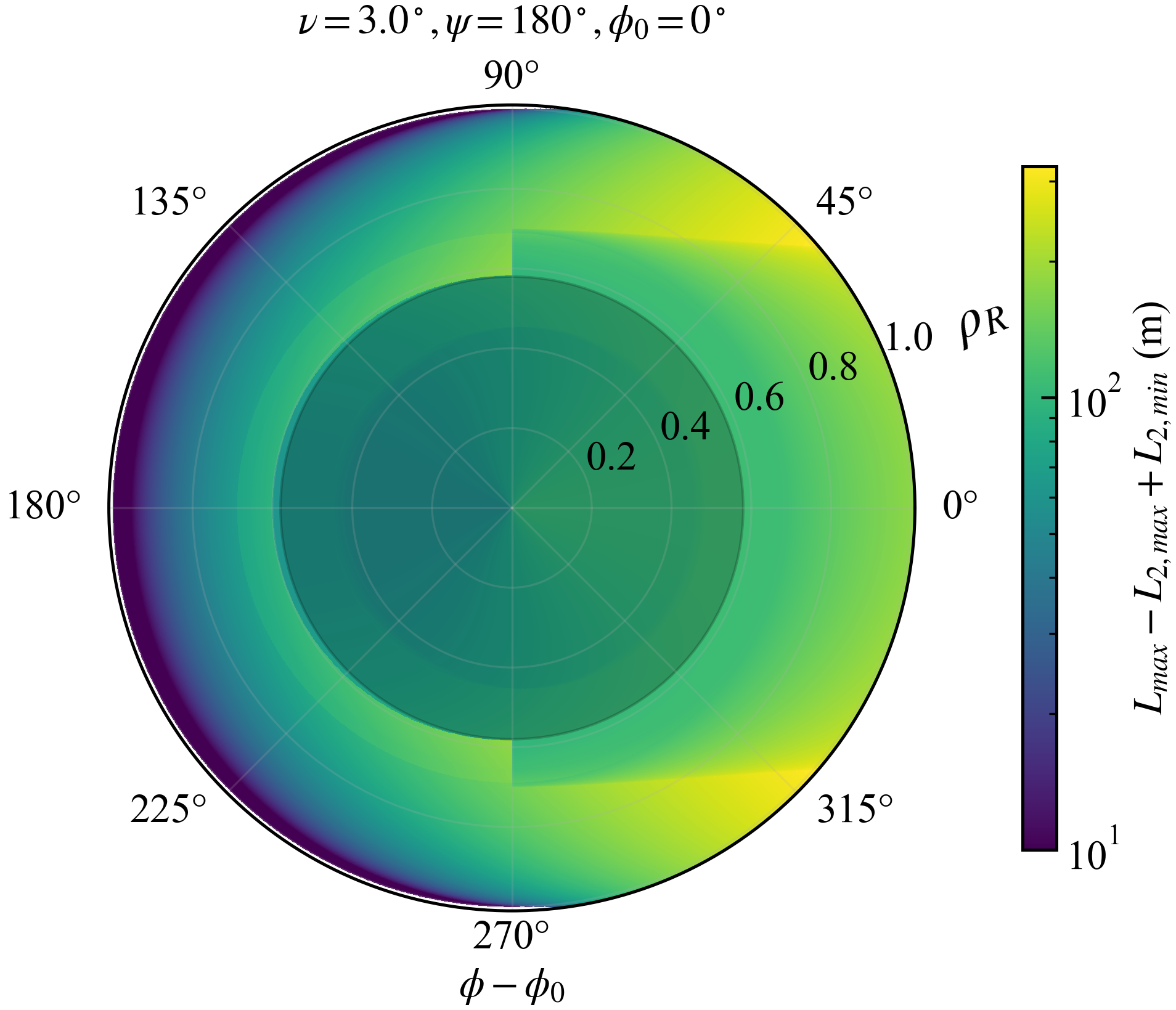} \\
\rotatebox[origin=c]{90}{\small Shadowed rel. to unshadowed}
& \includegraphics[width=0.28\textwidth]{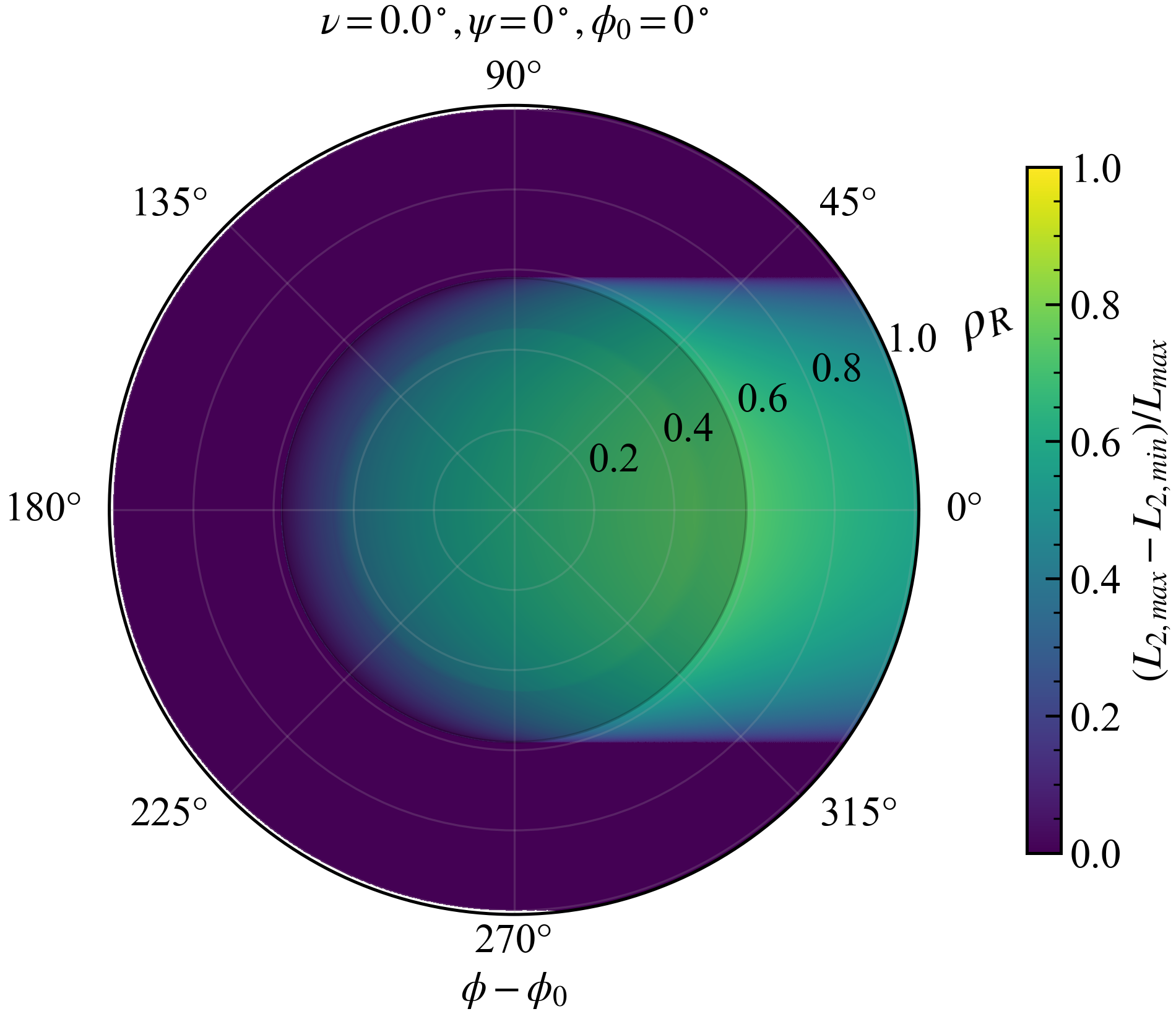}
& \includegraphics[width=0.28\textwidth]{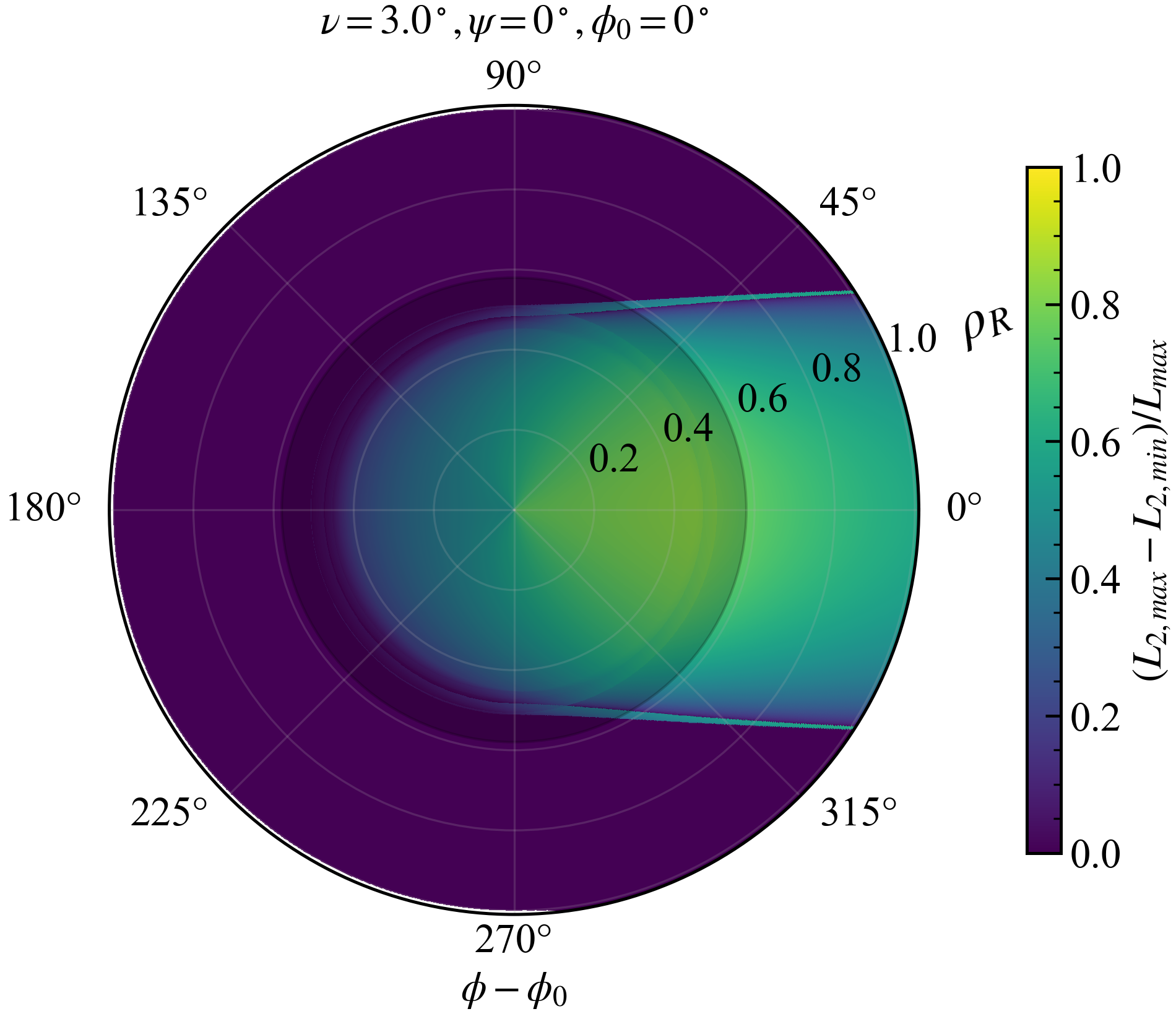}
& \includegraphics[width=0.28\textwidth]{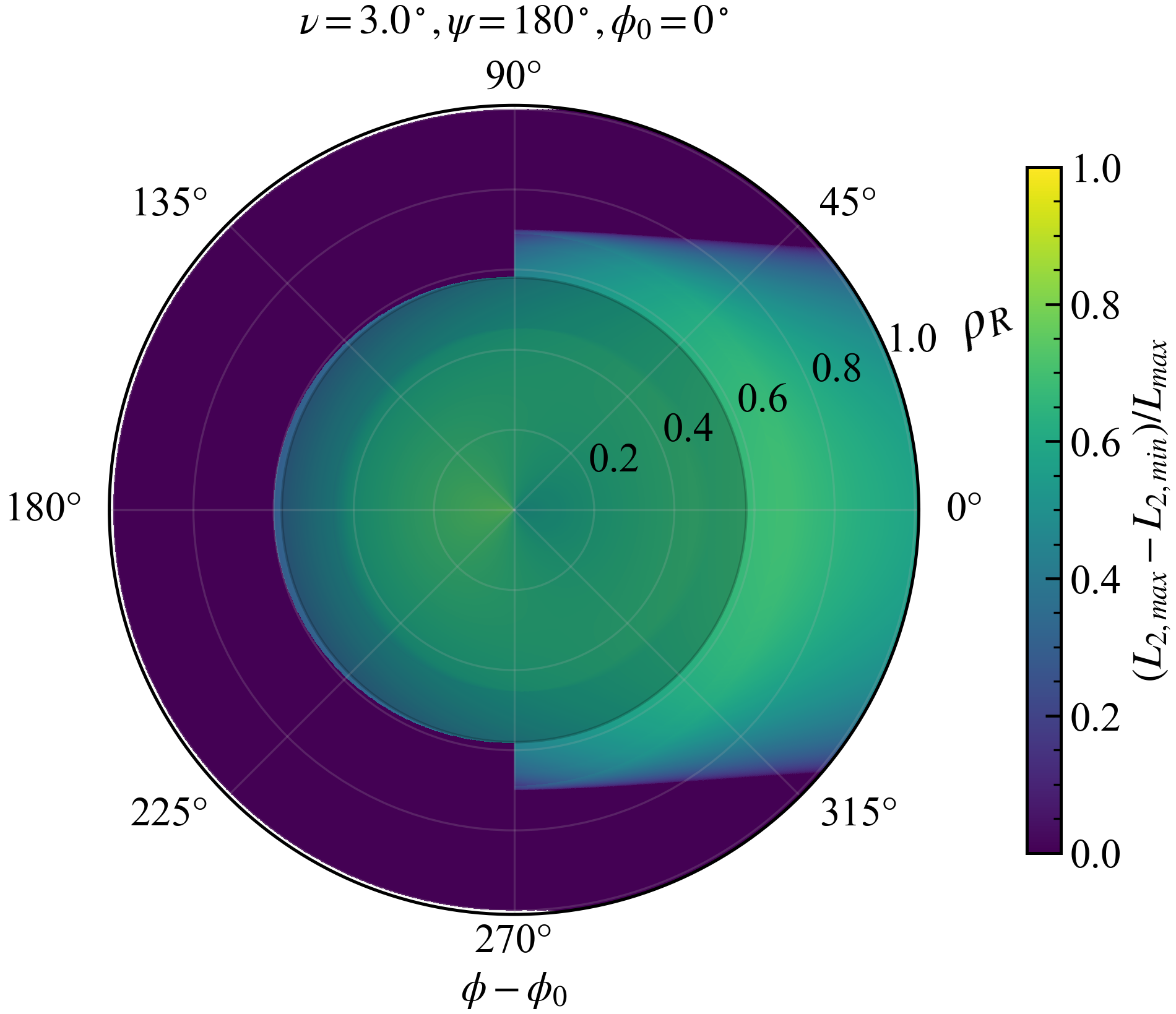} \\
\rotatebox[origin=c]{90}{\small Shadow rel. to Vacanti}
& \includegraphics[width=0.28\textwidth]{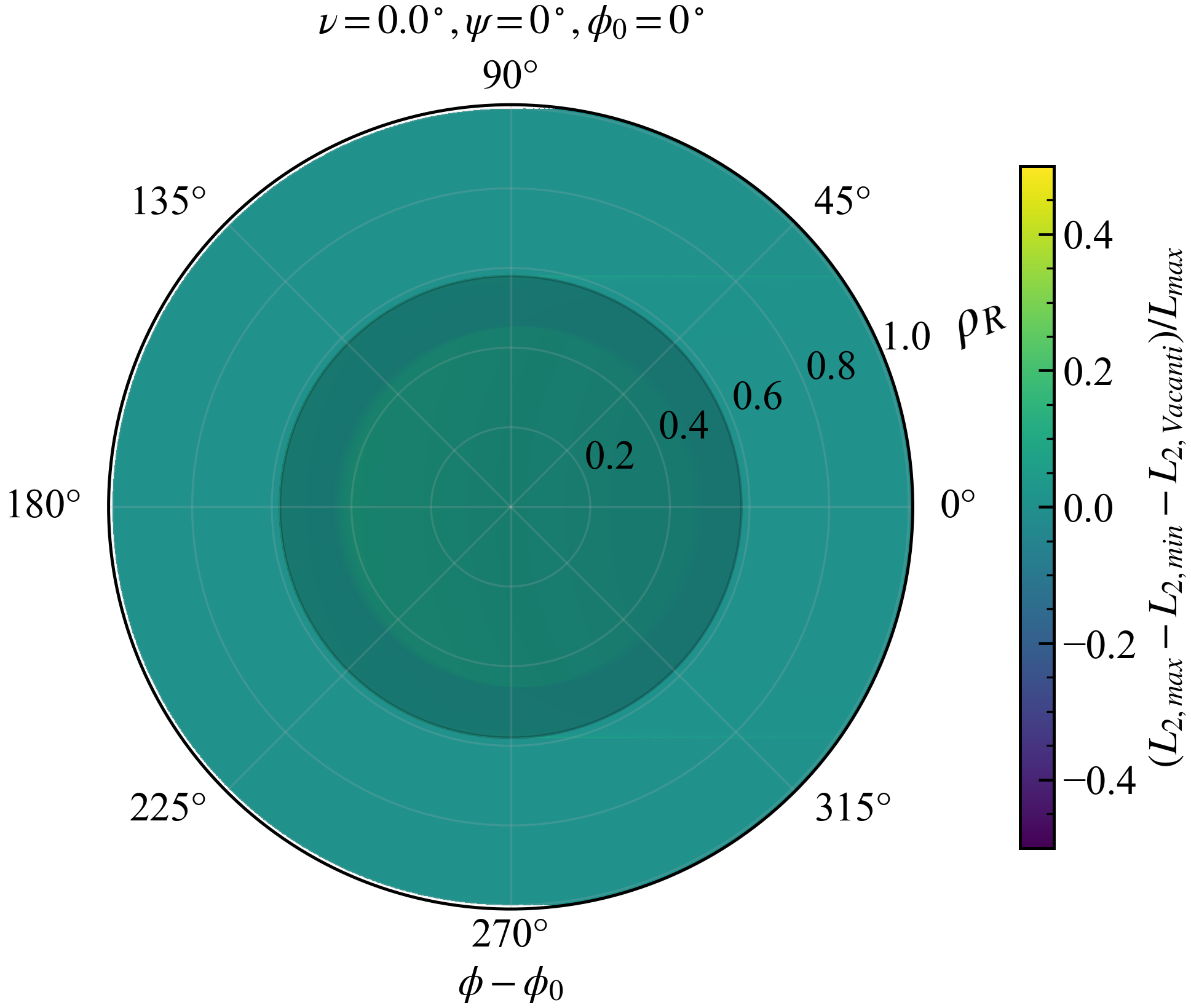}
& \includegraphics[width=0.28\textwidth]{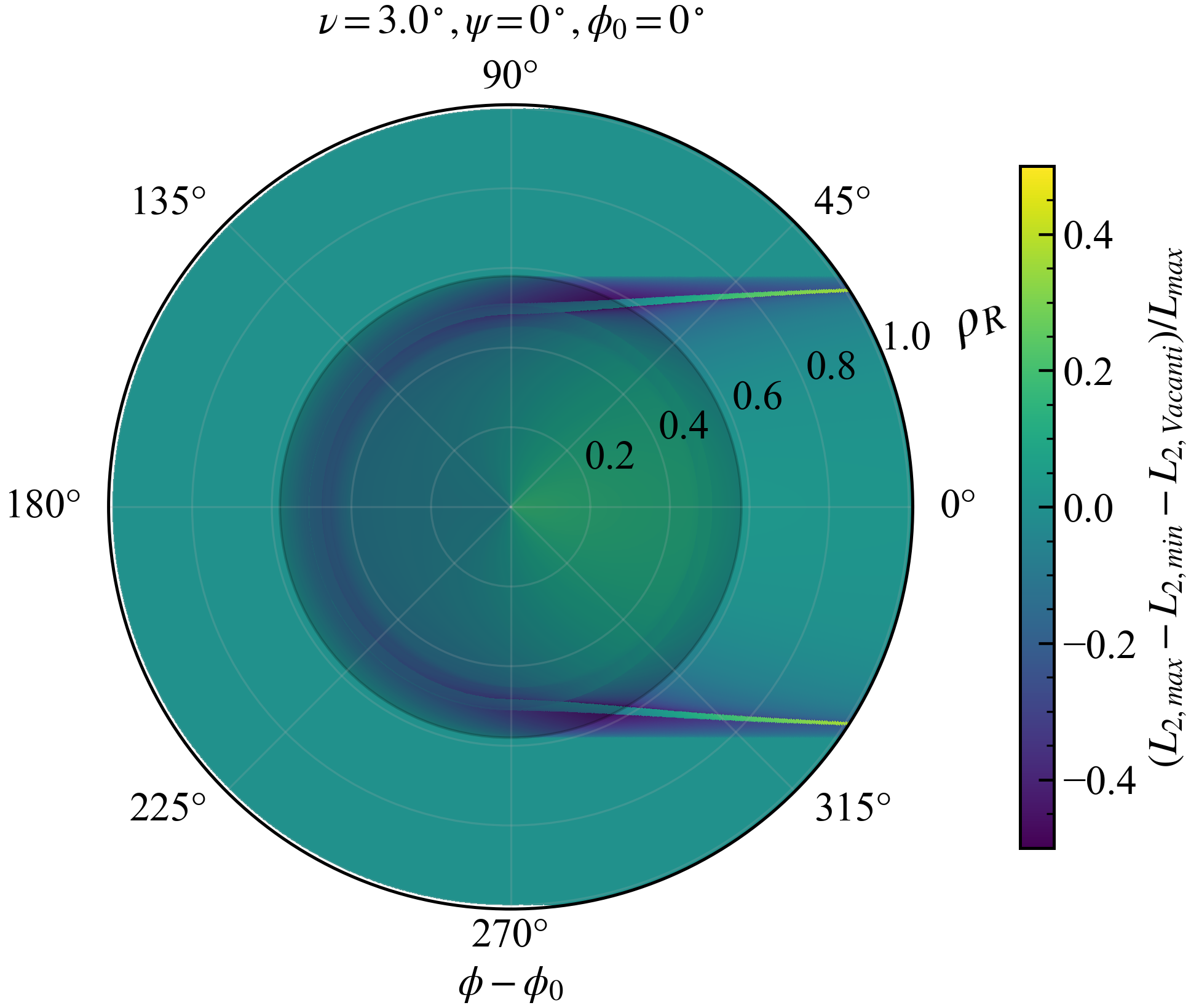}
& \includegraphics[width=0.28\textwidth]{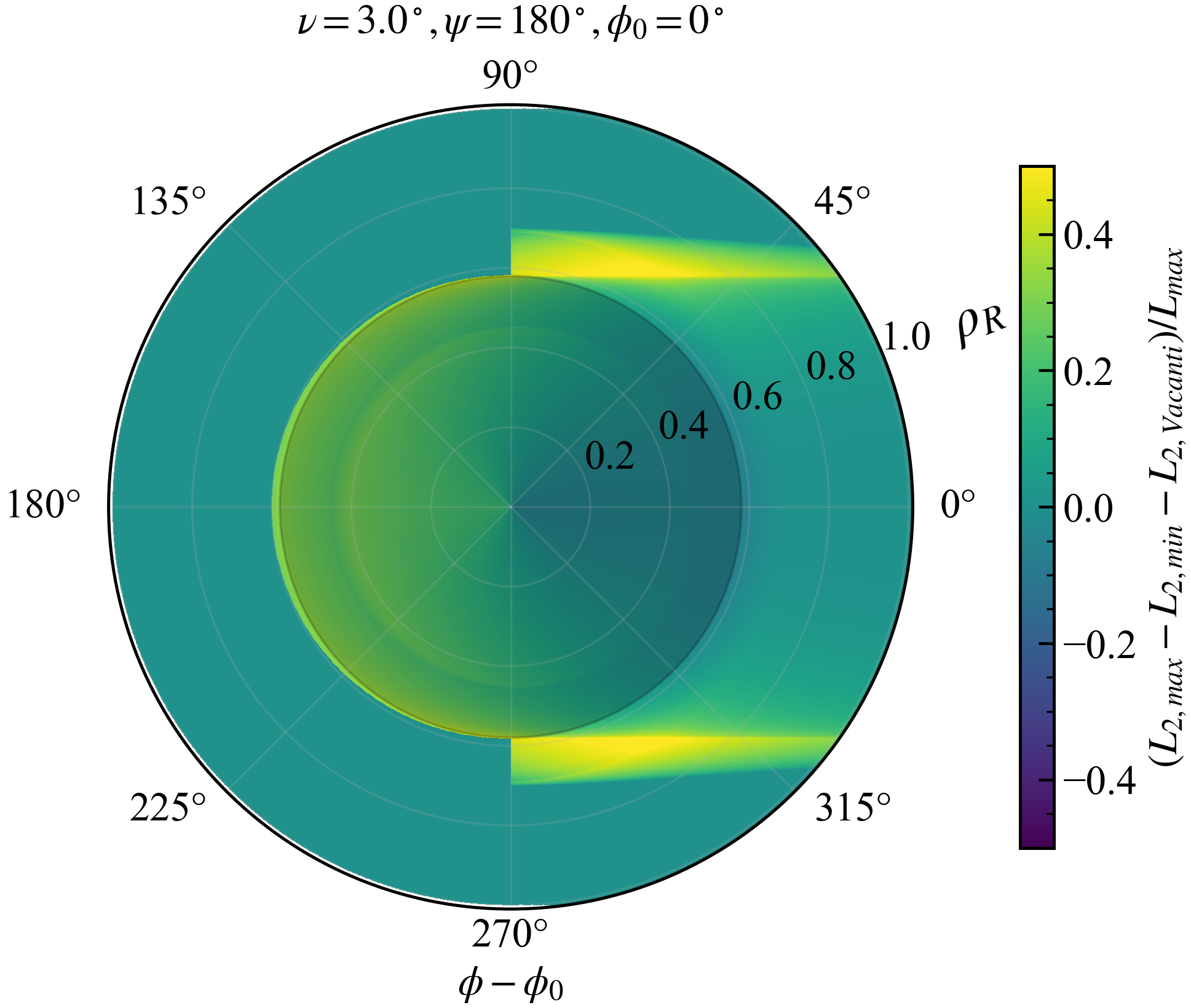} \\
\end{tabularx}
\caption{Shadow parameters shown for muons with different normalized impact distances $\rho_R$ and impact angles $\phi_0$, for the parameters of an SCT. In the left panels, the muon is not inclined, in the center, it is inclined toward the left ($\psi =0^\circ$), and in the right panels, it is inclined to the right ($\psi=180^\circ$), with a fixed inclination angle $\nu=3^\circ$.  Note that the azimuthal axis represents the photon emission angle $\phi-\phi_0$; see also Fig.~\ref{fig:Shadowcondition} for further details. 
From top to bottom, the panels show: the maximum shadow distance, $L_{2,\mathrm{max}}$; the minimum shadow distance, $L_{2,\mathrm{min}}$; the total imaged muon track length, $L_{\mathrm{max}}-L_{2,\mathrm{max}}+L_{2,\mathrm{min}}$;
the relative shadow contribution with respect to the unshadowed case; and the relative shadow contribution  compared with the simplified case of a hole of radius $R_\textit{sb}$ using Vacanti's formula. 
The grey shaded region in the center indicates the central hole in the primary mirror. 
}
\label{fig:L2SCT}
\end{figure}

\begin{figure}[tbh]
\setlength\tabcolsep{6pt}
  \adjustboxset{width=0.95\linewidth,valign=c}
\centering
\begin{tabularx}{1.0\linewidth}
{@{}c@{\hspace{6pt}}>{\centering\arraybackslash}X
      @{\hspace{6pt}}>{\centering\arraybackslash}X
      @{\hspace{6pt}}>{\centering\arraybackslash}X@{}}
\rotatebox[origin=c]{90}{\small Track length shadow start}
& \includegraphics[width=0.28\textwidth]{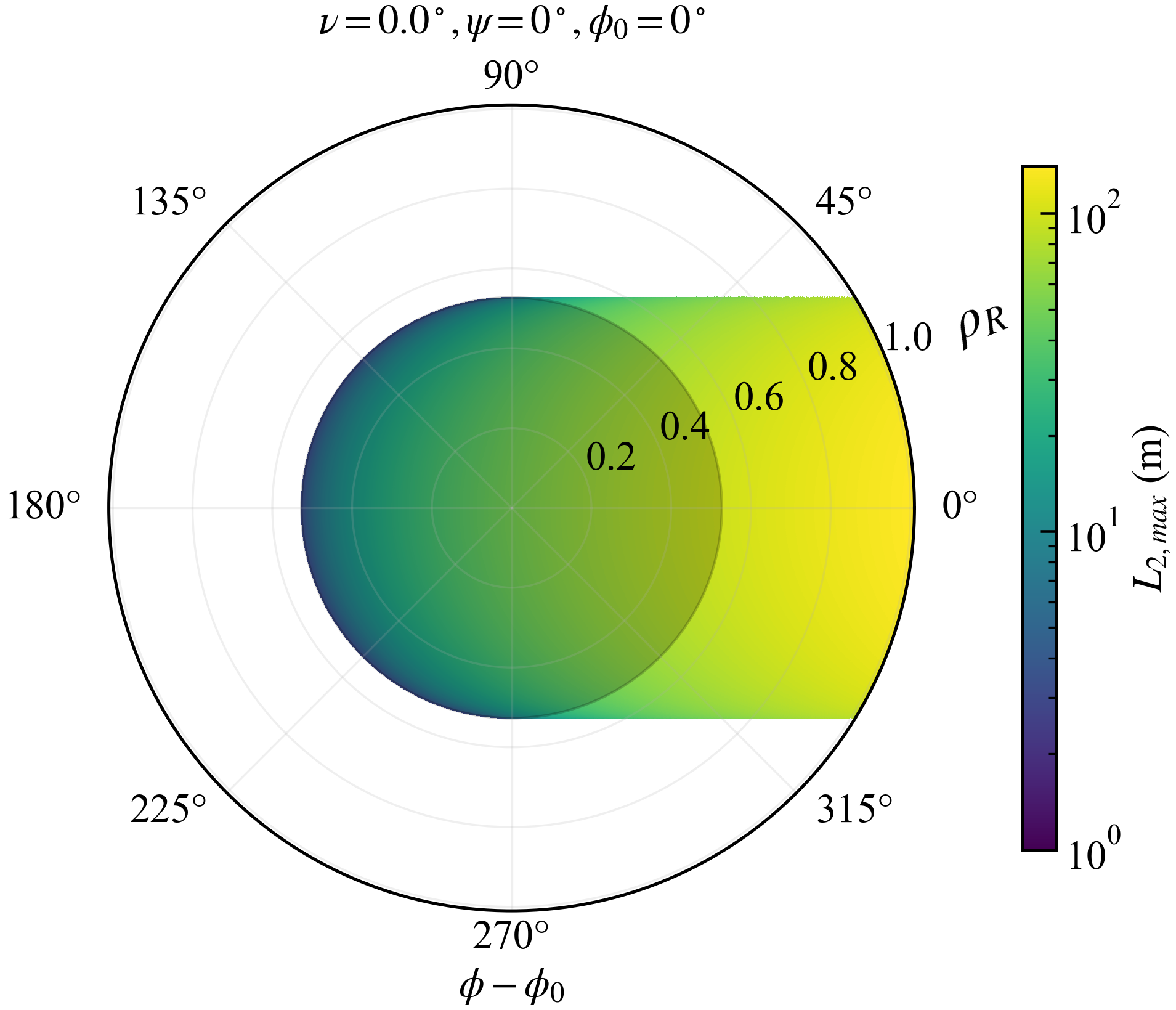}
& \includegraphics[width=0.28\textwidth]{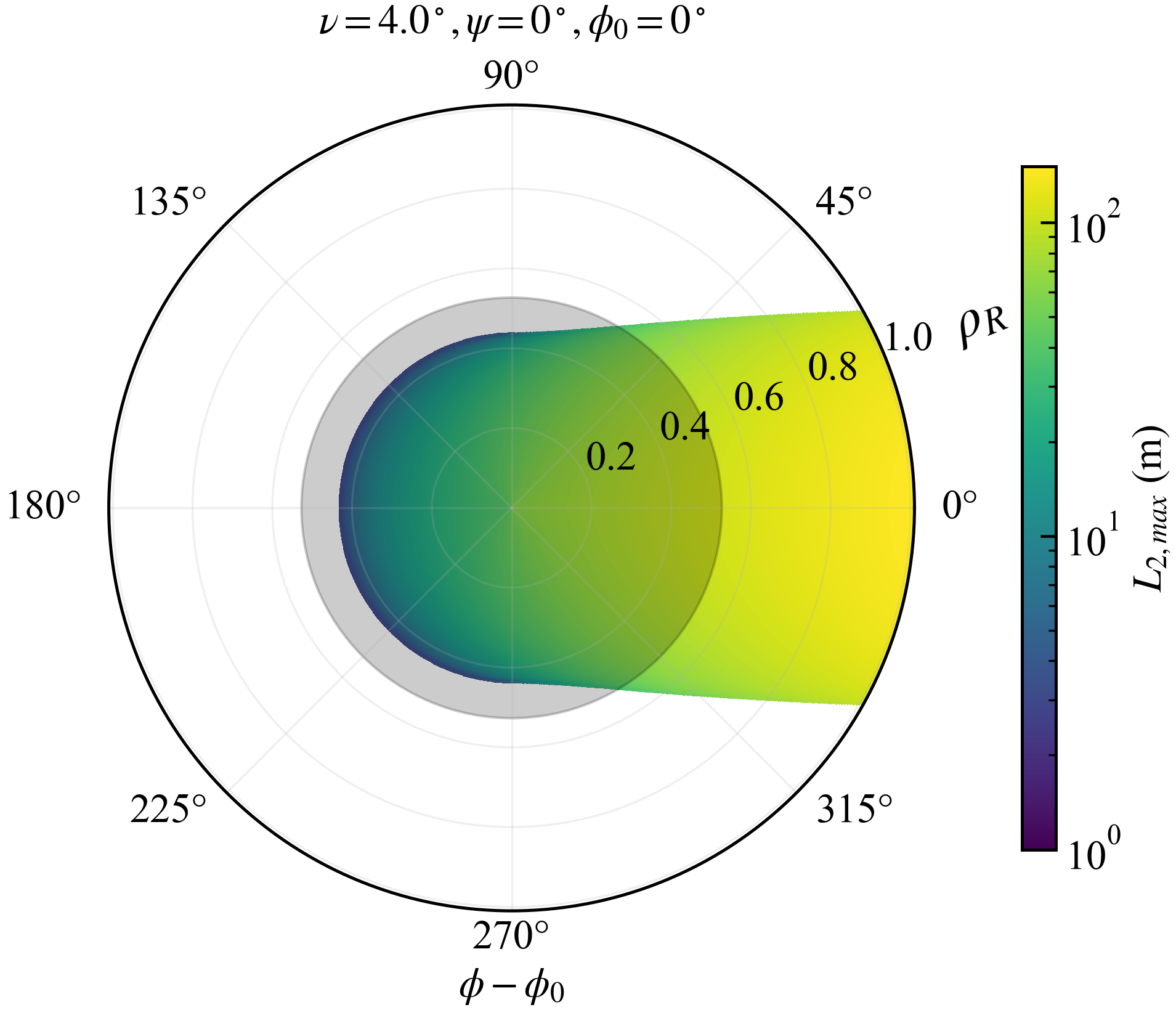}
& \includegraphics[width=0.28\textwidth]{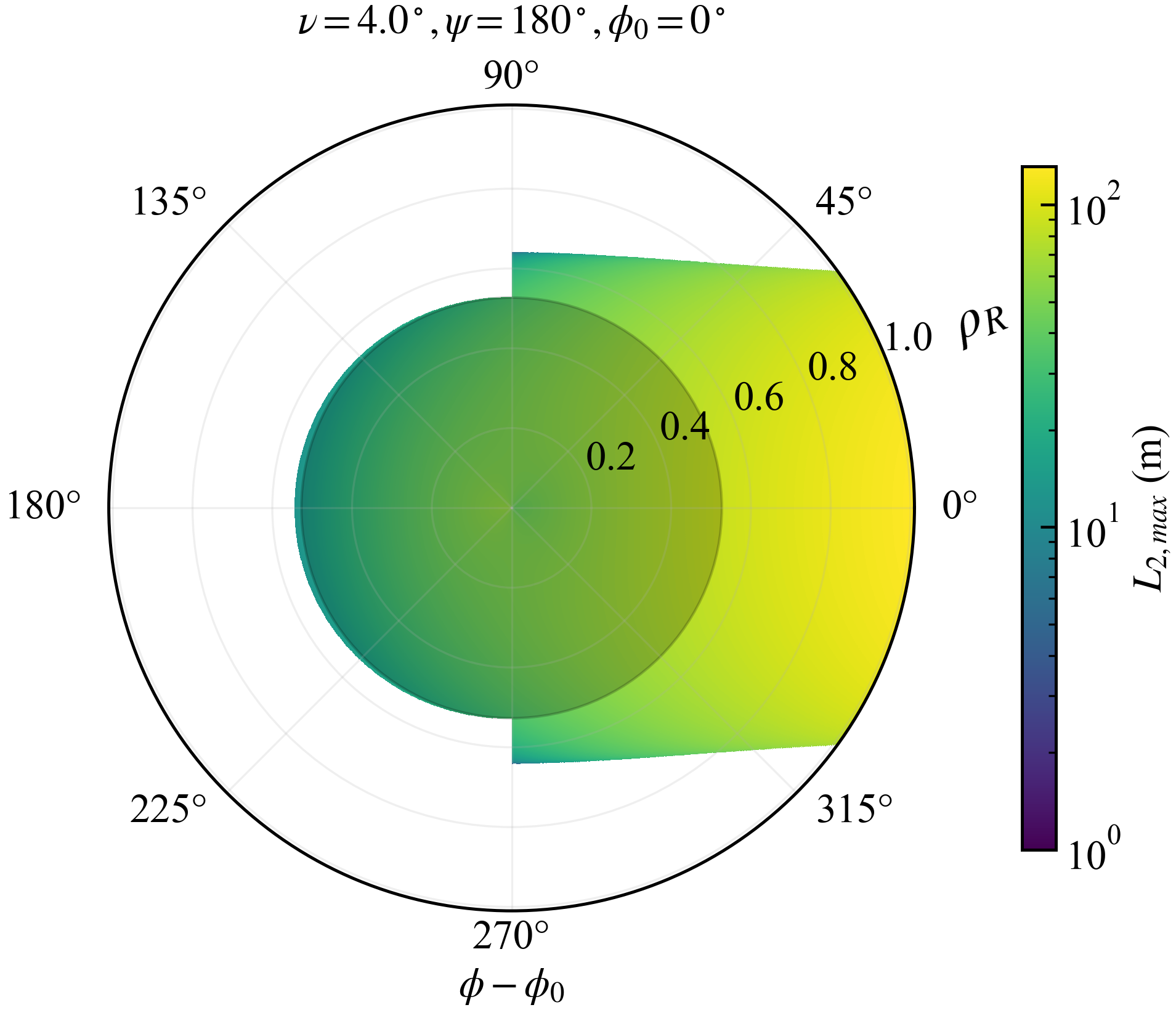} \\
\rotatebox[origin=c]{90}{\small Track length below shadow}
& \includegraphics[width=0.28\textwidth]{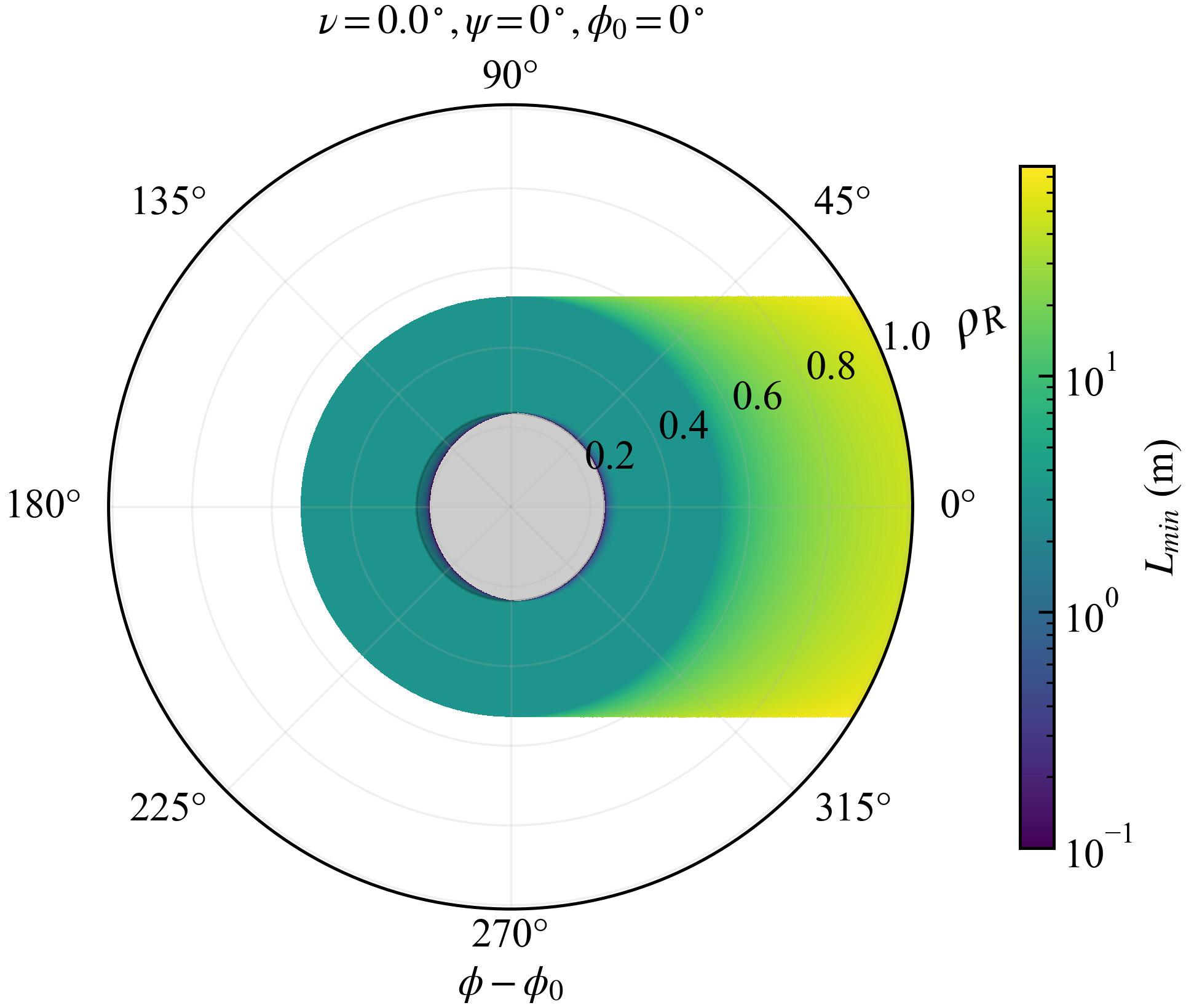}
& \includegraphics[width=0.28\textwidth]{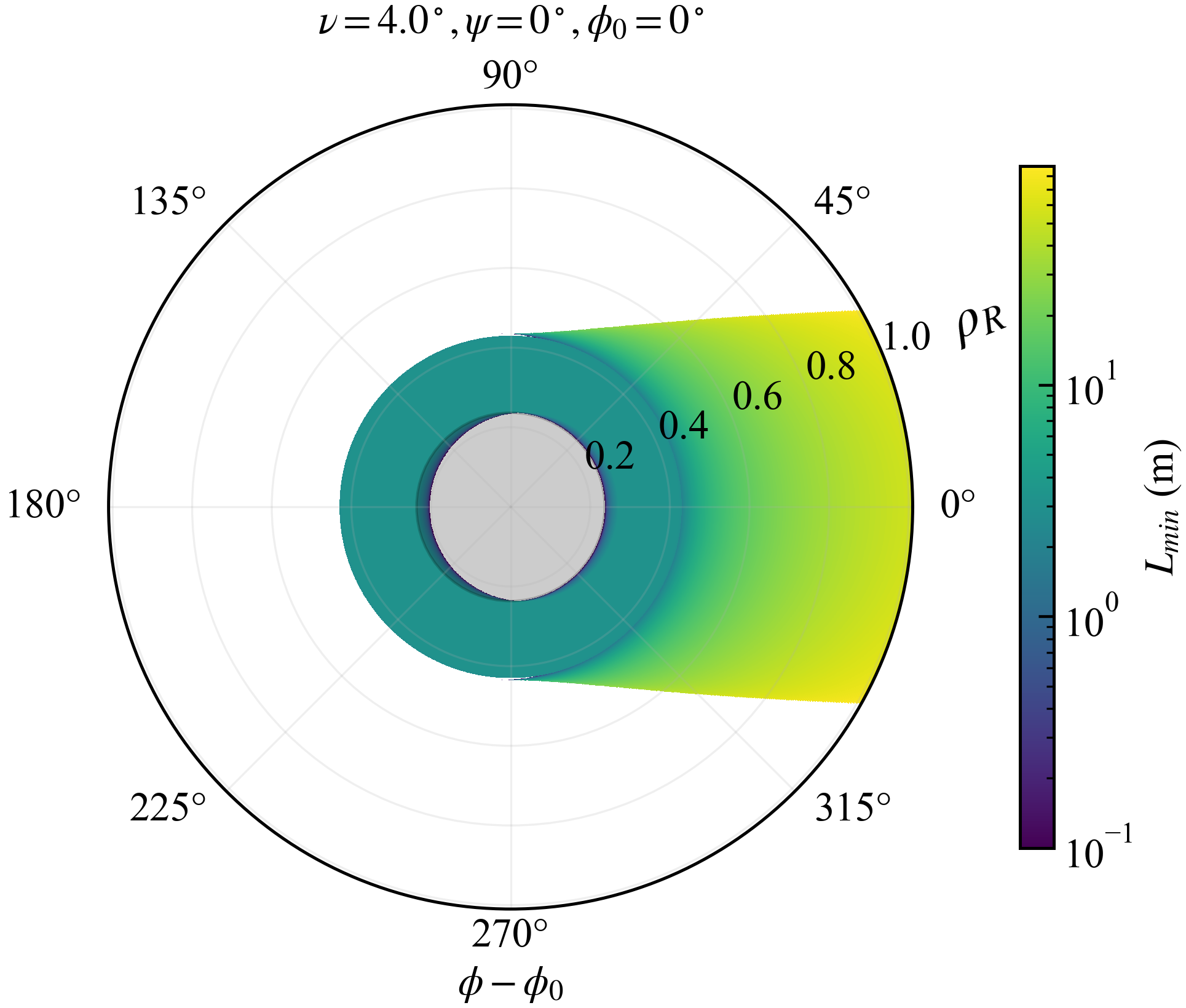}
& \includegraphics[width=0.28\textwidth]{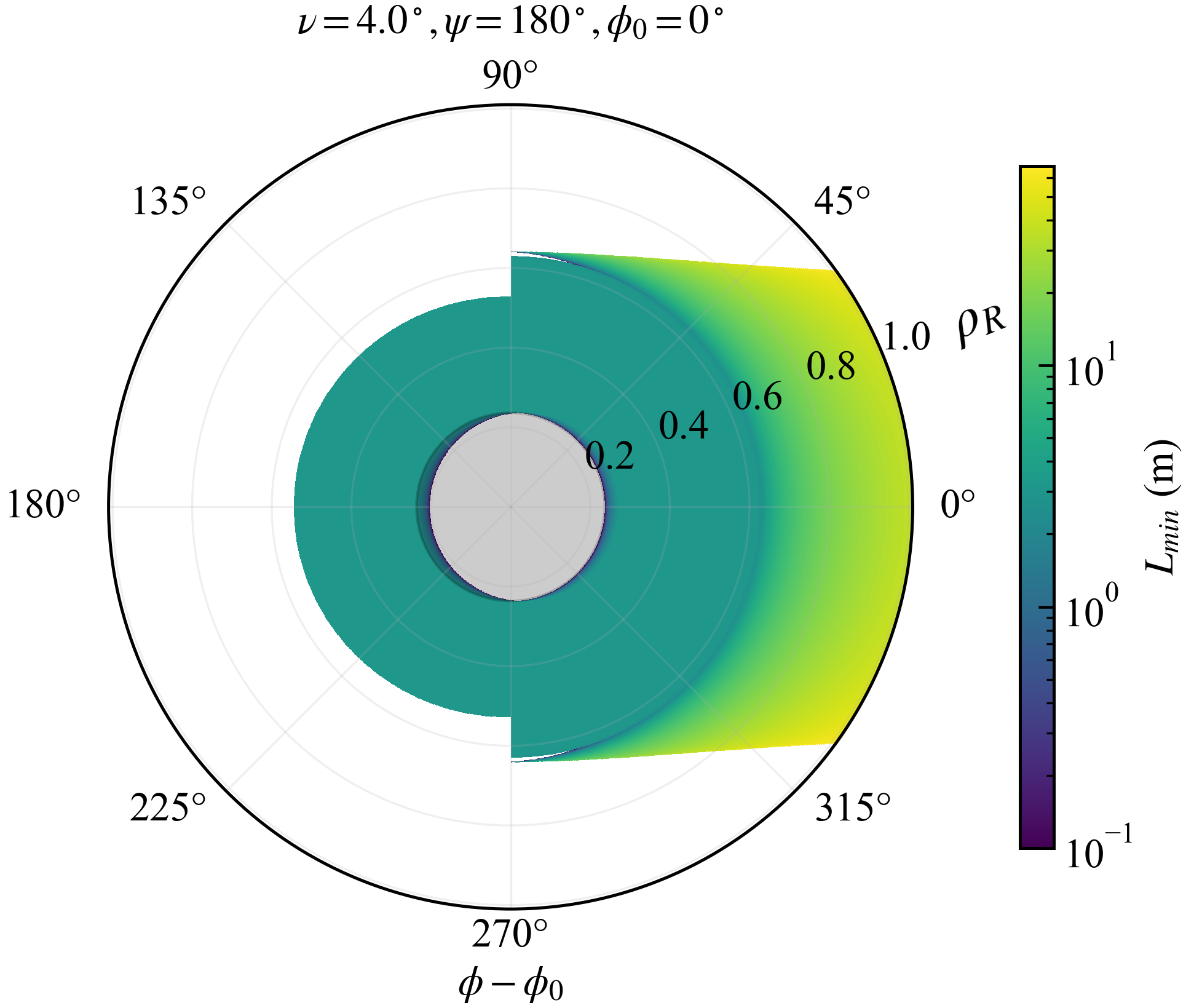} \\
\rotatebox[origin=c]{90}{{\small Imaged track length}}
& \includegraphics[width=0.28\textwidth]{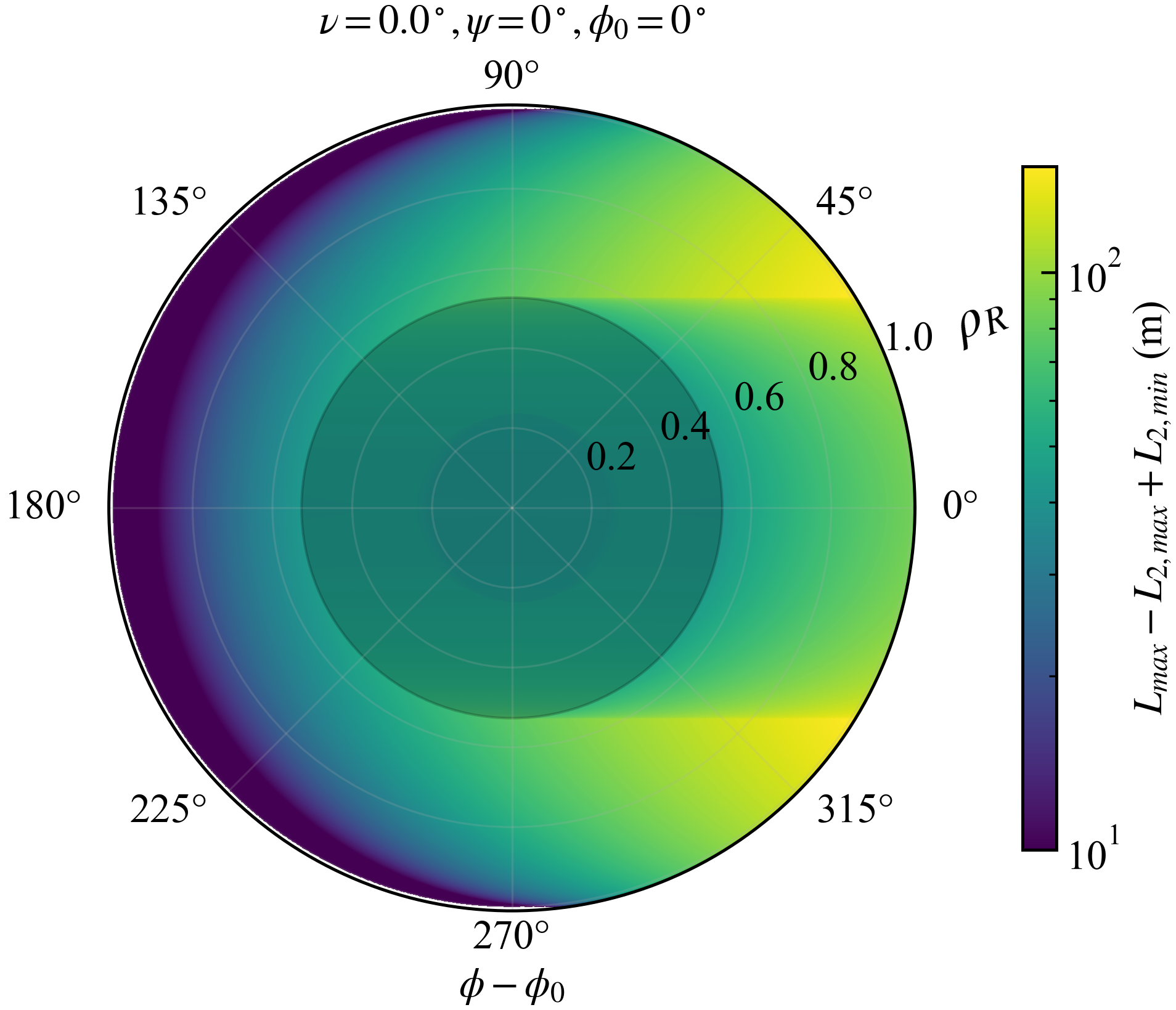} 
& \includegraphics[width=0.28\textwidth]{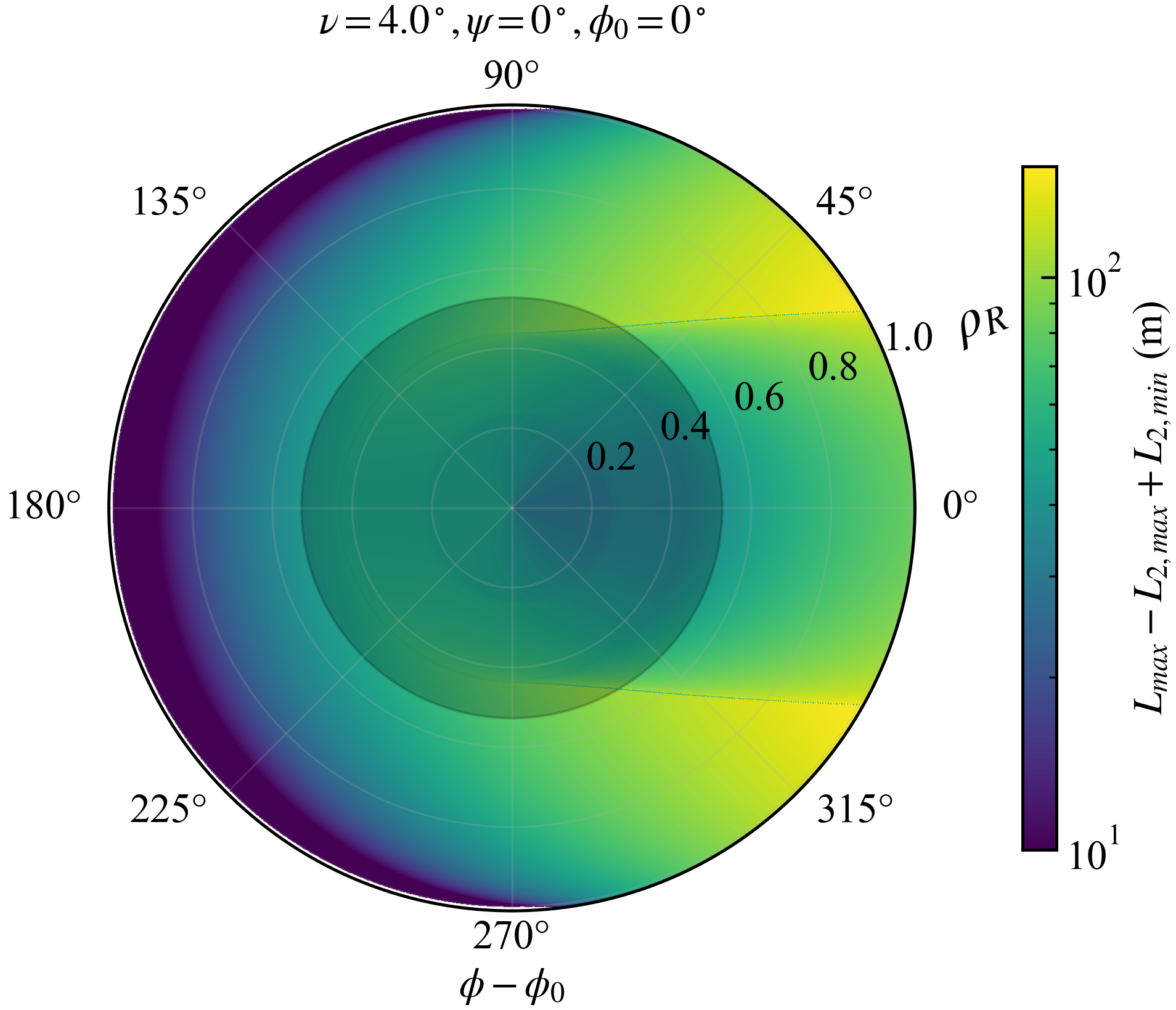}
& \includegraphics[width=0.28\textwidth]{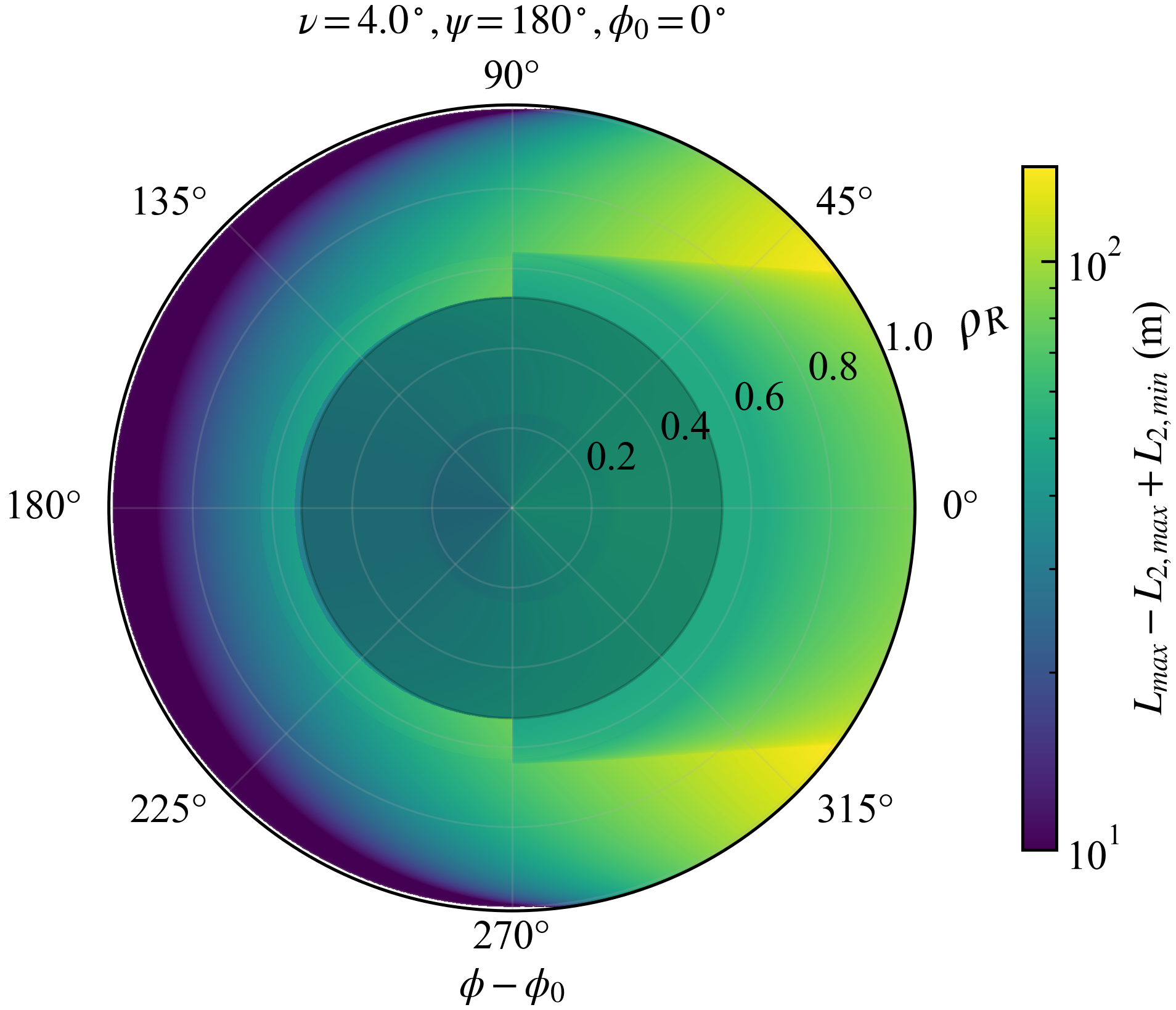} \\
\rotatebox[origin=c]{90}{\small Shadowed rel. to unshadowed}
& \includegraphics[width=0.28\textwidth]{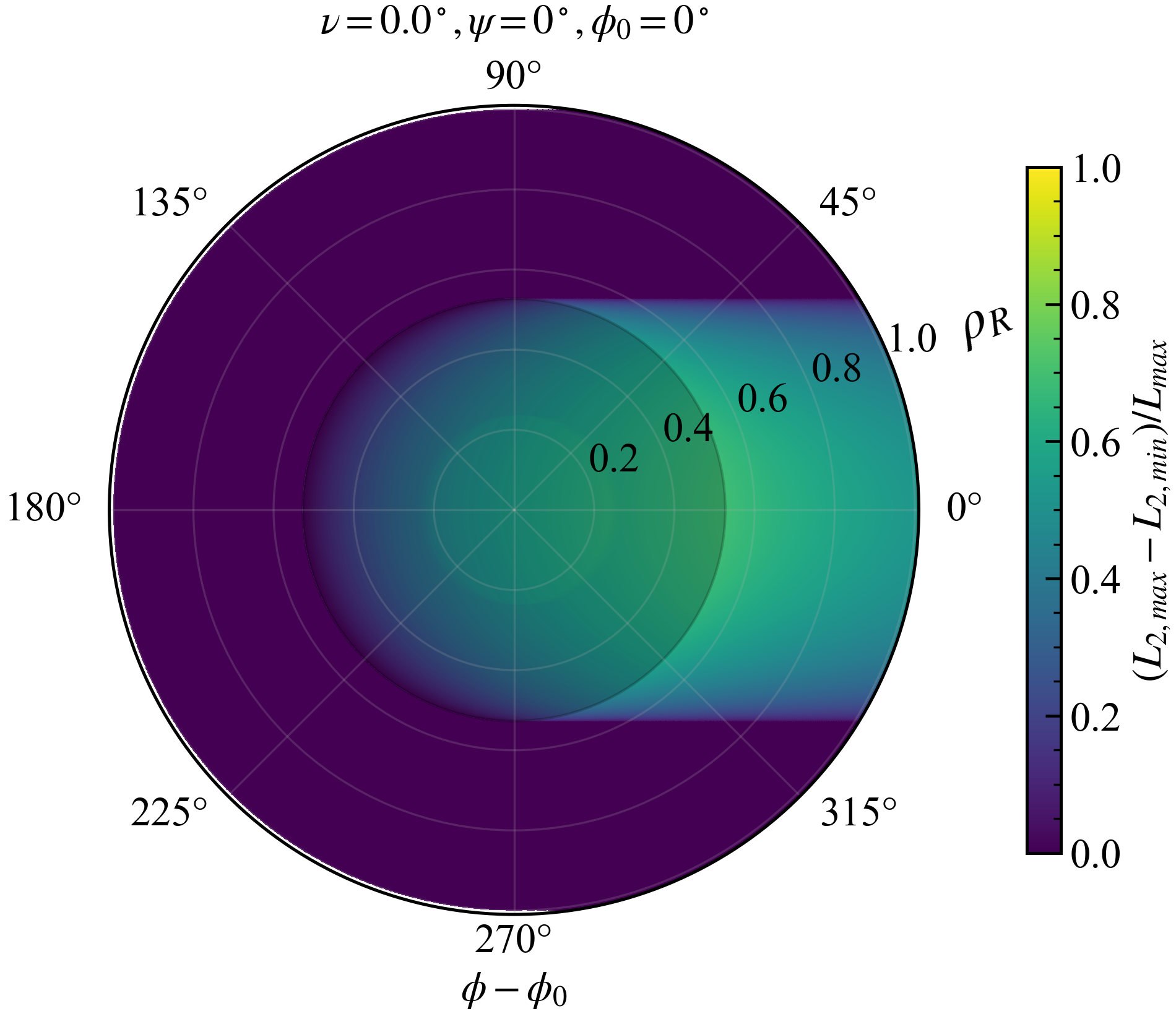}
& \includegraphics[width=0.28\textwidth]{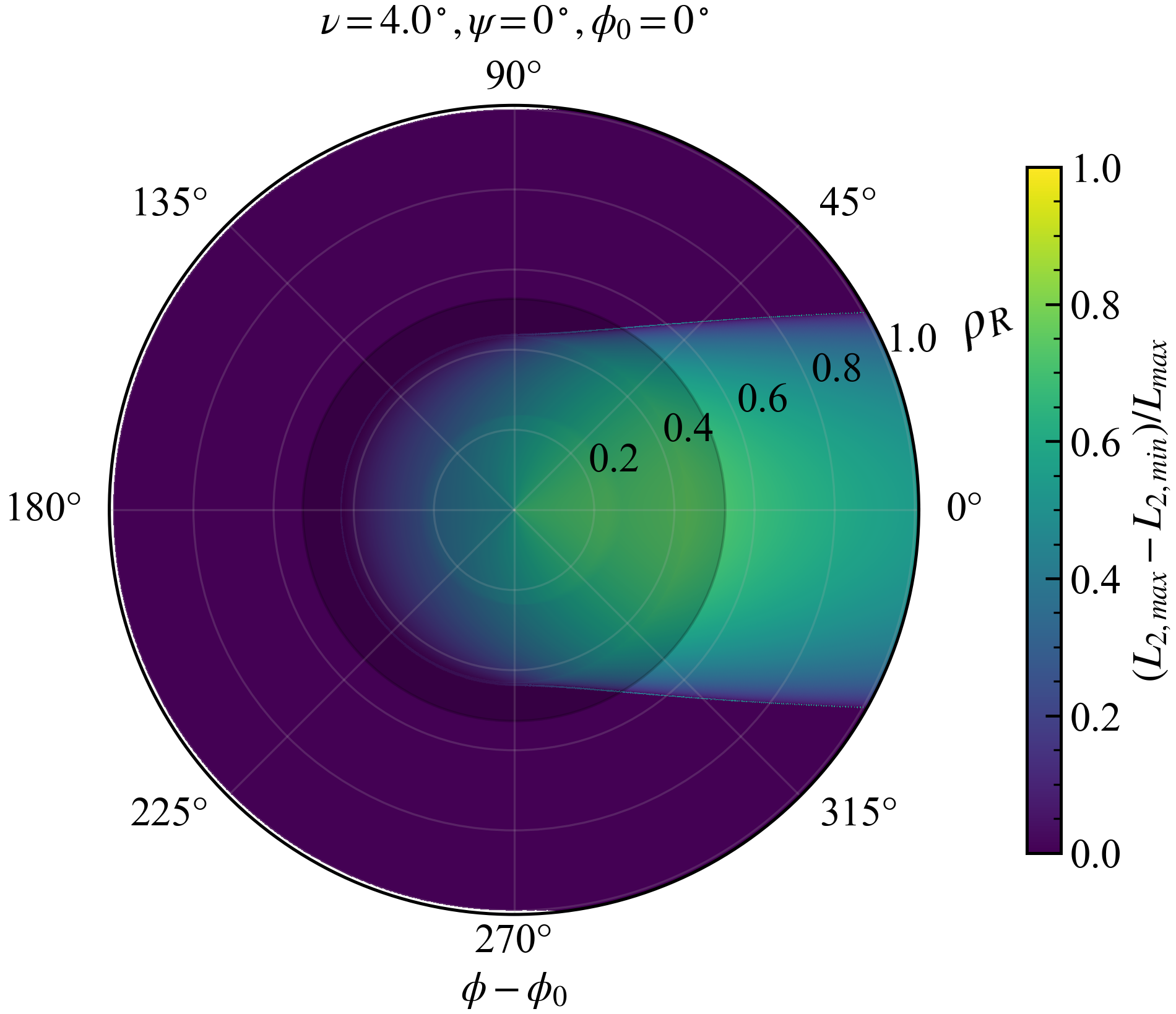}
& \includegraphics[width=0.28\textwidth]{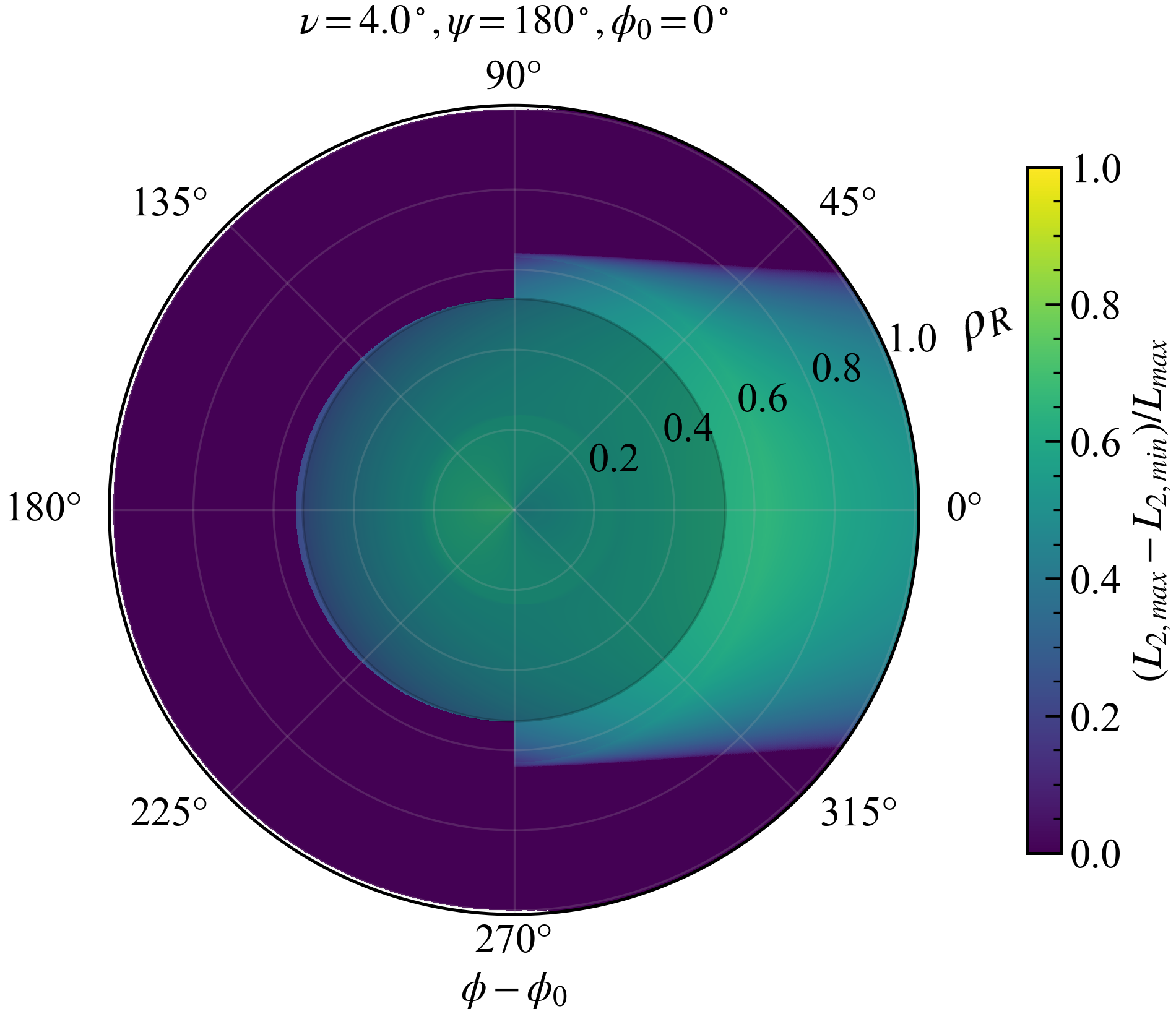} \\
\rotatebox[origin=c]{90}{\small Shadow rel. to Vacanti}
& \includegraphics[width=0.28\textwidth]{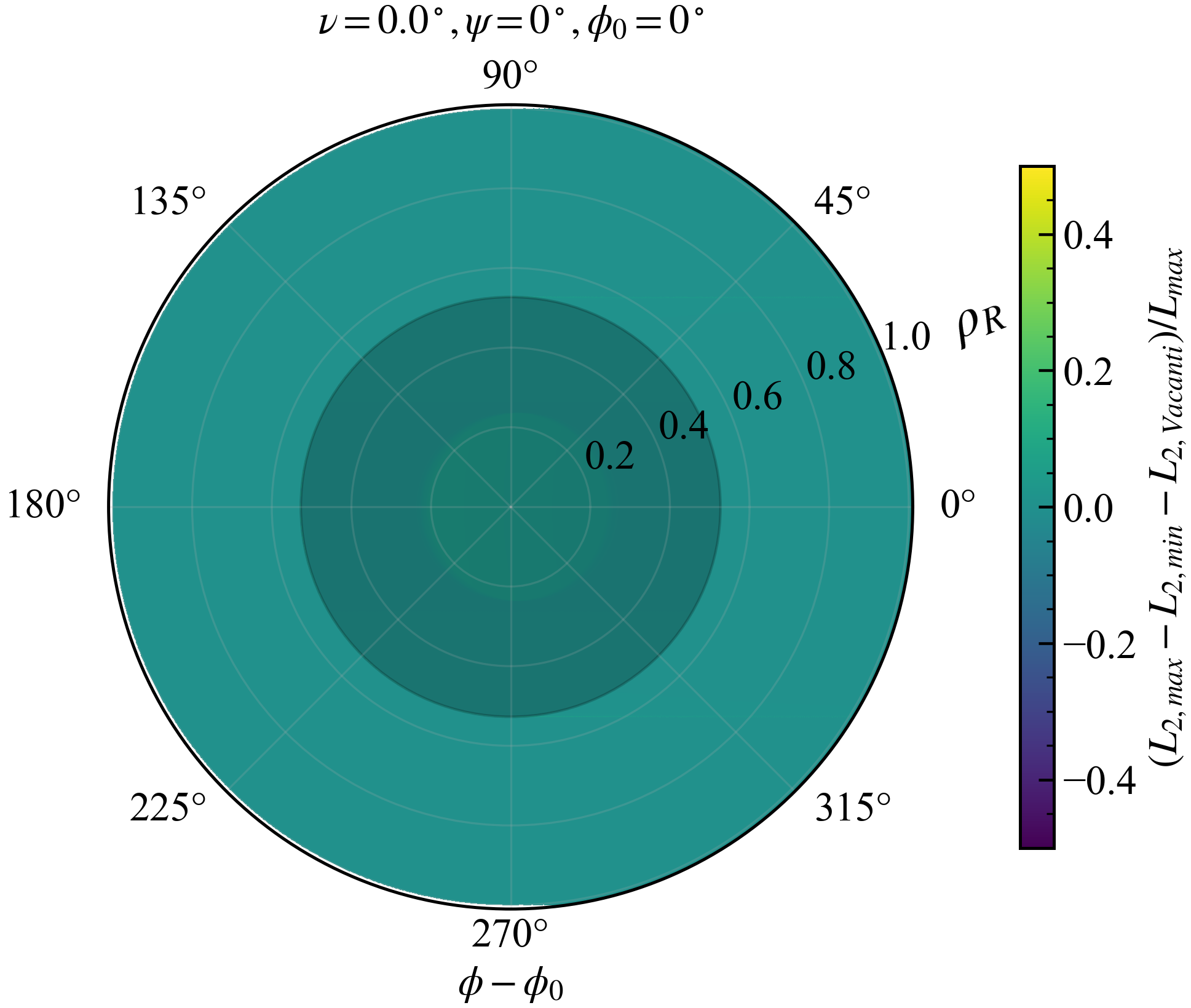}
& \includegraphics[width=0.28\textwidth]{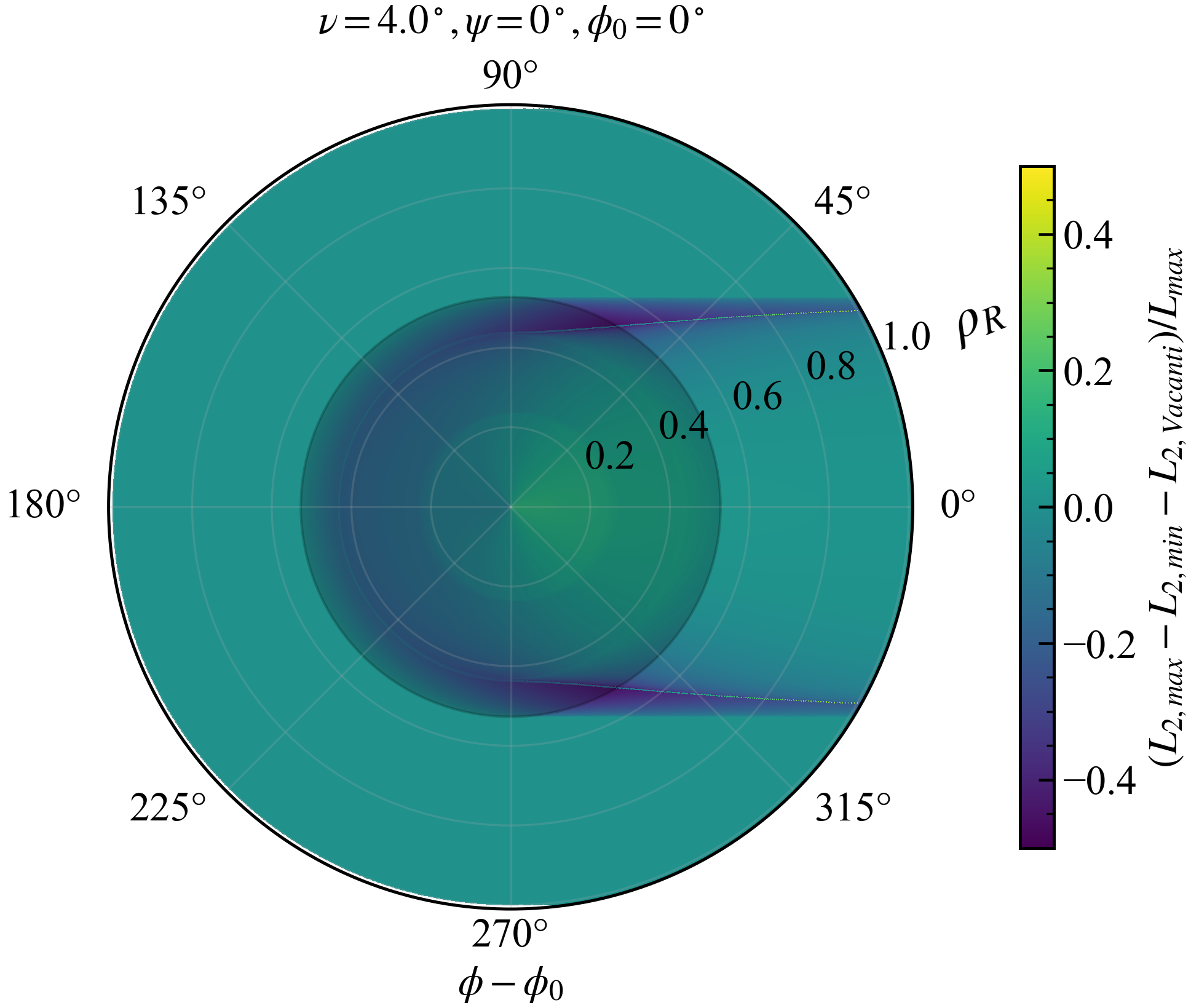}
& \includegraphics[width=0.28\textwidth]{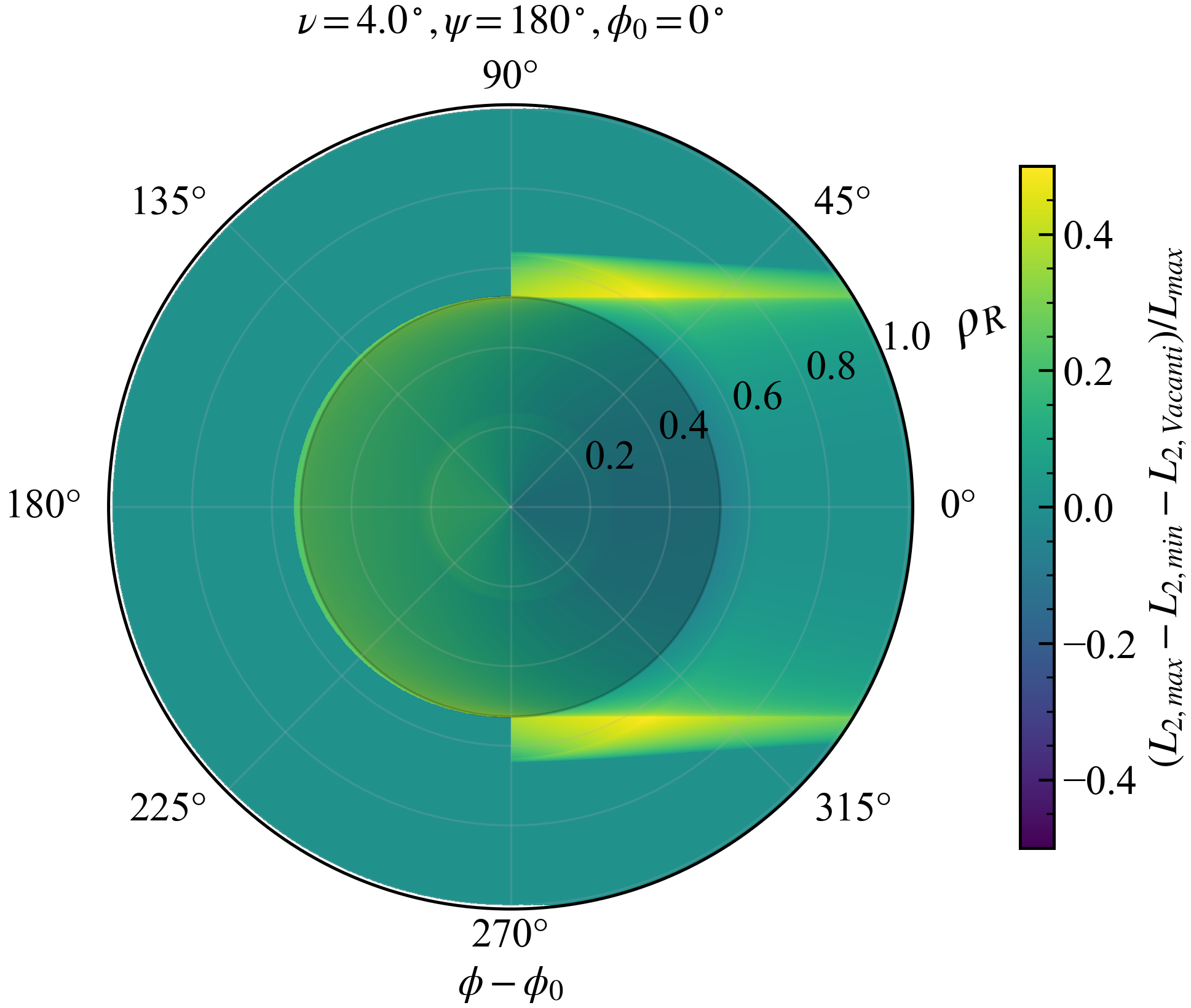}\\
\end{tabularx}
\caption{Shadow parameters shown for muons with different normalized impact distances $\rho_R$ and impact angles $\phi_0$, for the parameters of an SST. See Figure~\ref{fig:L2SCT} for details. 
}
\label{fig:L2SST}
\end{figure}

The combined minimum shadow point, $L_{2,\mathrm{min}}$, of photon emission along the muon track at angle $\phi$ can then be determined by combining Eqs.~\ref{eq:L2min} through~\ref{eq:Lminhole2}, as summarized  in the  flowchart in Fig.~\ref{fig:flow}. 

Its behaviour is shown in the second rows of  Figs.~\ref{fig:L2SCT} and~\ref{fig:L2SST} for an SCT and an SST, respectively, where the effect of the central mirror hole is clearly visible, along with a region of constant $L_{2,\mathrm{min}}$ around M2, corresponding to muons traversing the mirror. 

The third rows of Figs.~\ref{fig:L2SCT} and~\ref{fig:L2SST} show the behavior of the total imaged muon track length $L_{\mathrm{max}}-L_{2,\mathrm{max}}+L_{2,\mathrm{min}}$, including shadowing. This representation most clearly highlights the effect of muons traversing the baffles, particularly in the SCT case for an inclined muon with $\psi=0^\circ$. 

The fourth rows of Figs.~\ref{fig:L2SCT} and~\ref{fig:L2SST} show the relative shadow fraction with respect to the unshadowed image. The SCT reaches shadow fractions exceeding 
80\%, whereas the SST exceeds 75\%. 

Perhaps the most relevant results for this study are represented in the last rows of Figs.~\ref{fig:L2SCT} and~\ref{fig:L2SST}, which highlight the relative deviation of our solutions with respect to the previous algorithms based on the use of the solution of~\citet{vacanti}. In those approaches, the shadow was treated as if M2 were a primary mirror located at $z=0$, and its Cherenkov light contribution was subtracted from the image formed by the actual primary mirror. As expected, the largest deviations are found for inclined muons -- whose inclinations were not accounted for in previous algorithms -- reaching  differences of up to $\pm 40\%$. Depending on the inclination projection angle  $\psi$, either positive or negative relative deviations may dominate.

\clearpage
\section{Conclusions}

In this work we have derived, for the first time, a complete analytical description of the Cherenkov light produced by atmospheric muons and detected by a dual–mirror Imaging Atmospheric Cherenkov Telescope. The formalism is based on a vector–geometry approach combined with symbolic algebra techniques that allow the systematic expansion of the resulting expressions to the relevant orders in the Cherenkov angle, muon inclination, and mirror curvature in order to achieve a precision of better than 1\%. 

Starting from the general photon emission geometry, we obtained analytical solutions for the photon impact coordinates on the primary mirror and the focal plane. The method naturally reproduces previously known results for single–mirror telescopes, providing an important cross–check of the derivation. In particular, the well–known solution of \citet{vacanti} for flat mirrors is recovered when mirror curvature and inclination effects are neglected. To test the validity of our first-order corrections, we have been able to derive the analytical expectation of $3^\mathrm{rd}$-order Seidel coma aberration from a parabolic mirror and a flat camera. The formalism was then applied to evaluate the influence of coma aberrations on the reconstructed muon ring parameters. 
The resulting bias in the ring radius is generally small once standard plate–scale corrections are applied, but can reach several percent for extreme impact parameters. Averaged over realistic distributions of muon impact points, the effect is expected to remain at the level of $\mathcal{O}(1\%)$.
 Both validations demonstrate the internal consistency of the approach and its compatibility with existing muon calibration formalisms~\citet{GaugMuons:2019}.

Extending the analysis to dual–mirror telescope geometries introduces additional effects due to the shadowing produced by the secondary mirror, its supporting structure, and central hole in the primary mirror. We derived explicit analytical expressions describing the shadowing conditions and the corresponding limits on the photon emission heights for  arbitrary muon inclination, impact parameter, and photon emission angle. The resulting formalism provides a unified framework to determine the Cherenkov light yield recorded along the muon ring while accounting for all relevant geometric obstructions of the optical system. 
In particular, the formalism includes shadowing of Cherenkov light by secondary mirror baffles under various muon incidence angles, the emission of unshadowed Cherenkov light below the secondary mirror and if the muon crosses
 the protecting baffles. 

The derived expressions have been illustrated for representative parameters of CTAO dual–mirror telescopes. The analysis shows that the correctly simulated shadowing effects can significantly modify the amount of detected Cherenkov light depending on the muon impact distance, inclination, and azimuthal orientation. If compared with previous algorithms that treat M2 as a large hole in primary mirror, our solutions deviate by up to $\pm 40$\% for inclined muons, which have not been taken into account by previous studies. 

The formalism therefore enables a much more accurate prediction of the expected light distribution along muon rings in these instruments and will be directly applicable to the calibration of CTAO's Schwarzschild–Couder telescopes.  More generally, this work demonstrates that the complexity of dual–mirror systems can be treated analytically with sufficient precision for calibration purposes.

All formulae derived in this article have been implemented into suitable python scripts and libraries.  The Jupyter Notebook to execute the analysis in this paper and the python scripts to generate the figures in this paper are hosted at \url{https://github.com/mgaug/Muons_DualMirror}. 


\appendix
\section{Solution for a square camera shadow}

We introduce a plane located at $z=D_\textit{sq}$, where a square camera is positioned at a distance $D_\textit{sq}$ from the pole of the primary mirror. The Cherenkov photons intersect this plane at coordinates $x_\textit{sq}, y_\textit{sq},z=D_\textit{sq}$, obtained by solving  Eq.~\ref{eq:solver2}  in an analogous manner. This yields: 
\begin{subequations}
\begin{align}
x_\textit{sq} &=  \left(l-D_\textit{sq}\right) \cdot\theta_{c} \cdot\cos\left(\phi\right) - \rho \cdot\cos\left(\phi_{0}\right) - D_\textit{sq}\cdot\nu\cdot \cos(\psi) 
+   O(c\cdot\nu,\nu^2,c\cdot\theta_c)  \quad, \label{eq:xsq}\\
y_\textit{sq} &= \left(l-D_\textit{sq}\right) \cdot\theta_{c} \cdot\sin\left(\phi\right) - \rho \cdot\sin\left(\phi_{0}\right) - D_\textit{sq}\cdot\nu\cdot \sin(\psi) 
+  O(c\cdot\nu,\nu^2,c\cdot\theta_c) \quad, \label{eq:ysq}  
\end{align}
\end{subequations}
Imposing the condition that the Cherenkov light are obstructed by a square camera with side length $2\,A$,  $|x_\textit{sq}| \leq A, \quad |y_\textit{sq}| \leq A$,  
defines two intervals in $l$,   $L_\textit{x,min} \leq l \leq L_\textit{x,max}, \quad   
 L_\textit{y,min} \leq l \leq L_\textit{y,max}$, where 
\begin{subequations}
    \begin{align}
L_{x,0} &= D_\textit{sq} + \frac{D_\textit{sq}\cdot\nu\cdot\cos(\psi) + \rho\cdot\cos(\phi_0)}{\theta_c\cdot\cos(\phi)} \\
L_{y,0} &= D_\textit{sq} + \frac{D_\textit{sq}\cdot\nu\cdot\sin(\psi) + \rho\cdot\sin(\phi_0)}{\theta_c\cdot\sin(\phi)} 
\end{align}
\end{subequations}
and
\begin{subequations}
    \begin{align}
L_\textit{x,min} &= L_{x,0} - \frac{A}{\theta_c\cdot|\cos(\phi)|} \\
L_\textit{x,max} &= L_{x,0} + \frac{A}{\theta_c\cdot|\cos(\phi)|} \\
L_\textit{y,min} &= L_{y,0} - \frac{A}{\theta_c\cdot|\sin(\phi)|} \\
L_\textit{y,max} &= L_{y,0} + \frac{A}{\theta_c\cdot|\sin(\phi)|}
\end{align}
\end{subequations}
\noindent
A shadow is produced whenever the two intervals defined by the $x$ and $y$-constraints overlap, that is: $\max(L_\textit{x,min},L_\textit{y,min})\leq \min(L_\textit{x,max},L_\textit{y,max})$. 
This condition can be written as 
\begin{equation}
\left| \frac{D_\textit{sq}\cdot\nu\cdot\cos(\psi) + \rho\cdot\cos(\phi_0)}{\cos\phi} - \frac{D_\textit{sq}\cdot\nu\cdot\sin(\psi) + \rho\cdot\sin(\phi_0)}{\sin\phi}   \right| 
\leq A  \cdot \left( \frac{1}{|\cos(\phi)|} + \frac{1}{|\sin(\phi)|}\right) \quad,
\end{equation}
which, after multiplying both sides by $|\sin(\phi)\cdot\cos(\phi)|$ reduces to 
\begin{equation}
\big|\rho\cdot\sin(\phi-\phi_0) + D_\textit{sq} \cdot \nu \cdot \sin(\phi-\psi)\big| \leq A \cdot (|\sin\phi| + |\cos\phi|) \quad.  \label{eq:square_shadow_2}
\end{equation} 
Eq.~\ref{eq:square_shadow_2} defines the square camera shadow condition in terms of $\phi$.
Note that the right-hand side $A \cdot (|\sin\phi| + |\cos\phi|)$ is the support function of a square of half-side $A$. By comparison, for the M2 obstruction the corresponding support function is simply the constant radius $R_\textit{sb}$.

Finally, we consider the condition that the muon impact point $\mathbf{I}_{\mu}$ lies within the projection of the square camera onto the primary mirror plane.
The projected displacement of the muon to distance $D_\textit{sq}$ is $(D_\textit{sq}\cdot\nu\cdot\cos(\psi),D_\textit{sq}\cdot\nu\cdot\sin(\psi))$.
Consequently, a square camera centered at the origin in the plane 
$z=D_\textit{sq}$ and defined by $|x_\textit{sq}|\leq A,|y_\textit{sq}|\leq A$,  projects onto  the primary mirror as the translated square $|x+D_\textit{sq}\cdot\nu\cdot\cos(\psi)|\leq A,|y+D_\textit{sq}\cdot\nu\cdot\sin(\psi)|\leq A$.

Introducing the projected position vector 
\begin{equation}
\quad \boldsymbol{R}_{\textit{sq},\mathrm{proj}} := 
\begin{pmatrix}
\rho\cdot\cos(\phi_0)-D_\textit{sq}\cdot\nu\cdot\cos(\psi)\\
\rho\cdot\sin(\phi_0)-D_\textit{sq}\cdot\nu\cdot\sin(\psi)
\end{pmatrix} \quad, 
\end{equation}
the shadow interval is given by 
\begin{align}
 \mathrm{\quad for\quad} & (|R_{\textit{sq},\mathrm{proj},x}| \leq A \mathrm{~and~} 
 |R_{\textit{sq},\mathrm{proj},y}| \leq A) \mathrm{\quad or\quad} |\phi-\phi_0| < \pi/2 ~\mathrm{:} \nonumber\\[2mm]
& \big|\rho\cdot\sin(\phi-\phi_0) + D_\textit{sq} \cdot \nu \cdot \sin(\phi-\psi)\big| \leq A \cdot (\big|\sin\phi\big| + \big|\cos\phi\big|)
\quad. 
\end{align}
Under these conditions, 
\begin{align}
L_\mathrm{2,min} &= \max(L_\textit{x,min},L_\textit{y,min})  \\
L_\mathrm{2,max} &= \min(L_\textit{x,max},L_\textit{y,max})  
\end{align}

\acknowledgments

This project has received funding from the Spanish grant PID2022-139117NB-C43, funded by \\ MCIN/AEI/10.13039/501100011033/FEDER, UE, and the~Departament de Recerca i Universitats de la Generalitat de Catalunya (grant SGR2021 00607).

\clearpage

\bibliographystyle{elsarticle-harv_srt}
\bibliography{ccf}

\end{document}